\documentclass[sn-nature,pdflatex]{sn-jnl}

\usepackage{graphicx}%
\graphicspath{{figures/}}
\usepackage{multirow}%
\usepackage{amsmath,amssymb,amsfonts}%
\usepackage{amsthm}%
\usepackage{mathrsfs}%
\usepackage[title]{appendix}%
\usepackage{xcolor}%
\usepackage{textcomp}%
\usepackage{manyfoot}%
\usepackage{booktabs}%
\usepackage{algorithm}%
\usepackage{algorithmicx}%
\usepackage{algpseudocode}%
\usepackage{listings}%
\usepackage{xr}

\theoremstyle{thmstyleone}%
\theoremstyle{thmstyletwo}%

\theoremstyle{thmstylethree}%

\begin{document}

\title[Understanding Structural Representation in Foundation Models for Polymers]{Understanding Structural Representation in Foundation Models for Polymers}


\author*[1]{\fnm{Nathaniel H.} \sur{Park}}\email{npark@us.ibm.com}

\author[2]{\fnm{Eduardo} \sur{Soares}}

\author[2]{\fnm{Victor} \sur{Shirasuna}}

\author[1,3]{\fnm{Tiffany J.} \sur{Callahan}}

\author[1,3]{\fnm{Sara} \sur{Capponi}}

\author[2]{\fnm{Emilio Vital} \sur{Brazil}}

\affil[1]{\orgname{IBM Research Almaden}, \orgaddress{\street{650 Harry Rd.}, \city{San Jose}, \postcode{95120}, \state{CA}, \country{USA}}}

\affil[2]{\orgname{IBM Research Brazil}, \orgaddress{\city{Rio de Janeiro}, \postcode{10587}, \state{RJ}, \country{Brazil}}}

\affil[3]{\orgname{Center for Cellular Construction}, \orgaddress{\city{San Francisco}, \postcode{94158}, \state{CA}, \country{USA}}}

\abstract{From the relative scarcity of training data to the lack of standardized benchmarks, the creation of effective foundation models for polymers faces significant and multi-faceted challenges. At the core, many of these issues are tied directly to the structural representation of polymers. Here, we present a chemical language foundation model built on using a SMILES-based polymer graph representation (CPG) that incorporates polymer architectural features and connectivity that are often missing in other line notations. This foundation model exhibited excellent performance on 30 different polymer property benchmark datasets. Critical evaluation of the developed representation against other variations in control experiments reveals this approach to be a robust method of representing polymers in language-based foundation models. These experiments also reveal a strong invariance of structural representations to small perturbations, with many variations of structural representation exceeding or equaling state-of-the-art (SOTA) performance. Surprisingly, SMILES representations which are chemically or semantically invalid also provided near or SOTA performance in several instances--underscoring an unexamined blind spot in the development of chemistry language models. Examination of error sources and attention maps for the evaluated structural representations corroborate the findings of the control experiments, highlighting the ability of the model to interpolate SMILES sequence space in a manner that is loosely congruent to chemical and architectural space for polymers. Overall, this work highlights the surprising robustness of chemistry language models to structural representation perturbations and identifies the conditions under which CPG representation provides meaningful advantages.} 

\keywords{Polymers, Deep-Learning, Foundation Models}



\maketitle

\section{Introduction}\label{sec1}

Foundational artificial intelligence (AI) models hold immense promise for revolutionizing rational design of polymeric materials owing to their ability to interpolate over large regions of chemical space, providing considerable predictive and generative capabilities for a variety of downstream tasks. Despite numerous reports on the development of predictive and generative models for polymers\cite{gaoMLRev24,liAPLRev23,sustainableNatRev24,conjugatedRev24}, creation of foundation models for polymers still face major and frequently intractable challenges. Data for training and benchmarking deep-learning models for polymers is scarce\cite{zhaoDataRev25}, restrictively licensed\cite{polyinfo,polymerGenome,cript,bioplastic22}, and comes from only a handful of sources (Fig.~\ref{cpg_rep}c)—limiting both model development and their use in commercial applications. Only recently have larger synthetic datasets such as OPoly26 and PolyOmics become available for use\cite{oPoly2026,polyOmics2025}. Still, most existing datasets are often incomplete and lack many critical structural descriptors\cite{polyBERT,tranJAP2020,volgin2022,pi1M,xuPOINT2025}, such as dispersity, number-average molecular weight, or processing conditions that play important roles in determining polymer properties. Structural representations in these datasets are overwhelmingly based on variations of SMILES\cite{tranJAP2020,volgin2022,pi1M,xuPOINT2025}, which are featurized for consumption in machine learning models using several methods, including fingerprinting\cite{polymerGenome,taoCopolymerML2022,loRearrange2023}, conversion to graphs\cite{aldeghiGraph2022,antonGraph2022,gurnaniPolymerMGNN2023,bradfordConduct2023,vogelInverseDesignCopolymers2025}, tokenization\cite{polyBERT,transPolymer,mmPolymer,vogelInverseDesignCopolymers2025,zhangTransferringMolecularFoundation2023,polyBART2025,vogelInverseDesignCopolymers2025}, coarse graining\cite{doCGBrush2023,webbTargetedSequenceDesign2020}, image conversion\cite{qiuExpBigSMILES2025}, 3D conformations\cite{mmPolymer}, and others depending on the model architecture\cite{uniPoly2025}.  As a consequence of both data scarcity and narrow range of structural information, most of the reported predictive or generative model for polymers seem highly similar, as they are effectively trained on the same data and use nearly equivalent methods of polymer featurization. Yet, the lack of standardized benchmark datasets prevents disambiguation of the differences in performance as they relate to model architecture, structural representation, and training strategies--limiting rational design of improved models.

The core of many issues surrounding foundation models for polymers revolves around structural and architectural representations which dictate the quality of the latent representation used in downstream tasks. The inflexibility of SMILES to accurately represent critical architectural and compositional features of polymers has prompted development various extensions of SMILES to circumnavigate this issue, including modification of the SMILES grammar\cite{bigSMILES2019,bigSMInC2022,gBigSMI2024,qiuExpBigSMILES2025}, adjusting the connectivity of the SMILES string representing the repeat unit\cite{yuRingRepeatingUnit2023}, concatenation of additional text information to SMILES\cite{transPolymer,aldeghiGraph2022,vogelInverseDesignCopolymers2025}, or the conversion of SMILES to SELFIES\cite{polyBART2025}. Each approach has reported some success in terms of downstream model performance on selected benchmarks. Benchmark performance is an important indicator of the validity of a modeling approach, however, it alone is not enough to determine the effectiveness of a particular structural representation absent effective control experiments. Thus, in order to critically evaluate the influence of polymer structural representation, we surmised that a minimal extension to the SMILES grammar would allow for both encoding of common polymer architectures and facilitate introspection of possible underlying mechanisms accounting for any observed change in performance—providing valuable insight to further guide design of structural representations for polymers.

Previously, we have reported the use of a domain-specific programming language, the Chemical Markdown Language (CMDL), whose syntax represents polymers natively as graphs which can in turn be compiled to line notations for use in regression transformer models built for polymer design\cite{parkCMDL2023}. Although this approach was experimentally validated, the serialized output of the polymer graph representation was not easily interoperable with existing datasets that are overwhelmingly denoted in SMILES or PSMILES and necessitated a custom tokenizer\cite{parkCMDL2023}. However, the advantage of using a domain-specific language such as CMDL resides in its underlying mechanism for representing polymers as graphs of core structural components and flexibility to be compiled to a variety of serialized representations. We hypothesized that the CMDL polymer graph (CPG) representation could readily be applied to existing polymer datasets as a small extension of the SMILES grammar, allowing transmission of salient architectural information to the model affording an increase in performance on benchmark datasets.

\section{Results}\label{sec2}

\subsection{Polymer Representation}
\begin{figure}[H]
\centering
\includegraphics[width=\textwidth]{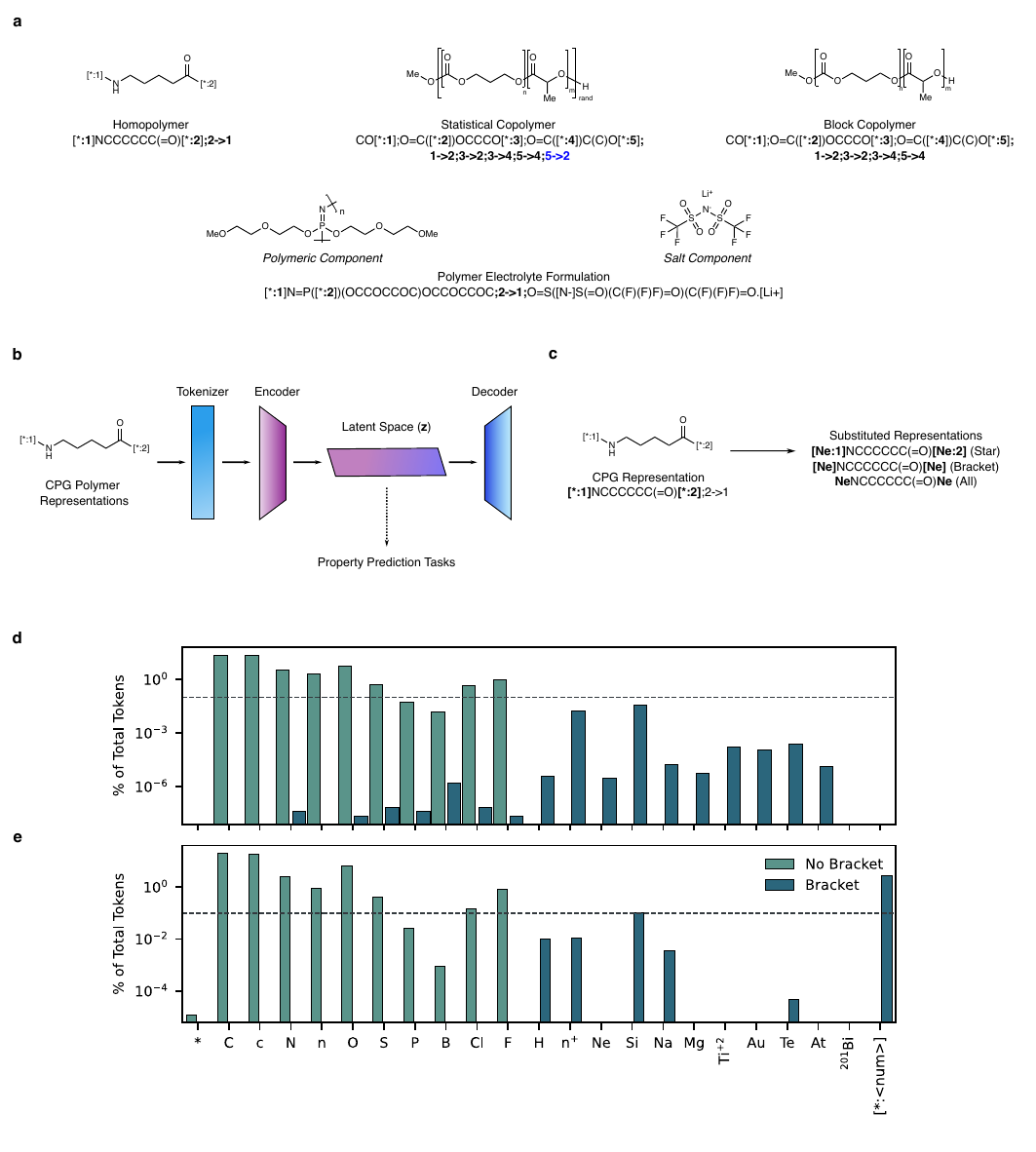}
\caption{\textbf{a}. Example structural representations using the CPG notation for a homopolymer, statistical copolymer, block copolymer, and polymer formulation. Bold text indicates additions from CPG relative to a typical PSMILES string; bold blue text indicates key edge difference between statistical and block copolymer representations. ~\textbf{b}. General model architecture for SMI-TED and SMI-TED-Polymer models.~\textbf{c}. Examples of atom substituted structural representations where a CPG representation is modified with a Ne atom.~\textbf{d}. Distribution of percentages of total SMI-TED pre-training tokens corresponding to atoms used in the atom substituted representations.~\textbf{e} Distribution of percentages of total SMI-TED-Polymer pre-training tokens corresponding to atoms used in the atom substituted representations. Dashed line for both \textbf{d} and \textbf{e} indicates 1\% of total tokens.}\label{cpg_rep}
\end{figure}

The core CPG representation is constructed on two principles: specifying the interconnections between core structural fragments of the polymer backbone and reduction of complexity via architectural symmetry. Each of the core polymer structural components was represented via a SMILES string containing non-atomic placeholders characters (Q, R) in brackets and architectural symmetry was indicated via weights attached to edges between different structural components\cite{parkCMDL2023}. Most polymer datasets denoted in PSMILES consist of relatively simple homo-, random, or block copolymers, with little to no architectural symmetry present\cite{volgin2022,tranJAP2020,yangGasSep2022,aldeghiGraph2022}. Thus, we felt that the core features of the CPG representation could easily be used in such datasets with some minor adaptations. Rather than using non-atomic placeholder characters, numbered asterisks in brackets—which are part of the base SMILES grammar—could be used to map the connections between different components or within a component, with the edge definitions appended to the end of the SMILES string (Fig.~\ref{cpg_rep}a). Analogous representations have been used with different models for polymer property prediction, although these either were converted to a different internal representation prior to consumption by the model\cite{aldeghiGraph2022} or were tokenized directly including floating-point numbers for stoichiometry and dispersity\cite{vogelInverseDesignCopolymers2025}.

The advantage of this approach to modeling polymer structure is that it inherently facilitates straightforward extension to cover a range of polymer architecture types, end groups, as well as blends or formulations. Importantly, this approach can readily distinguish between random and block copolymers via delineation of different edge connections (Fig.~\ref{cpg_rep}a). Such copolymers would appear identical to a chemistry language model using solely PSMILES representations, leading potentially erroneous associations between the copolymer structure and properties. Additionally, this representation approach can accommodate mixtures of polymers or formulations of polymers and small-molecules, such as found in many polymer electrolyte datasets (Fig.~\ref{cpg_rep}).

\subsection{Experiments}

\subsubsection{SMI-TED-Polymer Evaluation}
Training the SMI-TED-Polymer model followed an analogous protocol as the base SMI-TED model,\cite{smiTED2025} albeit using the assembled pre-training dataset of polymers in the CPG representation (see Methods). To evaluate the latent space generated by our methodology, we tested the SMI-TED-Polymer model on a suite of property prediction tasks on 30 different datasets. For each benchmark dataset, care was taken to not mix datasets for the same property but from different sources to both minimize error propagation and facilitate comparison with previous work. For the assessment, we adopted a randomized 80:10:10 train/validation/test splits for all tasks (see Methods). The property prediction tasks fell across four principal categories: electronic, physical, optical, and gas barrier properties (Table~\ref{tab1}). In each of the categories, the SMI–TED-Polymer model achieved or equaled state-of-the-art (SOTA) on given metrics (entries 2, 4, 11–19, 21, 24, and 29-30, Table~\ref{tab1}). In instances where the SMI-TED-Polymer model did not improve upon the SOTA values, it frequently provided results very close to them with only the copolymer electron affinity and ionization potential showing significant deviations (entries 6 and 7, Table~\ref{tab1}). Exceptions to this were the atomization energy (E\textsubscript{at}, entry 5) and the copolymer electron affinity and ionization potential (EA and IP, entries 7 and 8), where the SMI-TED-Polymer model performed 2-5x worse than the SOTA results. For the EA and IP datasets, the decrease in performance is likely a result of the model being unable to differentiate between copolymers of based on CPG representations alone without accounting for the relative repeat unit stoichiometry.\cite{aldeghiGraph2022} With the E\textsubscript{at} dataset, our results are analogous to what has been reported for some other models with transformer-based architectures.\cite{polyBERT,mmPolymer}

\begin{table}[tbp]
\caption{SMI-TED-Polymer Benchmark Experiments}\label{tab1}%
\begin{tabular}{@{}lllllcc@{}}
\toprule
Entry & Dataset\footnotemark[1] & Data Type\footnotemark[2] & Property Type & Metric & SOTA\footnotemark[3] & STP\footnotemark[4] \\
\midrule
1    & Chain Bandgap (E\textsubscript{gc})\cite{kuennethPolymerInformaticsMultitask2021} & Comp.  & Electronic &  RMSE ($\downarrow$) & \textbf{0.43}\cite{mmPolymer} & 0.49  \\
2    & Bulk Bandgap (E\textsubscript{gb})\cite{kuennethPolymerInformaticsMultitask2021} & Comp.  & Electronic &  RMSE ($\downarrow$) & 0.485\cite{shenMolecularTopologicalDeep2026} & \textbf{0.32}  \\
3    & Electron Affinity (E\textsubscript{ea})\cite{kuennethPolymerInformaticsMultitask2021} & Comp.  & Electronic &  RMSE ($\downarrow$) & \textbf{0.263}\cite{shenMolecularTopologicalDeep2026} & 0.29  \\
4    & Ionization Energy (E\textsubscript{i})\cite{kuennethPolymerInformaticsMultitask2021} & Comp. & Electronic &  RMSE ($\downarrow$) & 0.39\cite{mmPolymer} & \textbf{0.37}  \\
5    & Atomization Energy (E\textsubscript{at})\cite{kuennethPolymerInformaticsMultitask2021} & Comp. & Electronic & RMSE ($\downarrow$) & \textbf{0.050}\cite{zhangTransferringMolecularFoundation2023} & 0.10  \\
6    & Dielectric Constant (EPS)\cite{kuennethPolymerInformaticsMultitask2021} & Comp.  & Electronic &  RMSE ($\downarrow$) & \textbf{0.35}\cite{shenMolecularTopologicalDeep2026} & 0.39 \\
7    & Copolymer Electron Affinity (EA)\cite{aldeghiGraph2022} & Comp.  & Electronic &  RMSE ($\downarrow$) & \textbf{0.03}\cite{aldeghiGraph2022} & 0.15  \\
8    & Copolymer Ionization Potential (IP)\cite{aldeghiGraph2022} & Comp.  & Electronic &  RMSE ($\downarrow$) & \textbf{0.03}\cite{aldeghiGraph2022} & 0.16  \\
9    & Crystallization Tendency (X\textsubscript{c})\cite{kuennethPolymerInformaticsMultitask2021} & Comp.  & Physical &  RMSE ($\downarrow$) & \textbf{15.86}\cite{shenMolecularTopologicalDeep2026} & 17.82  \\
10    & Refractive Index–I (N\textsubscript{c})\cite{kuennethPolymerInformaticsMultitask2021} & Exp.  & Optical &  RMSE ($\downarrow$) & \textbf{0.07}\cite{shenMolecularTopologicalDeep2026} & 0.12  \\
11    & Refractive Index–II\cite{hataUsingGPT4Parameter2023} & Exp.  & Optical &  RMSE ($\downarrow$) & 0.031\cite{hataUsingGPT4Parameter2023} &  \textbf{0.021} \\
12    & Conductivity–I\cite{bradfordConduct2023} & Exp.  & Physical &  MAE ($\downarrow$) & 1.00\cite{bradfordConduct2023} &  \textbf{0.89} \\
13    & Conductivity–II\cite{schauserDB2021} & Exp.  & Physical &  RMSE ($\downarrow$) & \textbf{0.61}\cite{transPolymer} &  \textbf{0.61} \\
14    & CO\textsubscript{2} Permeability (P\textsubscript{CO$_2$})\cite{tiwariCreationPolymerDatasets2024} & Exp.  & Barrier &  MAE ($\downarrow$) & \textbf{0.29}\cite{tiwariCreationPolymerDatasets2024} &  \textbf{0.29} \\
15    & CH\textsubscript{4} Permeability (P\textsubscript{CH$_4$})\cite{tiwariCreationPolymerDatasets2024} & Exp.  & Barrier &  MAE ($\downarrow$) & 0.37\cite{tiwariCreationPolymerDatasets2024} &  \textbf{0.35} \\
16    & N\textsubscript{2} Permeability (P\textsubscript{N$_2$})\cite{tiwariCreationPolymerDatasets2024} & Exp.  & Barrier &  MAE ($\downarrow$) & 0.38\cite{tiwariCreationPolymerDatasets2024} &  \textbf{0.31} \\
17    & CO\textsubscript{2}:CH\textsubscript{4} Selectivity (S\textsubscript{CH$_4$})\cite{tiwariCreationPolymerDatasets2024} & Exp.  & Barrier &  MAE ($\downarrow$) & 5.34\cite{tiwariCreationPolymerDatasets2024} &  \textbf{4.71} \\
18    & CO\textsubscript{2}:N\textsubscript{2} Selectivity (S\textsubscript{N$_2$})\cite{tiwariCreationPolymerDatasets2024} & Exp.  & Barrier &  MAE ($\downarrow$) & 4.14\cite{tiwariCreationPolymerDatasets2024} &  \textbf{3.89} \\
19    & CO\textsubscript{2} Permeability–II\cite{yangGasSep2022} & Exp.  & Barrier &  R\textsuperscript{2} ($\uparrow$) & 0.90\cite{yangGasSep2022} &  \textbf{0.91} \\
20    & CH\textsubscript{4} Permeability–II\cite{yangGasSep2022} & Exp.  & Barrier &  R\textsuperscript{2} ($\uparrow$) & \textbf{0.89}\cite{yangGasSep2022} &  0.85 \\
21    & N\textsubscript{2} Permeability–II\cite{yangGasSep2022} & Exp.  & Barrier &  R\textsuperscript{2} ($\uparrow$) & \textbf{0.91}\cite{yangGasSep2022} & \textbf{0.91} \\
22    & H\textsubscript{2} Permeability\cite{yangGasSep2022} & Exp.  & Barrier &  R\textsuperscript{2} ($\uparrow$) & \textbf{0.90}\cite{yangGasSep2022} &  0.87 \\
23    & O\textsubscript{2} Permeability\cite{yangGasSep2022} & Exp.  & Barrier &  R\textsuperscript{2} ($\uparrow$) & \textbf{0.92}\cite{yangGasSep2022} &  0.89 \\
24    & He Permeability\cite{yangGasSep2022} & Exp.  & Barrier &  R\textsuperscript{2} ($\uparrow$) & 0.91\cite{yangGasSep2022} &  \textbf{0.92} \\
25    & CO\textsubscript{2} Permeability–III\cite{giroAIMembrane2023} & Sim. & Barrier &  R\textsuperscript{2} ($\uparrow$) & \textbf{0.90}\cite{giroAIMembrane2023} &  0.88 \\
26    & T\textsubscript{g}–I (polyimides)\cite{longLargeScaleTg2024} & Exp. \& Syn. & Thermal &  MAE ($\downarrow$) & 24.4\cite{longLargeScaleTg2024}  &  \textbf{9.56} \\
27    & T\textsubscript{g}–II (homopolymers)\cite{huPredictionInterpretabilityGlass2023} & Exp. & Thermal &  RMSE ($\downarrow$) & \textbf{19.4}\cite{huPredictionInterpretabilityGlass2023}  &  27.7 \\
28    & T\textsubscript{g}–III (homopolymers)\cite{giroAIMembrane2023} & Sim. & Thermal &  R\textsuperscript{2} ($\uparrow$) & \textbf{0.90}\cite{giroAIMembrane2023}  &  0.86 \\
29    & T\textsubscript{g}–IV (homopolymers)\cite{polymetrix2025} & Exp. & Thermal &  MAE ($\downarrow$) & 26.2\cite{kimmigMRC2026}  &  \textbf{12.8} \\
30    & T\textsubscript{d,50\%}\cite{giroAIMembrane2023} & Sim. & Thermal &  R\textsuperscript{2} ($\uparrow$) & 0.92\cite{giroAIMembrane2023}  & \textbf{0.96} \\
\botrule
\end{tabular}
\footnotetext[1]{Reference citation indicates origin of benchmark dataset.}
\footnotetext[2]{Indication of how data for polymers was collected, computational via DFT or DFTB methods (Comp.), experimentally (Exp.), simulation (Sim.), or generated synthetically (Syn.).}
\footnotetext[3]{Reference citation indicates current SOTA performance.}
\footnotetext[4]{STP = SMI-TED-Polymer}
\end{table}

\subsubsection{Representation Analysis}

The results in Table~\ref{tab1} show the efficacy of utilizing the CPG representation to extend SMILES-based models for polymer property prediction tasks as compared to existing SOTA models. However, it cannot be determined from such experiments alone that the CPG representation is solely responsible for the observed performance improvements nor could the case be convincingly made that CPG is indeed superior to PSMILES or other line notations. Critical evaluation of the CPG representations necessitates running extensive control experiments. Indeed, most benchmark experiments for polymer models focus exclusively on model architecture or featurization method, not the structural representation system. Thus, in order to allow for rational evolution in model and representation design, it is necessary to design control experiments to both systematically test the performance of the CPG representation against other variations. The advantage of using the SMI-TED model as base model for SMI-TED-Polymer, is that they share essentially identical overall architectures, tokenizers, SMILES vocabularies, and similar training histories\cite{smiTED2025}. This provides an ideal platform to perform controlled evaluation of the influence of structural representation on benchmark performance as only the SMI-TED-Polymer model has seen SMILES strings containing asterisk tokens based on analysis of the pre-training data for both models (Fig.~\ref{cpg_rep}d and e). Here, we envisioned evaluation of both the CPG representation and PSMILES on selected benchmark datasets using both the frozen weights of SMI-TED and SMI-TED-Polymer models as well as fine-tuned versions of both models using either CPG or PSMILES representations, affording six different models to use for comparison. 

Comparing the performance of the CPG and PSMILES representations across SMI-TED and SMI-TED-Polymer models alone is not illustrative; by default, there is an implicit assumption in both representations that asterisk containing tokens are important in modeling polymer structures using SMILES. Therefore, additional variations of the PSMILES representation must also be evaluated as negative controls wherein the asterisk character is systematically replaced with other atom tokens from the SMILES vocabulary (Supplementary Table S1).  In each substitution, all asterisk characters in the CPG representation were replaced with the same atom for each variation in addition to dropping the numbered edge notations. Three different approaches to replacing the asterisk character in the CPG representation were adopted for each atom type: i)~\textit{all} where the entire bracketed and numbered asterisk character was replaced, ii)~\textit{bracket} where the asterisk and the number were replaced inside the pair of brackets, and iii)~\textit{star} were only the asterisk character was substituted inside the bracket (Fig.~\ref{cpg_rep}c). Additionally, some substitutions resulted in new SMILES strings which could be canonicalized into a different token order. In such cases, a separate dataset with newly canonicalized representations was created to evaluate in addition to the original token order following the atom substitution. Not all substitution methods were viable for all atoms. Certain atoms used for substitution, such as metals, are only found inside of brackets in the SMILES vocabulary of both models. Non-bracketed versions are tokenized as having an \textit{unknown} token which is in turn replaced by a \textit{padding} token. As the insertion of the \textit{padding} token resulted in mutually equivalent representations, only one example of such an input representation was evaluated (Supplementary Table S1). The SMI-TED model tokenizes the CPG representation in an identical manner as the SMI-TED-Polymer model albeit the numbered asterisk tokens are replaced with the \textit{padding} token as they are not part of the SMI-TED model vocabulary. The structural modifications via atom-substitution (Fig.~\ref{cpg_rep}c) are also symmetric in terms that all numbered asterisk tokens are replaced with the same atom and could potentially introduce a source of systemic bias. Hence evaluation of shuffled and randomized SMILES strings were performed as positive controls.

\begin{figure}[H]
\centering
\includegraphics[width=\textwidth]{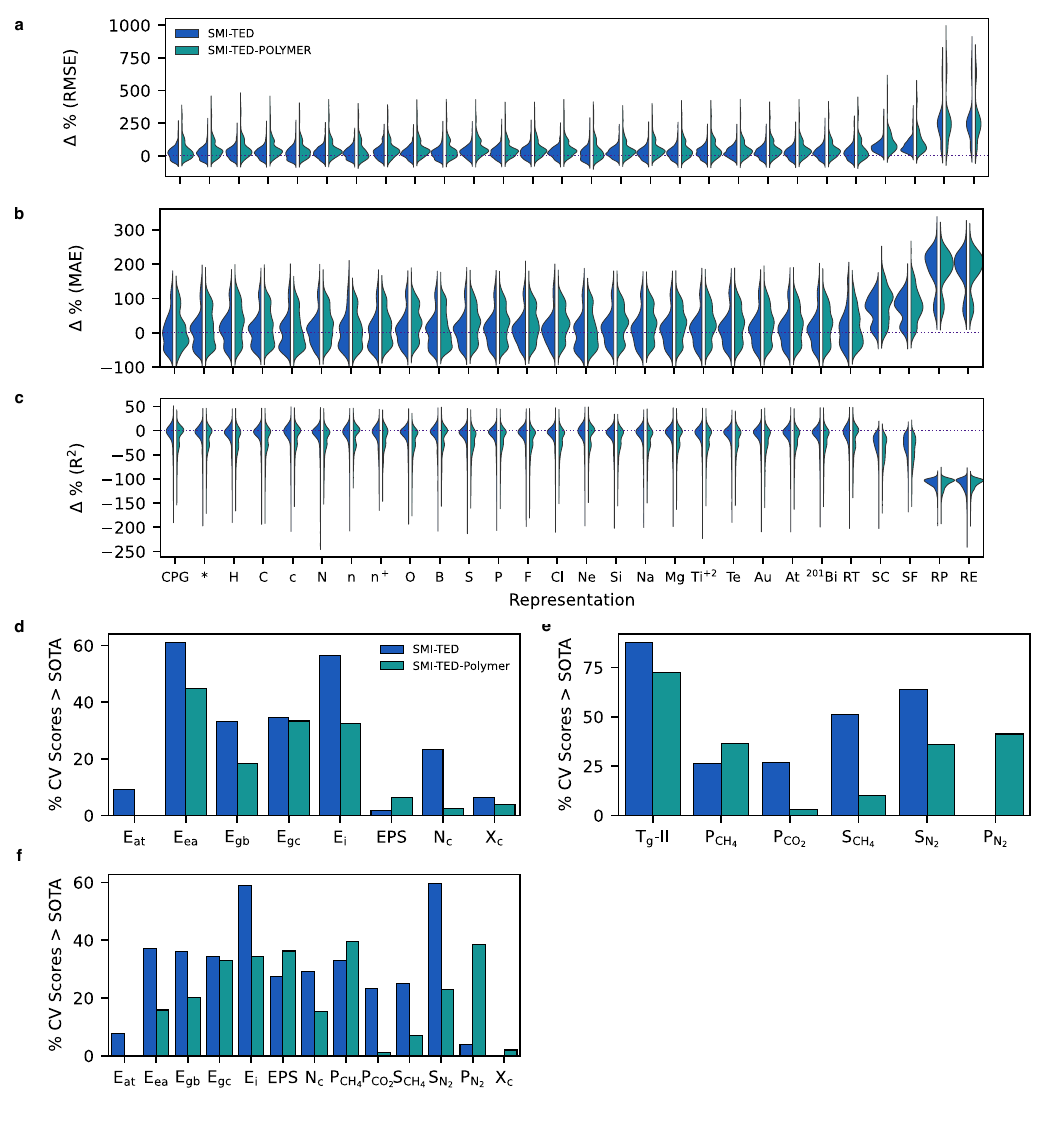}
\caption{Aggregated comparison of performance of structural representations across all benchmarks from repeated, 5-fold cross-validation.~\textbf{a}-\textbf{c} Plot of distribution of individual cross-validation scores as their percent change from SOTA values for RMSE, MAE, and R\textsuperscript{2}, respectively. Below zero (MAE, RMSE) or above zero (R\textsuperscript{2}) portions of the distribution indicates an improvement upon SOTA values.~\textbf{d}-\textbf{f} Plots of the percent of individual cross-validation scores that improve upon SOTA values for each tested dataset for RMSE, MAE, and R\textsuperscript{2}, respectively. PSMILES representations are indicated by the "*" character. RT = randomized token substitutions, SC = shuffled CPG, SF = shuffled SMILES with random token substitutions, RP = random SMILES with probabilistically weighted token selections, RE = random SMILES with equally weighted token selections.}\label{char_mod}
\end{figure}

The different structural representations were evaluated against the two frozen and four fine-tuned SMI-TED and SMI-TED-Polymer models for 14 selected benchmark datasets from Table~\ref{tab1}. Repeated five-fold cross-validation experiments, with each fold being repeated five times, were performed for each structural representation. Aggregated individual measurements from the cross-validation experiments from all models are shown in Figure~\ref{char_mod}. The measurements are normalized to their individual dataset's SOTA metric indicating the relative percent change from the SOTA values with zero being equivalent to the SOTA performance—allowing all of the benchmark data to be visualized on a single plot. Qualitatively, the plots in Fig.~\ref{char_mod}a-c show the largely equivalent aggregate performance across all token substitutions, with only the randomized controls showing noticeably worse performance as anticipated. Minimal differences between the performance of the SMI-TED and SMI-TED-Polymer models were observed in the aggregated results, with the SMI-TED-Polymer model results skewing slightly towards worse performance. This skew is much more noticeable in the frozen weight models on the un-aggregated results for individual benchmark datasets (see Supplementary Information), where the SMI-TED-Polymer model performed worse than the frozen weight SMI-TED model. Fine-tuning typically resulted in comparable performance between SMI-TED and SMI-TED-Polymer models across all benchmark datasets. In all benchmark datasets, a percentage of measurements across all structural representations resulted in improvements upon SOTA values. These results again show a small edge towards SMI-TED for providing more measurements exceeding the SOTA values on each of the three metrics (RMSE, MAE, R\textsuperscript{2}, Fig.~\ref{char_mod}d-f).

\begin{figure}[H]
\centering
\includegraphics[width=\textwidth]{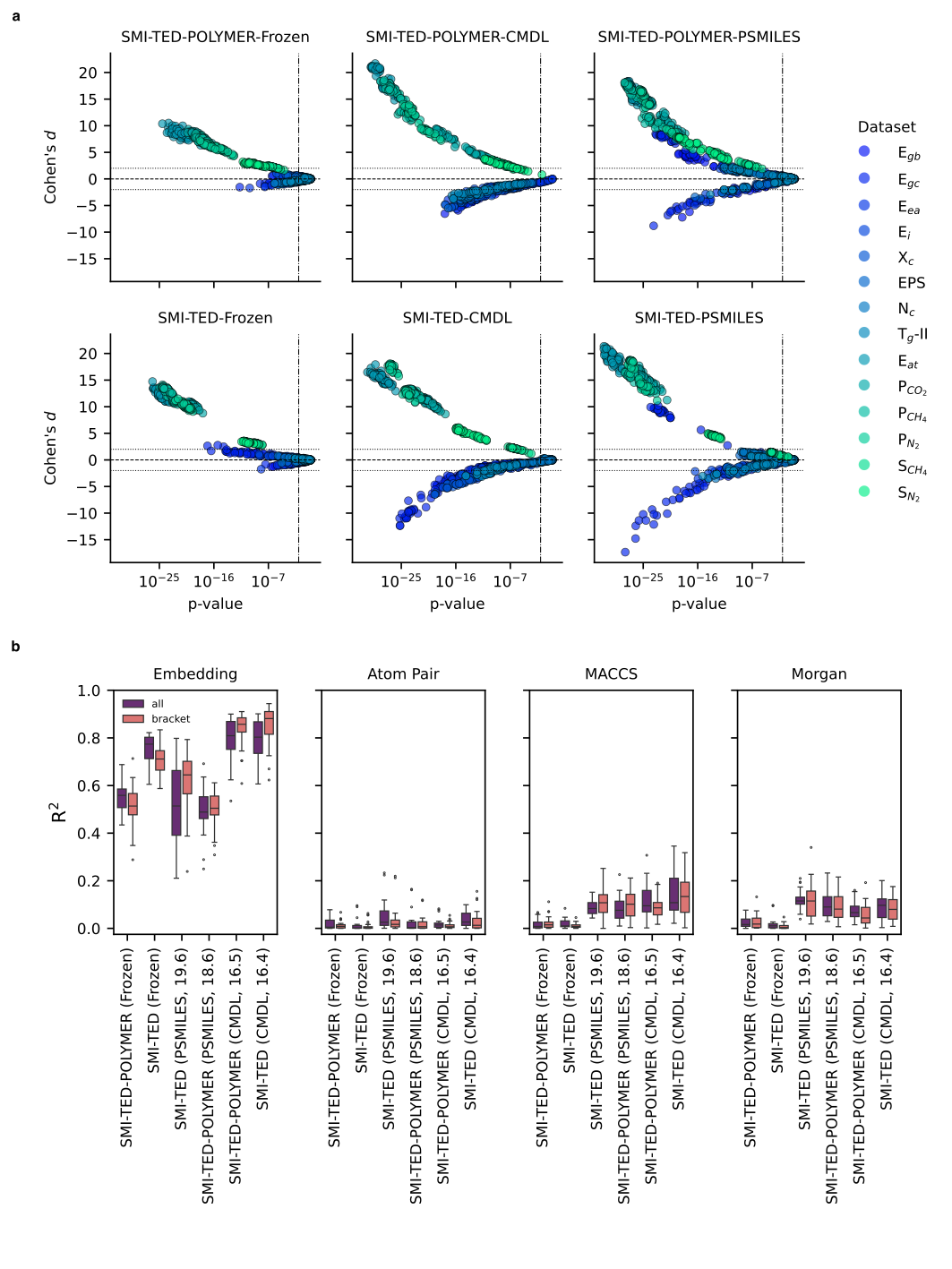}
\caption{\textbf{a} Plot of Cohen's \textit{d} versus p-value from t-test for cross-validation scores aggregated from all tested datasets for each model type evaluated. Cohen's \textit{d} is computed between CPG and other representations. Values less than zero indicate improved performance relative the the CPG representation.~\textbf{b} Plots of the R\textsuperscript{2} values from the linear regression between the predicted property value in the hold-out test set and the average true value from the the top five nearest neighbors in the training set. Nearest neighbors are computed based off of the latent embedding from the SMI-TED or SMI-TED-Polymer models or a fingerprint (atom pair, MACCS, Morgan). Data shown is from the X\textsubscript{c} dataset, plots of the other data may be found in the Supplementary Information.}\label{knn_rsq}
\end{figure}

In order to better understand the effect of the change in structural representation from CPG to PSMILES or other substituted SMILES representation we compared the data from the repeated, five-fold cross-validation experiments between the CPG representation and every other structural representation using a paired t-test and Cohen's \textit{d}. The p-value from the t-test and the Cohen's \textit{d} are plotted in Fig.~\ref{knn_rsq}a for each comparison between CPG and another structural representation for 14 benchmark datasets, separated by the respective model used. In all plots, the effect sizes can be dramatic in either improved performance versus CPG (negative values) or decreased performance versus CPG (positive values). The corresponding p-values for such effect sizes are vanishingly small, implying the effects are both large and significant for many structural variations. Correspondingly, the p-values increase as the effect sizes decrease as substitutions yield less and less difference in performance from the CPG representation. Intriguingly, this analysis reveals some clustering as a function of the benchmark dataset itself, suggesting some datasets are more sensitive or resilient to perturbations in the structural representation. These effects appear to be reinforced through the fine-tuning process, regardless of which representation or base model is used (Fig.~\ref{knn_rsq}).

The results from the representation analysis and the apparent invariance regarding the structural input made it imperative to better understand if prediction error is in any way correlated with structural representation. Prediction errors with deep-learning models can frequently be attributed their interpolative nature as well as systemic biases introduced from training and benchmark data. The interpolative nature of deep-learning models is well-understood and hence such models frequently struggle with out-of-distribution (OOD) tasks\cite{liuOODRev2023}—spurring efforts to ameliorate these performance deficits\cite{bradshawRxnGenPred2025,toniatoFastRxn2023,tangUQPoly2025}. Additionally, extensive work has been done to understand the effect of different dataset splitting techniques\cite{landrumSIMPDAlgorithmGenerating2023,guoScaffoldSplitsOverestimate2024,steshinLoHi2023} on model performance, quantifying the uncertainty of model predictions\cite{hirschfeldUQNN2020,tangUQPoly2025,scaliaEvaluatingScalableUncertainty2020,varivodaMaterialsPropertyPrediction2023}, and assessing the roughness of structure–property relationships as a function of representation\cite{graffROGIXD2023}. In the context of this work, we were first interested in understanding the potential underlying sources of error as a function of random dataset splits, which are most commonly used in benchmark evaluations for polymer deep-learning models\cite{polyBERT,transPolymer,mmPolymer,zhangTransferringMolecularFoundation2023}.  Here, we opted to evaluate the potential correlation between training and test set predictions as dictated by latent embedding distance which is a function of chemical similarity. Chemical similarity based on atom-pair fingerprints has been shown previously to be highly explanatory of prediction error in previous studies on small-molecule drug datasets\cite{sheridanExperimentalErrorKurtosis2020,sheridanSimilarityMoleculesTraining2004}. 

For each representation system, we examined each candidate in the test set and compared its predicted property value from the model to the average known property values of the top five nearest neighbors in the training set based on Euclidean distance between their embedding vectors. Each dataset of predicted versus averaged values was fitted with via linear regression and the resulting aggregated R\textsuperscript{2} values are shown in Fig.~\ref{knn_rsq} for the X\textsubscript{c} benchmark datasets and with the remainder in the Supplementary Information (see Supplementary Information). The results demonstrate a strong correlation between the predicted value for a test set candidate and the known values of its closest neighbors within the training dataset. Notably, the strength of the correlation increases for all representation types when more performant, fine-tuned models are used for the prediction (Fig.~\ref{knn_rsq}b). Switching from the Euclidean distance between the latent embeddings and using Morgan, MACCS, or atom-pair fingerprints and Dice similarity as the ranking metric dramatically reduced or eliminated any correlation between predicted and the average neighbor values (Fig.~\ref{knn_rsq}b).

\begin{figure}[H]
\centering
\includegraphics[width=\textwidth]{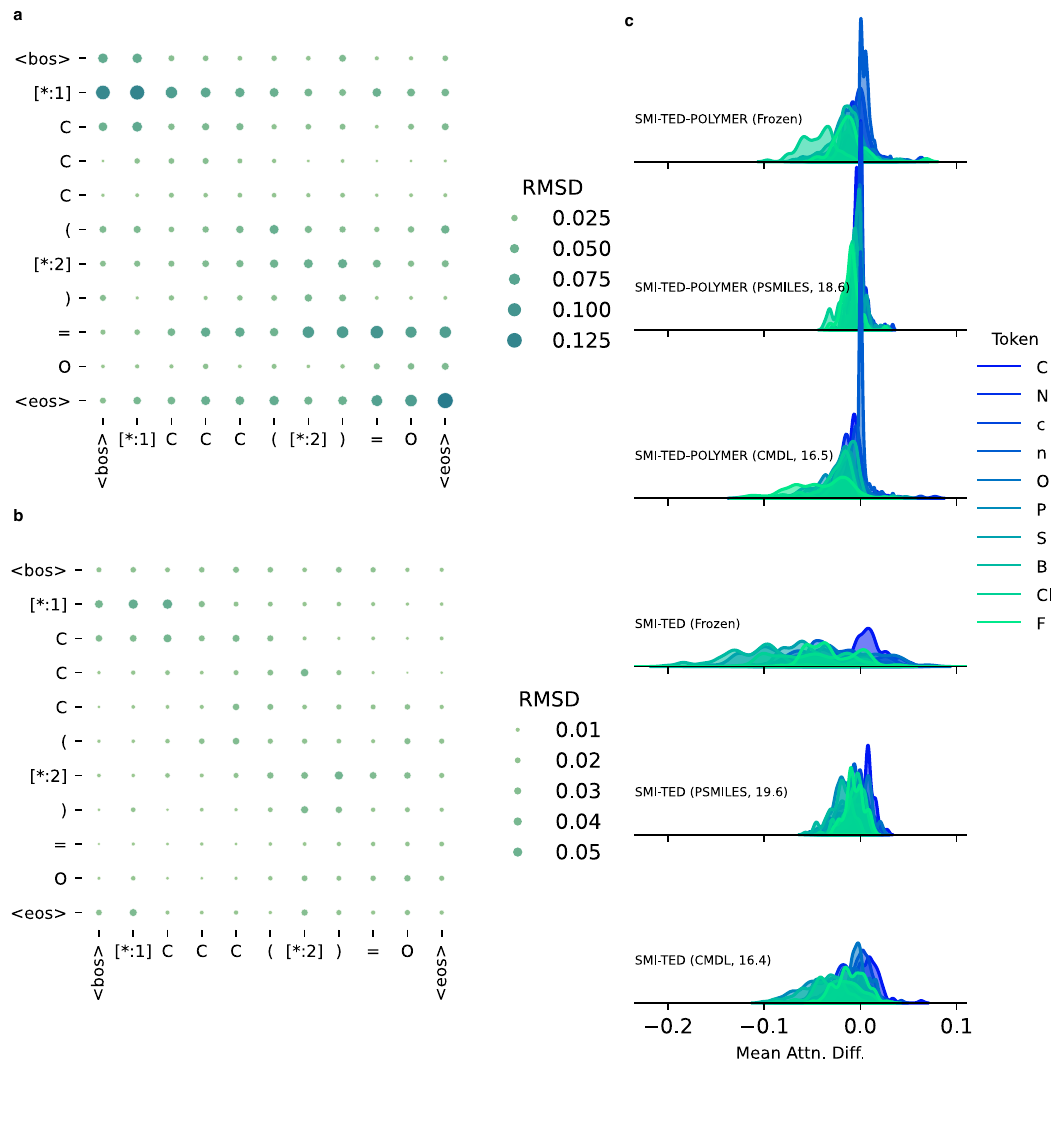}
\caption{\textbf{a}-\textbf{b} Example RMSD attention maps from the X\textsubscript{c} dataset for the CPG fine-tuned SMI-TED-Polymer and SMI-TED models, respectively.~\textbf{c} Aggregated mean attention difference of \textit{sub} tokens for the X\textsubscript{c} dataset for the \textit{all} substitution strategy.}\label{attn_maps}
\end{figure}

Attention mechanisms in deep-learning models assist in providing models a means to better capture long-range relationships and improving context in a particular task. For transformer-based chemistry language models, analysis of attention maps are typically accompanied by chemistry-focused interpretations of the observed distribution of attention weights as an indication of the model learning the salient features of a given input line notation\cite{dollarAttentionbasedGenerativeModels2021,polyBERT,qiuExpBigSMILES2025}. Here, we sought to examine how the averaged attention maps changed as a function of the structural representation with the hypothesis that the largest absolute changes in attention weights would be observed for tokens that were substituted. Given the large number of chemically and semantically invalid structural representations—many of which retained strong performance in benchmark evaluations—a chemistry-focused interpretation of attention weights is not necessarily warranted. Additionally, examination of individual attention maps for the sheer number of different structural representations evaluated would be impractical. Instead, we first averaged the attention weights across all 12 layers for each different input representation type for a given entity, then computed root mean squared difference (RMSD) between the truncated CPG representation's--where the edge definition tokens are removed (e.g. 1->2, Fig.~\ref{cpg_rep})-- average attention weights ($X_p$) and all other non-randomized structural representations ($X_i$) (Eq.~\ref{eq1}). It should be noted that this approach was done for tractability on the large volume of data collected and may discard important, layer-specific information.
\begin{equation}\label{eq1}
    \text{RMSD}(X_p, X) = \sqrt{\frac{\sum_{i=0}^{N - 1} {(X_p - X_i)}^2}{N}}
\end{equation}
Example RMSD attention maps from the X\textsubscript{c} benchmark dataset for fine-tuned SMI-TED and SMI-TED-Polymer models are shown in Fig.~\ref{attn_maps}a and b, respectively. The aggregated changes in attention weights show a stark difference between the SMI-TED and SMI-TED-Polymer models, with the SMI-TED model showing larger changes in attention upon changing the structural representation. The change in attention weights is not only reserved for the substituted token, but affects surround tokens as may be anticipated. With the SMI-TED example (Fig.~\ref{attn_maps}), the double bond token for a carbonyl has larger overall magnitude changes in attention compared to the nearby substituted token. However, the same effect is not observed in the attention map for SMI-TED-Polymer, which has overall weaker differences in weights (Fig.~\ref{attn_maps}). The effect of the identity of the model on the relative difference in attention between the CPG representation and individual token substitutions for the X\textsubscript{c} benchmark dataset is shown in Fig.~\ref{attn_maps}c. Here, the distribution of the attention weight differences can vary strongly both as a function of the model and the individual token used in the substitution.

\begin{table}[htbp]
\caption{Representation Benchmarks Summary}\label{tab2}%
\begin{tabular}{@{}lllllccc@{}}
\toprule
Entry & Dataset\footnotemark[1] & Metric & SOTA\footnotemark[2] & Score & Model\footnotemark[3] & Token & Format \\
\midrule
1    & E\textsubscript{gc}\cite{kuennethPolymerInformaticsMultitask2021} & RMSE ($\downarrow$) & 0.44\cite{mmPolymer} & \textbf{0.23$\pm$0.017} & ST (CPG) & CPG & —  \\
2    & E\textsubscript{gb}\cite{kuennethPolymerInformaticsMultitask2021} & RMSE ($\downarrow$) & 0.485\cite{shenMolecularTopologicalDeep2026} & \textbf{0.38$\pm$0.034} & ST (CPG) & CPG & —  \\
3    & E\textsubscript{ea}\cite{kuennethPolymerInformaticsMultitask2021} & RMSE ($\downarrow$) & 0.263\cite{shenMolecularTopologicalDeep2026} & \textbf{0.18$\pm$0.032} & ST (PS) & Au & \textit{bracket}  \\
4    & E\textsubscript{i}\cite{kuennethPolymerInformaticsMultitask2021} & RMSE ($\downarrow$) & 0.39\cite{mmPolymer} & \textbf{0.25$\pm$0.065} & ST (PS) & B & \textit{bracket}  \\
5    & E\textsubscript{at}\cite{kuennethPolymerInformaticsMultitask2021} & RMSE ($\downarrow$) & 0.050\cite{mmPolymer} & \textbf{0.050$\pm$0.01} & ST (PS) & Te & \textit{bracket} \\
6    & EPS\cite{kuennethPolymerInformaticsMultitask2021} & RMSE ($\downarrow$) & \textbf{0.35}\cite{shenMolecularTopologicalDeep2026} & \textbf{0.35$\pm$0.041} & STP (PS) & P & \textit{bracket} \\
7    & X\textsubscript{c}\cite{kuennethPolymerInformaticsMultitask2021} & RMSE ($\downarrow$) & 15.86\cite{shenMolecularTopologicalDeep2026} & \textbf{14.0$ \pm $1.2} & ST (CPG) & CPG & —  \\
8    & N\textsubscript{c}\cite{kuennethPolymerInformaticsMultitask2021} & RMSE ($\downarrow$) & 0.068\cite{shenMolecularTopologicalDeep2026} & \textbf{0.042$\pm$0.007} & ST (PS) & Ne & \textit{bracket}  \\
9    & P\textsubscript{CO$_2$}\cite{tiwariCreationPolymerDatasets2024} & MAE ($\downarrow$) & 0.29\cite{tiwariCreationPolymerDatasets2024} &  \textbf{0.18$\pm$0.017}  & ST (PS) & C & \textit{bracket}\\
10   & P\textsubscript{CH$_4$}\cite{tiwariCreationPolymerDatasets2024} & MAE ($\downarrow$) & 0.37\cite{tiwariCreationPolymerDatasets2024} &  \textbf{0.22$\pm$0.02}  & STP (PS) & * & \textit{star} \\
11    & P\textsubscript{N$_2$}\cite{tiwariCreationPolymerDatasets2024} & MAE ($\downarrow$) & 0.38\cite{tiwariCreationPolymerDatasets2024} &  \textbf{0.14$\pm$0.015} &  ST (PS) & * & \textit{all} \\
12    & S\textsubscript{CH$_4$}\cite{tiwariCreationPolymerDatasets2024} & MAE ($\downarrow$) & 5.34\cite{tiwariCreationPolymerDatasets2024} &  \textbf{4.89$\pm$0.59} & STP (PS) & B & \textit{bracket}\\
13    & S\textsubscript{N$_2$}\cite{tiwariCreationPolymerDatasets2024} & MAE ($\downarrow$) & 4.14\cite{tiwariCreationPolymerDatasets2024} &  \textbf{2.05$\pm$0.59} & STP (PS) & B & \textit{bracket} \\
14    & T\textsubscript{g}–II\cite{huPredictionInterpretabilityGlass2023} & RMSE ($\downarrow$) & 19.4\cite{huPredictionInterpretabilityGlass2023}  &  \textbf{13.6$\pm$4.2} & STP (CPG) & CPG & — \\
\botrule
\end{tabular}
\footnotetext[1]{Reference citation indicates source of benchmark dataset.}
\footnotetext[2]{Reference citation indicates current SOTA performance.}
\footnotetext[3]{Indicates model used for cross-validation experiments. Base model (ST or STP) followed by representation used for fine-tuning in parenthesis (CPG or PS). ST = SMI-TED, STP = SMI-TED-Polymer, PS = PSMILES}
\end{table}

The top-performing structural representations and the respective model used from the cross-validation experiments are presented in Table~\ref{tab2}. While the CPG representation was the best representation in several benchmarks, the remaining was divided among several different structural substitutions, including PSMILES. In all cases, the results from these experiments were better than the current SOTA values. These results again convey the resilience of the latent representation given by these models to perturbations in structural representation. However, as indicated above, the large majority of structural perturbations lead to worse performance.

\begin{figure}[H]
\centering
\includegraphics[width=\textwidth]{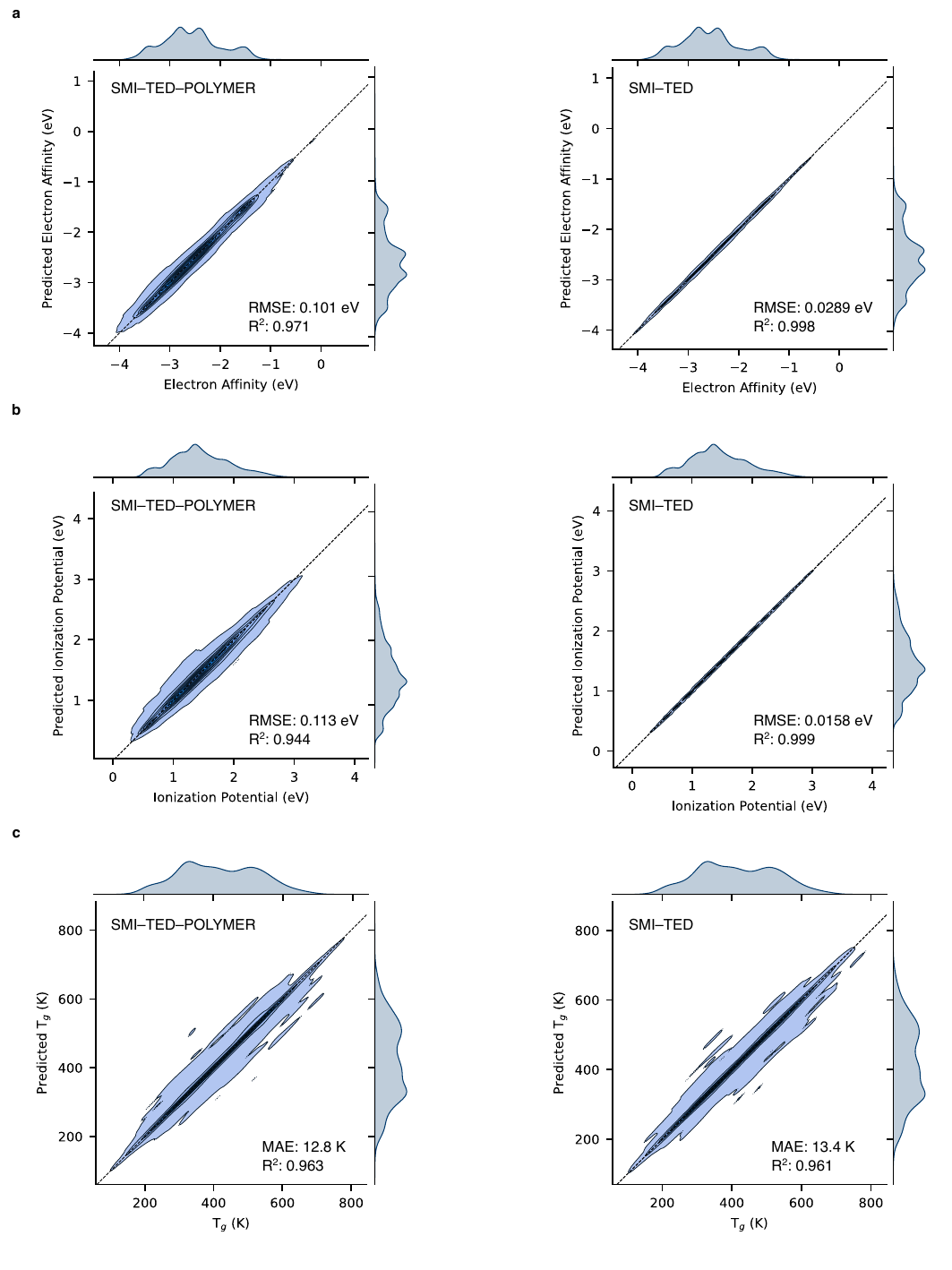}
\caption{\textbf{a} Kernel density estimate plots of true versus predicted property values for the copolymer DFT electron affinity dataset using stoichiometric weighted averaging of the latent embeddings of copolymer components using the CPG representation.~\textbf{b} Kernel density estimate plots of true versus predicted property values for the copolymer DFT ionization potential dataset using stoichiometric weighted averaging of the latent embeddings of copolymer components using the CPG representation.~\textbf{c} Kernel density estimate plots of true versus predicted property values for Tg-IV dataset.}\label{copolymer_bench}
\end{figure}

Given the improved performance of the SMI-TED model in many of the selected benchmark datasets used in the control experiments, we sought to examine whether leveraging SMI-TED model would improve performance on the EA and IP copolymer benchmark dataset (Entries 7 and 8, Table~\ref{tab1})\cite{aldeghiGraph2022}, where the CPG copolymer representations failed to equal SOTA of performance. Based on our prior work building scalar-weighted polymer representations for vector data stores for retrieval augmented generation tasks\cite{parkRAG}, we opted to utilize a similar strategy to improve results on the copolymer benchmark. This involved splitting the CPG representation into its component repeat units and computing the weighted average latent representation during fine-tuning. For the SMI-TED-Polymer model this approach yielded improved, but not SOTA performance on both the EA and IP datasets (Fig.~\ref{copolymer_bench}a and b), while using the SMI-TED model SOTA performance was achieved in both cases. This result demonstrates that scalar information, such as monomer stoichiometry, can easily be incorporated during fine-tuning and applied directly to the latent representations. Prior work has focused on incorporating such information into the pre-training phase\cite{aldeghiGraph2022} or in the fine-tuning phase with a using a custom string representation with encoded numeric scalar values\cite{zohairChemicalFoundationModel2025,transPolymer}; both approaches require additional customizations to the tokenization or featurization process of the model. Finally, a comparison of the SMI-TED and SMI-TED-Polymer model was also performed for a new homopolymer T\textsubscript{g} dataset, where both models provided nearly equivalent MAE for the dataset, an improvement on prior SOTA (Fig.~\ref{copolymer_bench}c).

\section{Discussion}

The success of developing foundation models for polymers and polymer containing materials will hinge greatly on the development of improved datasets, accurate structural representations, and a thorough understanding of how a foundation model handles such information. While our initial hypothesis surrounding the effect on performance of the CPG representation has been partially validated, our results also show that under commonly used benchmark and split protocols, strong performance can be achieved without a chemically meaningful structural encoding. This implies that current practice often measures interpolation within the training distribution more than mechanistic ‘understanding’ of chemistry or chemical structure. This effect is further confirmed by the strong observed correlation between predicted values and the nearest-neighbor average for all representation types, demonstrating that model performance is arising from interpolation within latent space. These results further suggest that other dataset spitting techniques,\cite{landrumSIMPDAlgorithmGenerating2023,guoScaffoldSplitsOverestimate2024,steshinLoHi2023} some of which may be further adapted to polymer domains, could provide a more realistic assessment of model performance in polymer tasks than commonly used random splits. As a consequence of these findings, the incorporation of scalar information into the latent representation during fine-tuning demonstrates how the latent space may be reshaped in a task dependent manner, allowing better interpolation over a dataset of otherwise closely related materials with large variances in calculated properties. Moreover, this also more clearly demonstrates the foundational nature of such pre-trained chemistry language models; scalar metadata can be used for specific downstream tasks without the need to be uniformly present in the pre-training data--greatly expanding the amount of otherwise scarce data available for use in pre-training.  Overall, we have demonstrated the effectiveness of the CPG in enhancing the performance of a polymer foundation model as well as provided critical analysis of the performance of the CPG relative to other SMILES representations which can guide further improvements in the development of representations and datasets in AI for polymers.

\section{Methods}

\subsection{Datasets}

Pre-training and benchmark datasets were sourced from openly available datasets\cite{aldeghiGraph2022,tranJAP2020,yangGasSep2022,bradfordConduct2023,huPredictionInterpretabilityGlass2023,longLargeScaleTg2024,transPolymer,reisML19F2021,aroraRFPredictor2021,giroAIMembrane2023,aldeghiGraph2022,tiwariCreationPolymerDatasets2024}. The pre-training dataset size for SMI-TED-Polymer was $1M$ polymers. For each benchmark dataset, the SMILES-based structural representation was converted to the CPG representation before splitting into train/valid/test datasets.

\subsection{Pre-training}

Pre-training of the SMI-TED-Polymer model was conducted over 150 epochs using the curated CPG dataset, with a fixed learning rate of 1.6e-4 and a batch size of 256 molecules. The training was distributed across 4 NVIDIA V100 (16G) GPUs, parallelized into 4 nodes using DDP and \textit{torch run}. The process involves two key phases: i. Learning polymer token embeddings through a masking mechanism. ii. Mapping these embeddings into a unified latent space that represents the entire CPG string. This latent space not only captures the structural representation of the CPG but also enables the reconstruction of both individual polymer tokens and the complete CPG strings.

Accordingly, the pre-training process utilizes two distinct loss functions: one associated with the token embeddings, driven by the masking process, and another targeting the encoder-decoder layer, focusing on token reconstruction. 

\begin{itemize}
     \item In phase 1, the polymer token encoder is initially pre-trained using 95\% of the available samples, while the remaining 5\% is reserved for training the encoder-decoder layer. This partitioning is necessary as the polymer token embeddings may encounter convergence difficulties in the initial epochs, which could adversely affect the training of the encoder-decoder layer. 

    \item In Phase 2, after the polymer token embeddings layer has converged, the pre-training process is scaled to leverage 100\% of the available samples for both phases. This approach significantly enhances the performance of the encoder-decoder layer, particularly in improving polymer token reconstruction accuracy.
 \end{itemize}

For encoder pre-training we use the masked language model method defined previously\cite{devlinBERTPretrainingDeep2019}. Initially 15\% of the tokens are selected for possible learning. From that selection, 80\% of the tokens are randomly selected and replaced with the [MASK] token, 10\% of the tokens are randomly selected to be replaced with a random token, while the remaining 10\% of the tokens will be unchanged. 

\subsection{Experiments}

\paragraph{Benchmark Experiments}
Benchmark experiments for Table~\ref{tab1} for SMI-TED-Polymer were conducted using a randomized 80:10:10 train/valid/test split. For each benchmark dataset, the SMI-TED-Polymer model was fine-tuned using the training and validation datasets. Using the fine-tuned model, the XGBoost regressor was created using the latent embedding (dim = 768) of the structure input and the regressor hyperparameters were optimized using the Optuna python package with 300 trials. Final performance of the XGBoost regressor was evaluated using the holdout test dataset.

\paragraph{Representation Analysis}
Selected benchmark datasets from Table~\ref{tab1} were converted into a variety of alternative input representations, via replacement of asterisk tokens or randomization of token order. For consistency, the representation conversion was done from the CPG, not the original representation. This is due to the fact that the CPG representations were canonicalized once the numbered asterisks were inserted. For meaningful comparisons, this token order was maintained in all other input representations. Re-canonicalization of newly created input representation was also performed as a separate comparative experiment, such that both the identical and different token orders could be compared. Depending on the atom used, re-canonicalization resulted in some or all the token orders of the input representation to be changed.

For the evaluation of the new representations, both the frozen weight base model of SMI-TED and SMI-TED-Polymer were used in addition to fine-tuned versions for each benchmark dataset. Each base model (SMI-TED or SMI-TED-Polymer) was fine-tuned for 300 epochs using either the CPG or PSMILES input representation, resulting in four fine-tuned models for each benchmark dataset. Each input representation type for each benchmark dataset was evaluated against all six models (two frozen weight and four fine-tuned), wherein the latent embeddings from the respective model were used in XGBoost regressor evaluated using a repeated 5-fold cross-validation, repeated for five times. Results from both the cross-validation and test set evaluation were used to compare the efficacy of each input representation.

\backmatter{}

\section*{Declarations}

\subsection{Data Availability}
The benchmark data used to asses property prediction tasks is available from their respective published sources\cite{aldeghiGraph2022,tranJAP2020,yangGasSep2022,bradfordConduct2023,huPredictionInterpretabilityGlass2023,longLargeScaleTg2024,transPolymer,reisML19F2021,aroraRFPredictor2021,giroAIMembrane2023,aldeghiGraph2022,tiwariCreationPolymerDatasets2024}.

\subsection{Code Availability}
Code used for generating the CPG graphs will be released upon final publication of the manuscript.

\bibliography{sn-bibliography}

\end{document}


\title[Understanding Structural Representation in Foundation Models for Polymers]{Understanding Structural Representation in Foundation Models for Polymers}

\author*[1]{\fnm{Nathaniel H.} \sur{Park}}\email{npark@us.ibm.com}

\author[2]{\fnm{Eduardo} \sur{Soares}}

\author[2]{\fnm{Victor} \sur{Shirasuna}}

\author[1, 3]{\fnm{Tiffany J.} \sur{Callahan}}

\author[1, 3]{\fnm{Sara} \sur{Capponi}}

\author[2]{\fnm{Emilio Vital} \sur{Brazil}}

\affil*[1]{\orgname{IBM Research Almaden}, \orgaddress{\street{650 Harry Rd.}, \city{San Jose}, \postcode{95120}, \state{CA}, \country{United States of America}}}

\affil[2]{\orgname{IBM Research Brazil}, \orgaddress{\street{Street}, \city{Rio de Janeiro}, \postcode{10587}, \state{RJ}, \country{Brazil}}}

\affil[3]{\orgname{Center for Cellular Construction}, \orgaddress{\city{San Francisco}, \postcode{94158}, \state{CA}, \country{USA}}}

\maketitle

\tableofcontents

\newpage

\section{Supplemental Methods}

\subsection{Representation Experiments}\label{rep_exp_section}

\paragraph{Atom Replacement Methods}As noted in the main text, we utilized three different approaches to transform the CPG representation into other SMILES strings, including PSMILES. The transformation was performed from benchmark data which had been converted into the CPG representation as opposed to the original data in order to ensure identical sequences of tokens for each entity were maintained. During conversion to the CPG notation, the SMILES strings had been canonicalized using RDKit, which resulted in new ordering in the SMILES strings as compared to the original data. The three replacement methods are:

\begin{itemize}
    \item \textit{All} – which replaces all characters including the brackets.
    \item \textit{Bracket} – which replaces all characters inside the brackets.
    \item \textit{Star} – which replaces only the asterisk character.
\end{itemize}

Following substitution, a separate copy of the converted dataset was canonicalized again for comparison, as some atom substitutions resulted in different canonical atom orders than the original.

\paragraph{Replacement Atoms}Below is a list of all the replacement atoms, which substitution methods were used, and whether a separate, re-canonicalized dataset was also created.

\begin{table}[hbtp]
\caption{Replacement Atoms for Input Representations}\label{atomtab}
\begin{tabular}{ccc}
\toprule
Atom & Methods & Canonicalization\\
\midrule
\textbf{*} & \textit{All}\footnotemark[1], \textit{Star} & Yes\\
\textbf{*} & \textit{Bracket}\footnotemark[1] & No\footnotemark[2] \\
\textbf{H} & \textit{Bracket} & Yes\footnotemark[3] \\
\textbf{C} & \textit{All}, \textit{Bracket}, \textit{Star} & Yes\\
\textbf{c} & \textit{All}, \textit{Bracket} & No\\
\textbf{N} & \textit{All}, \textit{Bracket} & Yes\\
\textbf{n} & \textit{All} & No\\
\textbf{n\textsuperscript{+}} & \textit{Bracket} & No\\
\textbf{O} & \textit{All}, \textit{Bracket} & Yes\\
\textbf{S} & \textit{All}, \textit{Bracket} & Yes\\
\textbf{P} & \textit{All}, \textit{Bracket} & Yes\\
\textbf{B} & \textit{All}, \textit{Bracket} & Yes\\
\textbf{Cl} & \textit{All}, \textit{Bracket} & Yes\\
\textbf{F} & \textit{All}, \textit{Bracket} & Yes\\
\textbf{Ne} & \textit{Bracket} & No\\
\textbf{Si} & \textit{Bracket} & Yes\\
\textbf{Na} & \textit{Bracket} & Yes\\
\textbf{Mg} & \textit{Bracket} & Yes\\
\textbf{Ti\textsuperscript{2+}} & \textit{Bracket} & Yes\\
\textbf{Au} & \textit{Bracket} & Yes\\
\textbf{Te} & \textit{Bracket} & Yes\\
\textbf{At} & \textit{Bracket} & Yes\\
\textbf{\textsuperscript{201}Bi} & \textit{Bracket} & Yes\\
\botrule
\end{tabular}
\footnotetext[1]{These replacement methods on the using the asterisk will afford the corresponding PSMILES representation}
\footnotetext[2]{Canonicalization of the asterisk inside a bracket removes the bracket, giving an equivalent representation as the \textit{all} method}
\footnotetext[3]{Canonicaliztion of SMILES with the [H] token, removes the token entirely as hydrogen atoms are implicit in SMILES.}
\end{table}

\pagebreak

\section{Supplemental Figures}

\subsection{Error Analysis}
\subsubsection{Latent Embedding Distance}

Below are the remaining plots of the R\textsuperscript{2} from linear regression between the predicted property value of each candidate in the hold-out test set and the average true value from its top five nearest neighbors in the training set as a function of the model used not included in the main manuscript. As in the main text, fine-tuned models are labeled by their respective base model followed by the representation with which it was fine-tuned with and the associated loss for the checkpoint. 

\subsubsection{Fingerprint Similarity}

Below are the plots of the R\textsuperscript{2} from linear regression between the predicted property value of each candidate in the hold-out test set and the average true value from its top five nearest neighbors based on fingerprint chemical similarity in the training set as a function of the model used. Not all representations could be used to compute viable chemical fingerprints using the RDKit package. Substituted representations that resulted in an invalid SMILES or any of the randomly shuffled forms were not included in this analysis. Chemical similarity was computed using Dice similarity as reported previously. Fine-tuned models are labeled by their respective base model followed by the representation with which it was fine-tuned with and the associated loss for the checkpoint. All losses for fine-tuned models are RMSE unless indicated otherwise.

\subsection{Attention Figures}

\begin{figure}[H]
\centering
\includegraphics[width=1\textwidth]{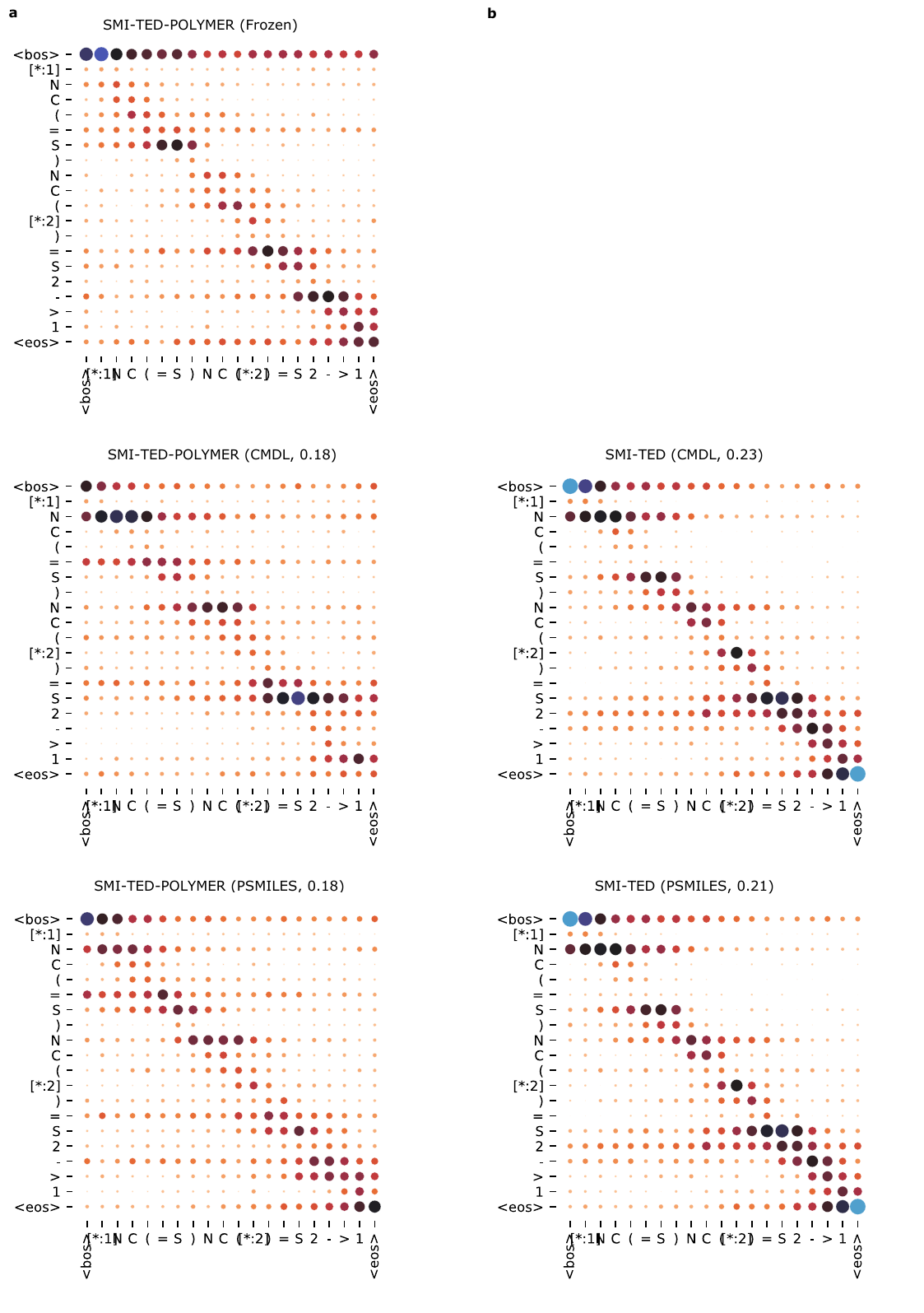}
\caption{Attention maps of CPG representations from the E\textsubscript{ea} dataset.}\label{eea_attn}
\end{figure}

\begin{figure}[H]
\centering
\includegraphics[width=1\textwidth]{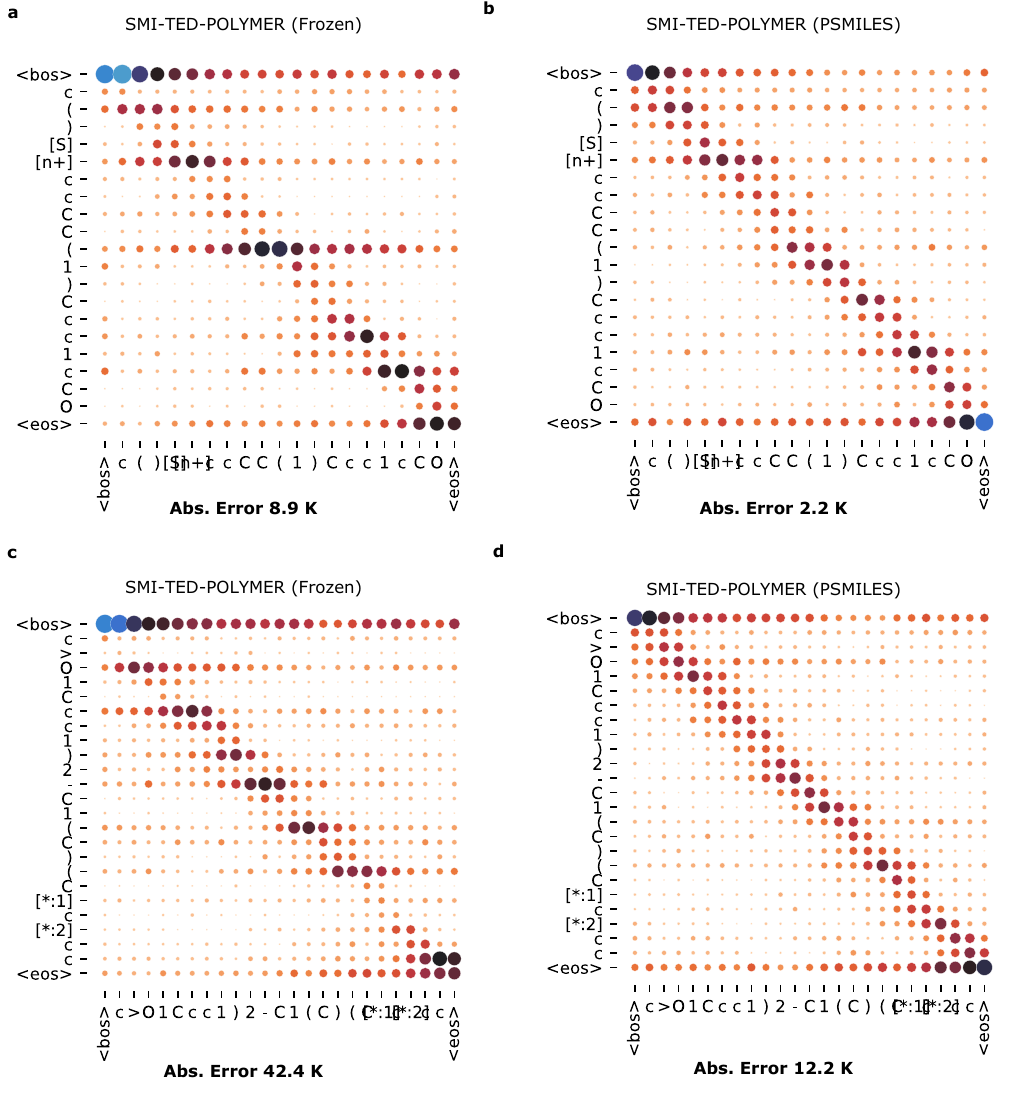}
\caption{Comparison of average attention probabilities for random representations}\label{rand_attn}
\end{figure}

\subsection{Representation Evaluation Results}\label{rep_eval}

\subsubsection{Polymer Chain Bandgap (E\textsubscript{gc})}\label{cvEgc}

\begin{figure}[H]
\centering
\includegraphics[width=1\textwidth]{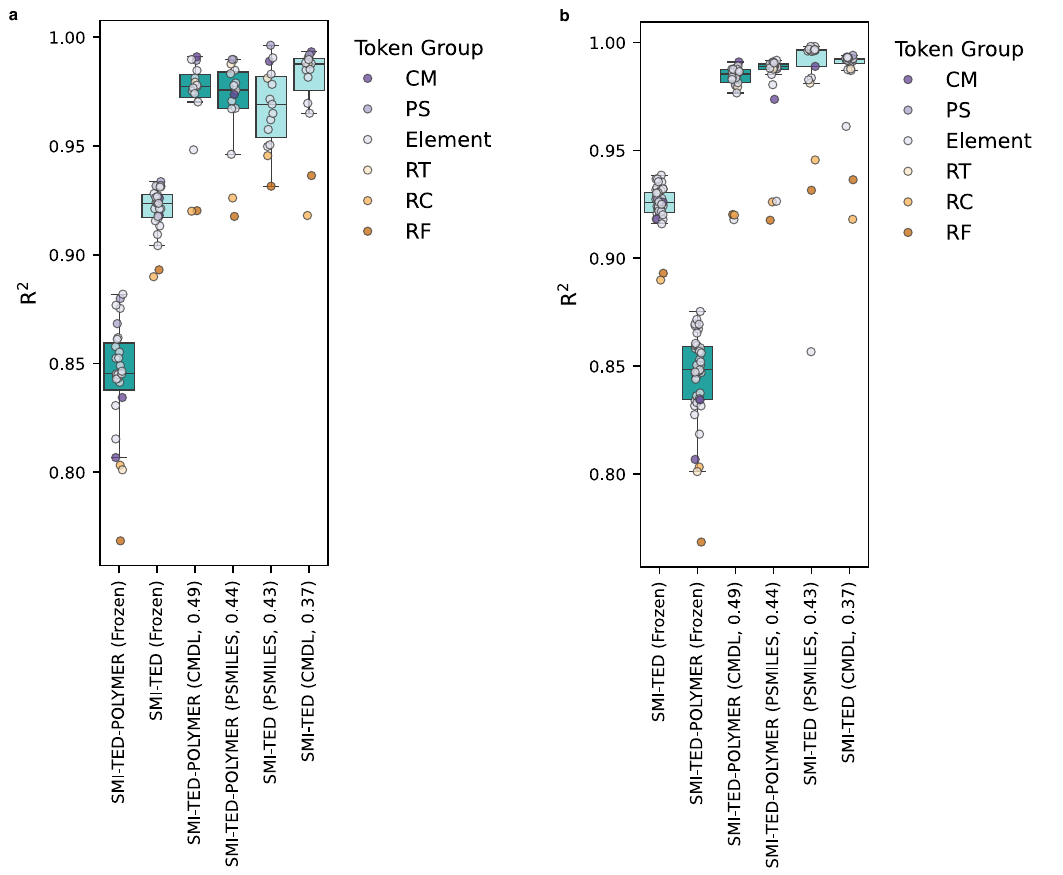}
\caption{Plots of embedding R\textsuperscript{2} value distributions for the E\textsubscript{gc} benchmark dataset using the \textit{bracket} (\textbf{a}) and \textit{all} (\textbf{b}) substitution strategy.}\label{egc_knn}
\end{figure}

\begin{figure}[H]
\centering
\includegraphics[width=1\textwidth]{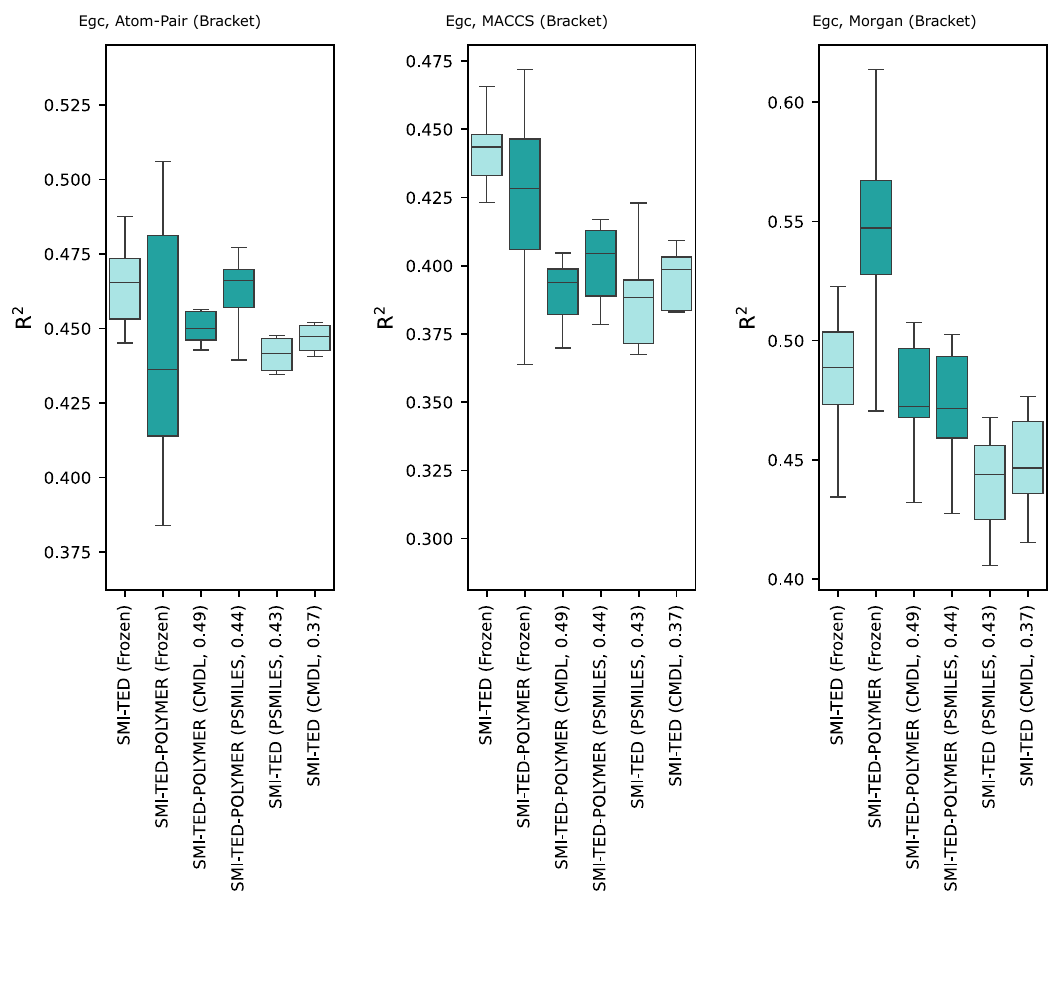}
\caption{Plots of fingerprint R\textsuperscript{2} value distributions for the E\textsubscript{gc} benchmark dataset using the \textit{bracket} substitution strategy.}\label{egc_knn_fp_bracket}
\end{figure}

\begin{figure}[H]
\centering
\includegraphics[width=1\textwidth]{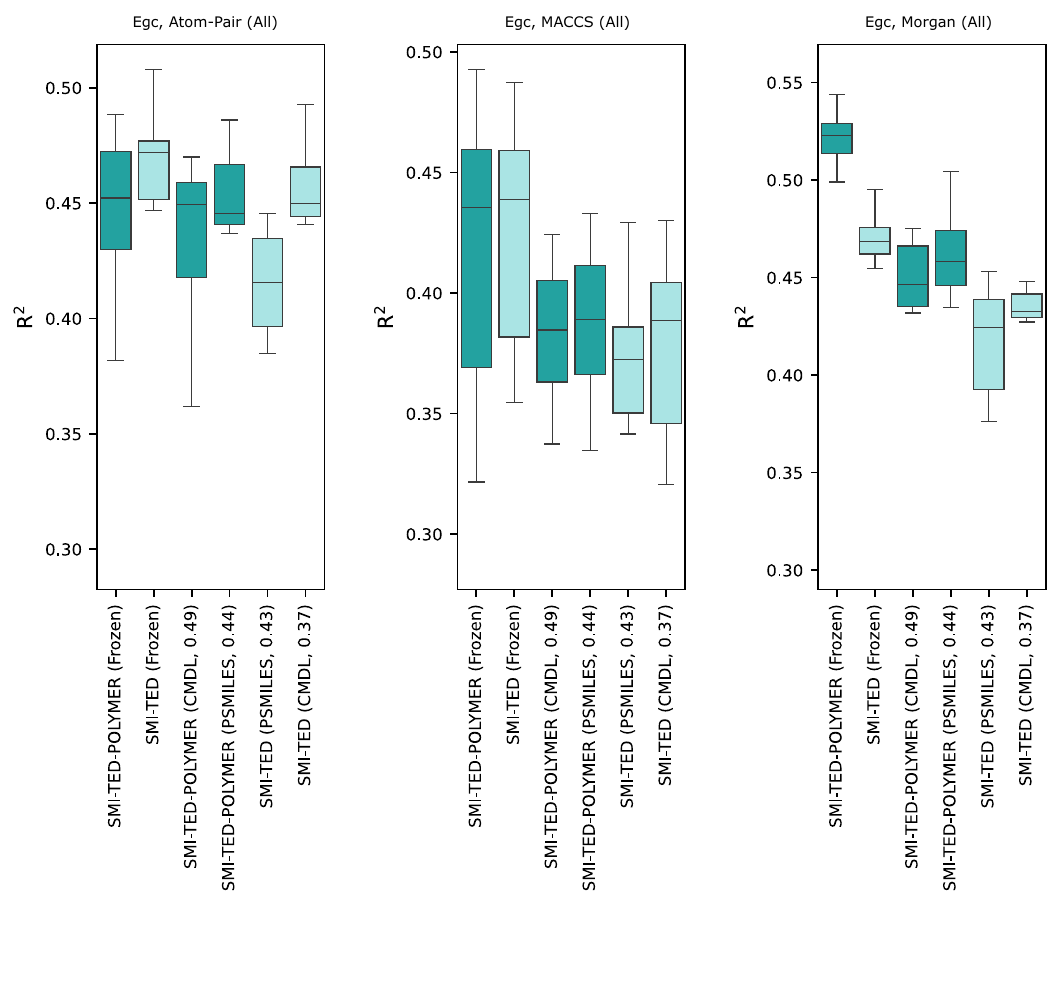}
\caption{Plots of fingerprint R\textsuperscript{2} value distributions for the E\textsubscript{gc} benchmark dataset using the \textit{all} substitution strategy.}\label{egc_knn_fp_all}
\end{figure}

\begin{figure}[H]
\centering
\includegraphics[width=1\textwidth]{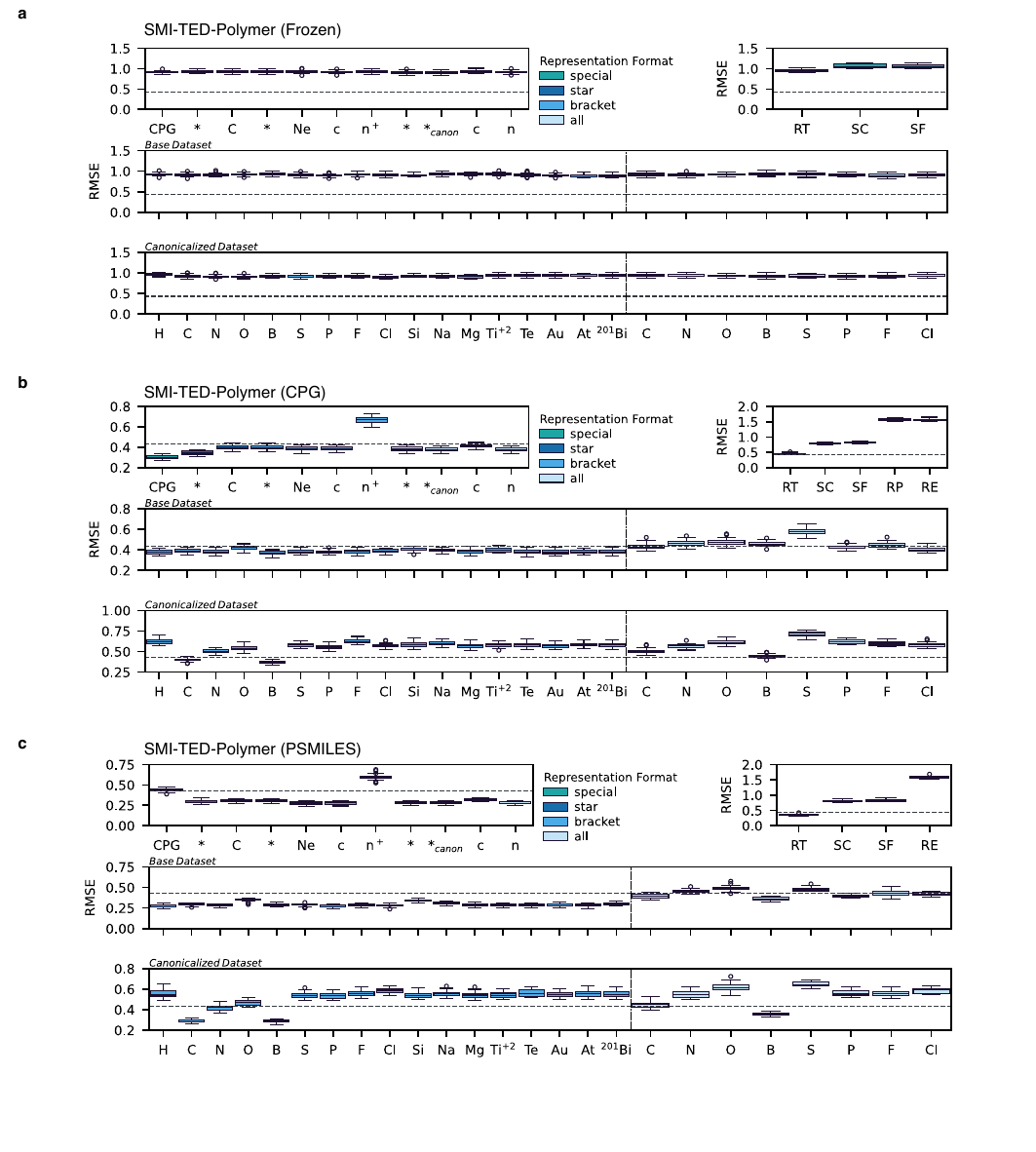}
\caption{RMSE values from repeated 5-fold cross-validation experiments with the E\textsubscript{gc} dataset using SMI-TED-Polymer models with frozen weights (\textbf{a}) or fine-tuned with CPG (\textbf{b}) and PSMILES (\textbf{c}).}\label{egc_stp_rmse}
\end{figure}

\begin{figure}[H]
\centering
\includegraphics[width=1\textwidth]{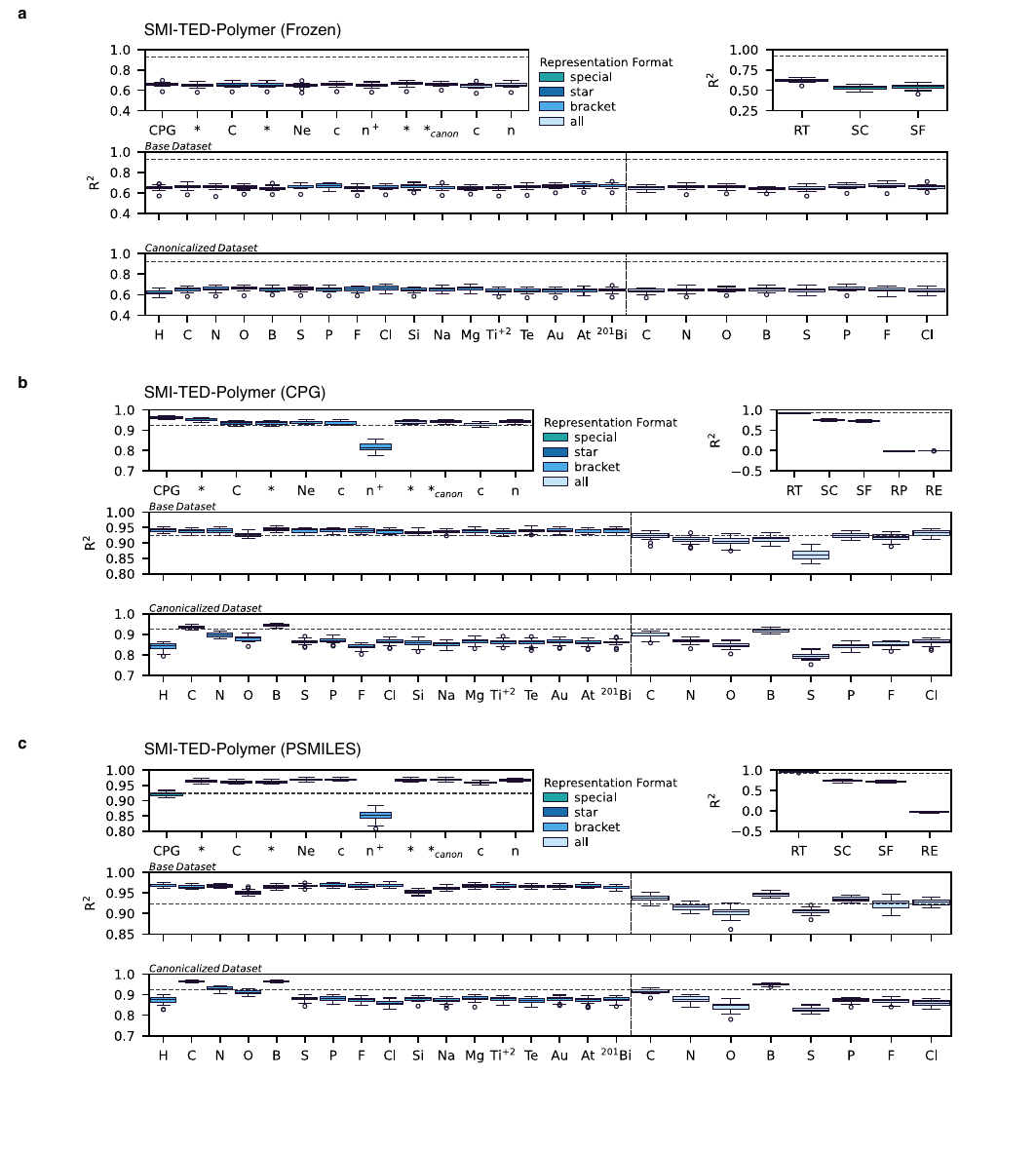}
\caption{R\textsuperscript{2} values from repeated 5-fold cross-validation experiments with the E\textsubscript{gc} dataset using SMI-TED-Polymer models with frozen weights (\textbf{a}) or fine-tuned with CPG (\textbf{b}) and PSMILES (\textbf{c}).}\label{egc_stp_r2}
\end{figure}

\begin{figure}[H]
\centering
\includegraphics[width=1\textwidth]{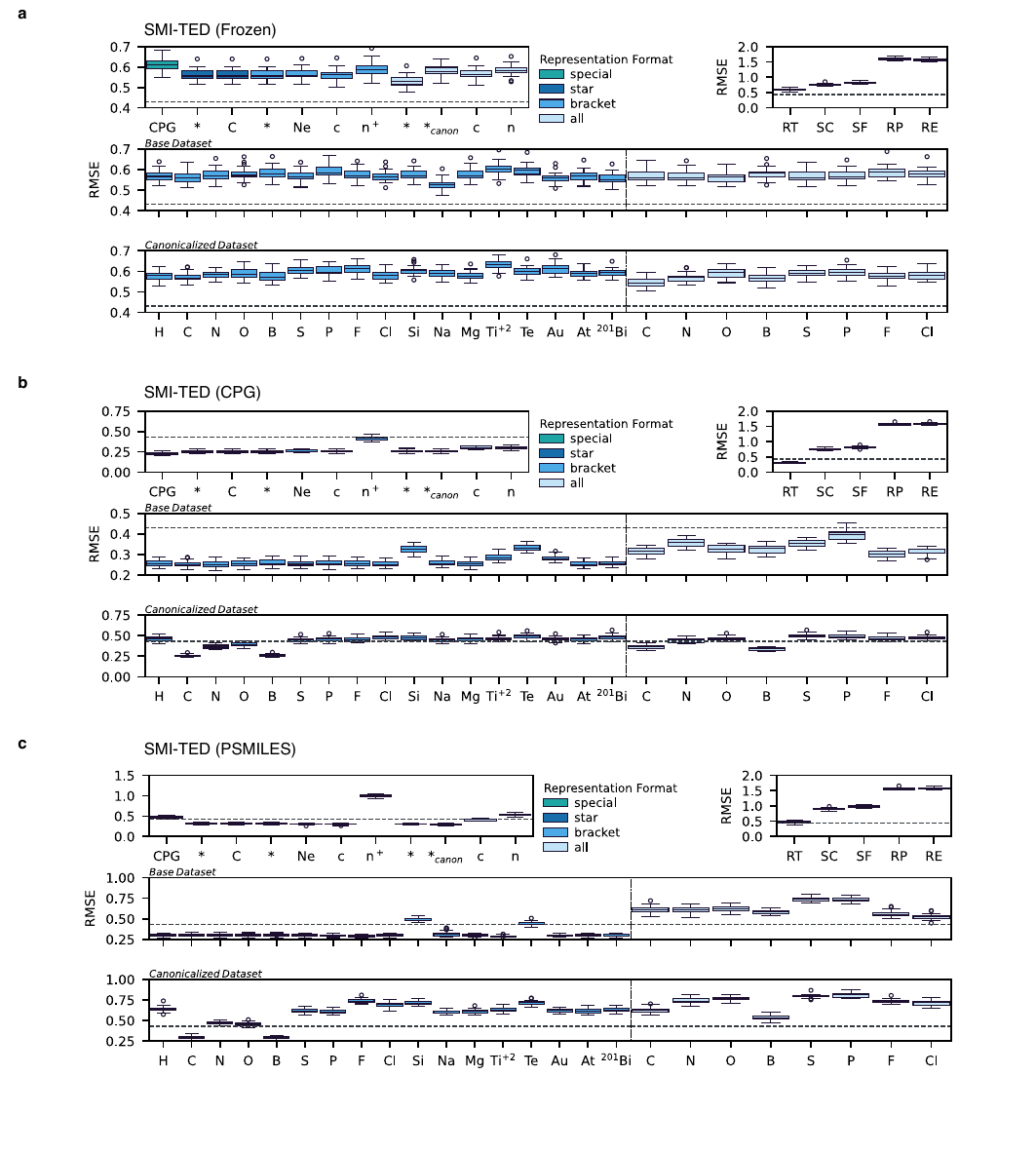}
\caption{RMSE values from repeated 5-fold cross-validation experiments with the E\textsubscript{gc} dataset using SMI-TED models with frozen weights (\textbf{a}) or fine-tuned with CPG (\textbf{b}) and PSMILES (\textbf{c}).}\label{egc_st_rmse}
\end{figure}

\begin{figure}[H]
\centering
\includegraphics[width=1\textwidth]{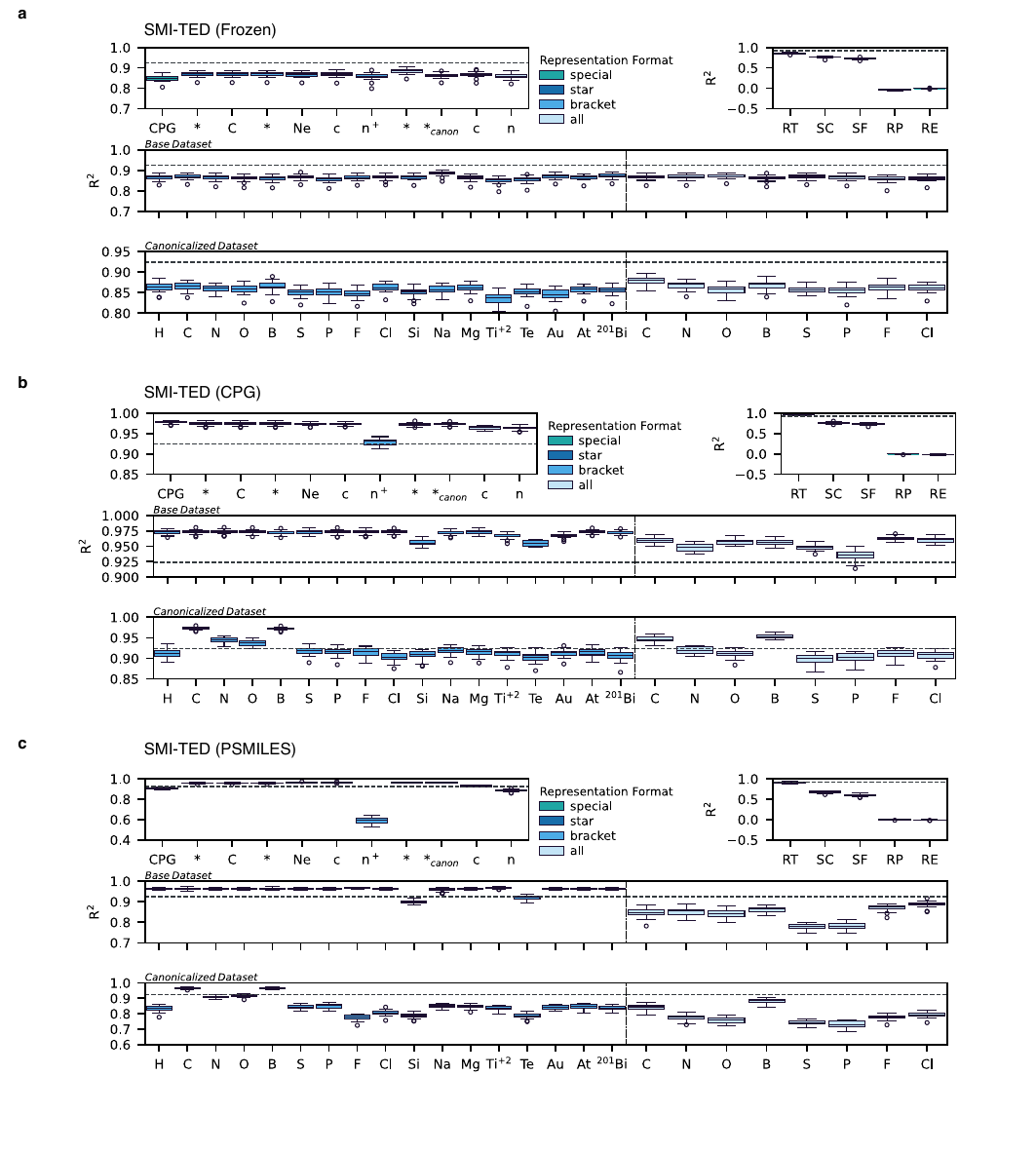}
\caption{R\textsuperscript{2} values from repeated 5-fold cross-validation experiments with the E\textsubscript{gc} dataset using SMI-TED models with frozen weights (\textbf{a}) or fine-tuned with CPG (\textbf{b}) and PSMILES (\textbf{c}).}\label{egc_st_r2}
\end{figure}

\subsubsection{Polymer Bulk Bandgap (E\textsubscript{gb})}\label{cvEgb}

\begin{figure}[H]
\centering
\includegraphics[width=1\textwidth]{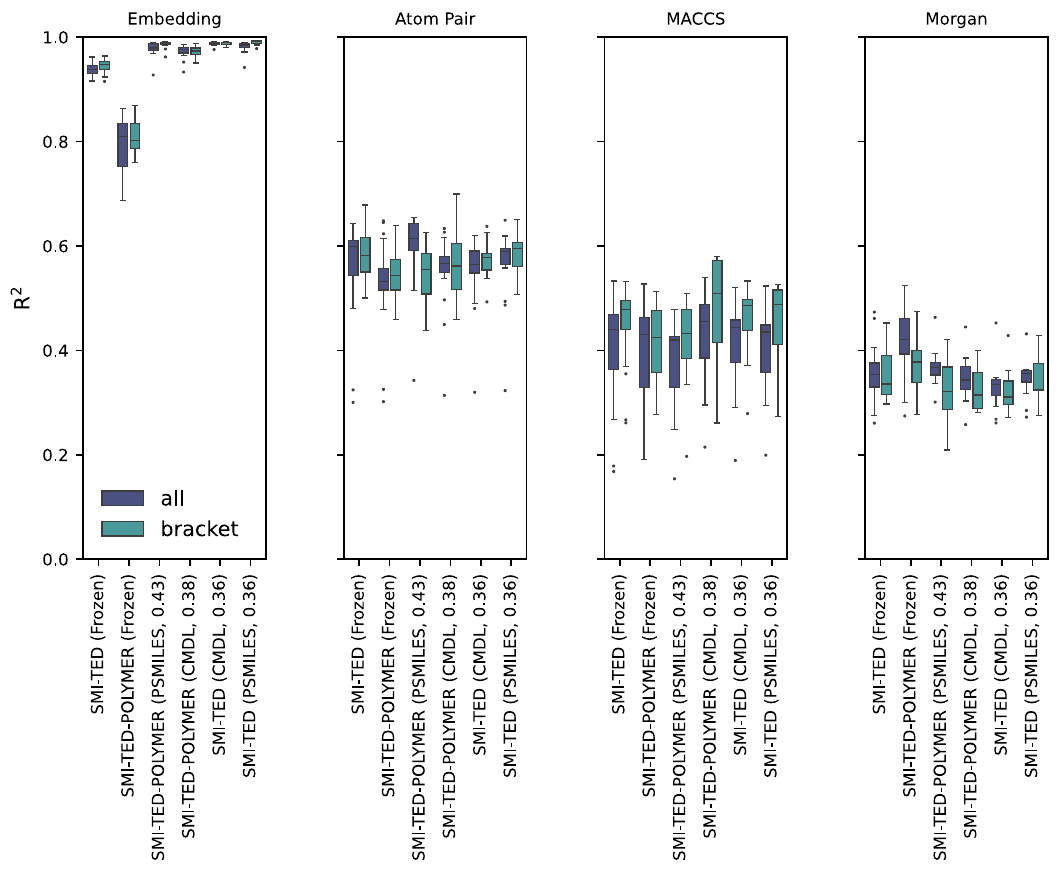}
\caption{Plots of embedding and fingerprint R\textsuperscript{2} value distributions for the E\textsubscript{gb} benchmark dataset using the \textit{bracket} and \textit{all} substitution strategy.}\label{egb_knn_fp_bracket}
\end{figure}

\begin{figure}[H]
\centering
\includegraphics[width=1\textwidth]{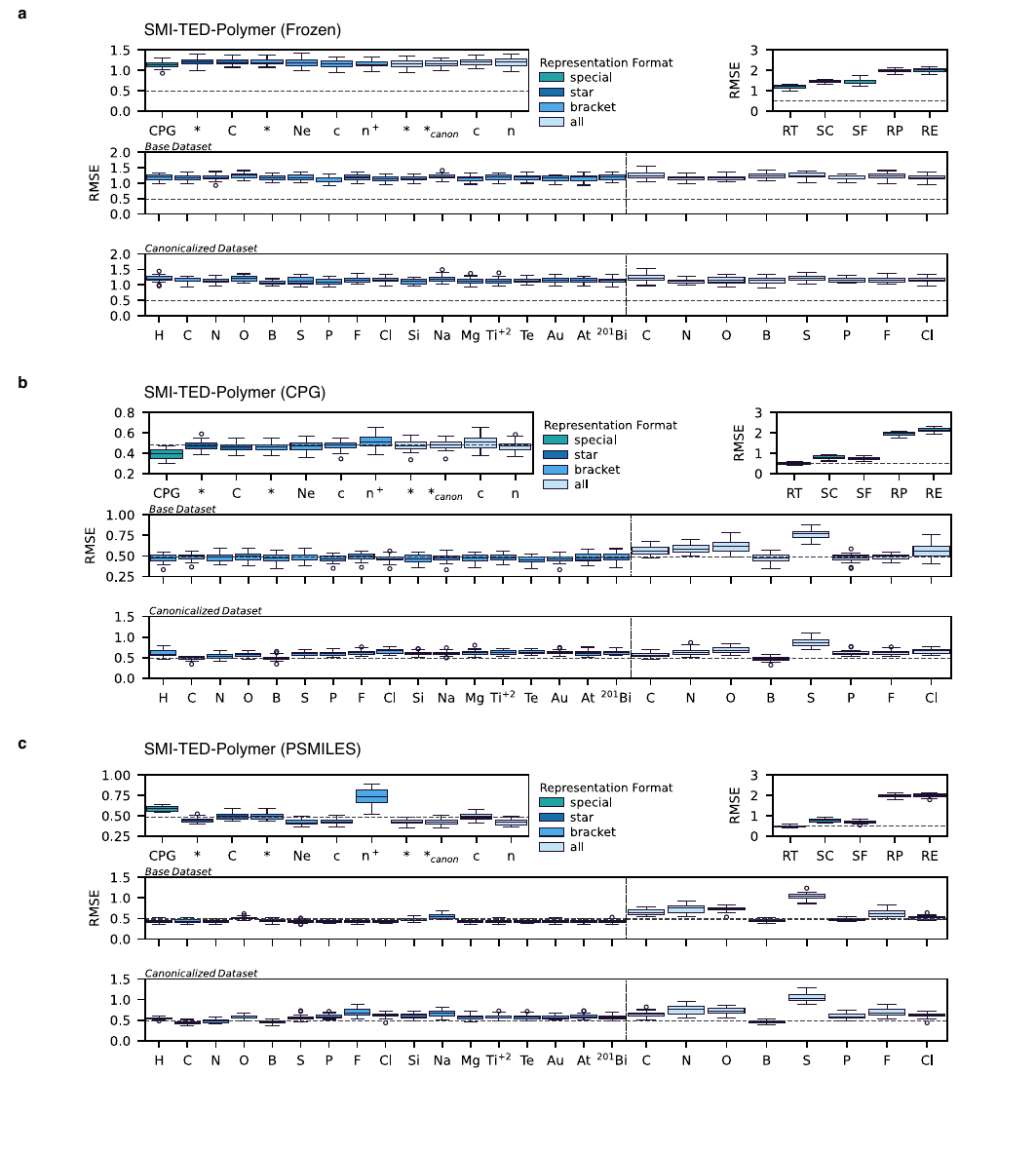}
\caption{RMSE values from repeated 5-fold cross-validation experiments with the E\textsubscript{gb} dataset using SMI-TED-Polymer models with frozen weights (\textbf{a}) or fine-tuned with CPG (\textbf{b}) and PSMILES (\textbf{c}).}\label{egb_stp_rmse}
\end{figure}

\begin{figure}[H]
\centering
\includegraphics[width=1\textwidth]{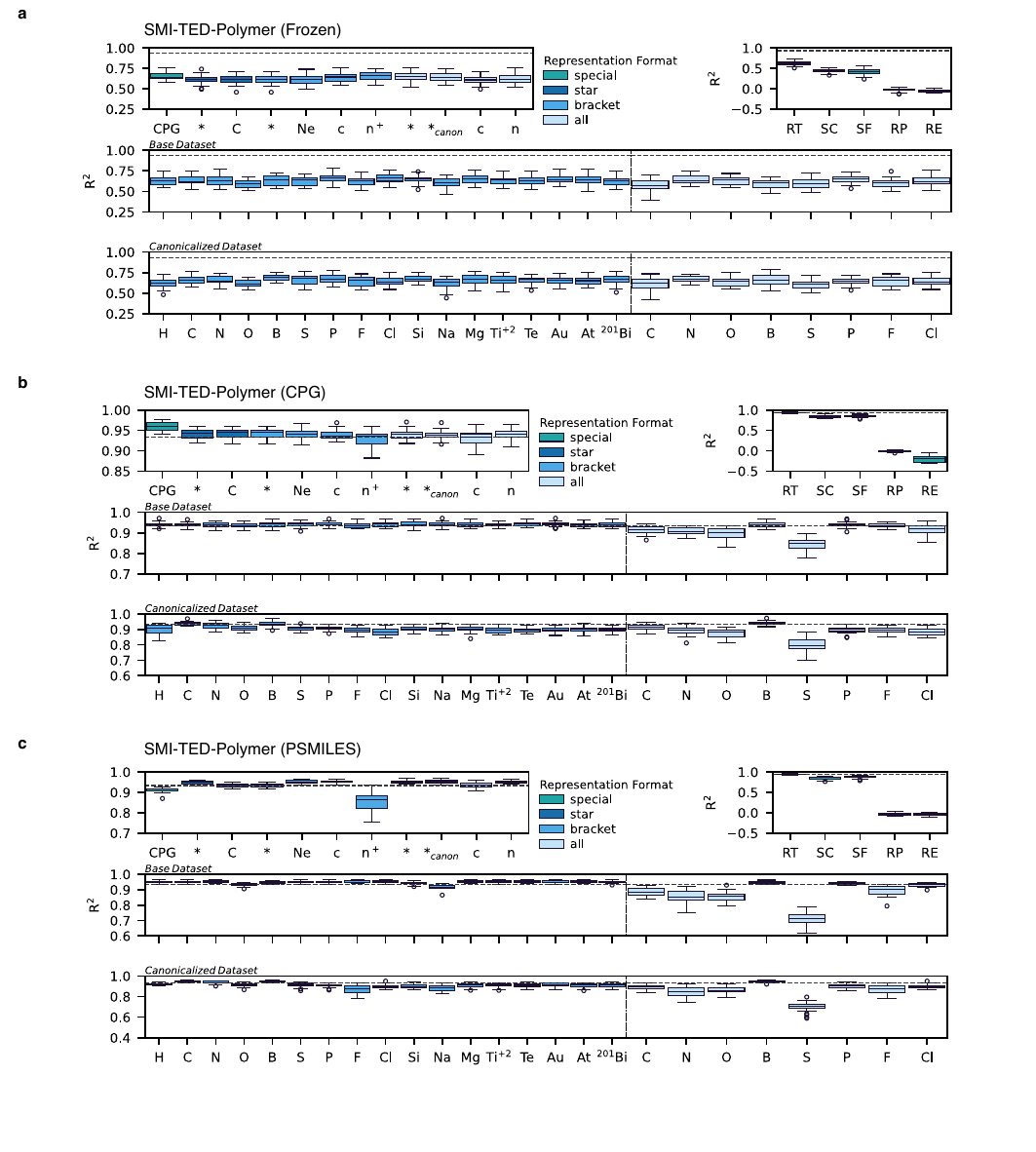}
\caption{R\textsuperscript{2} values from repeated 5-fold cross-validation experiments with the E\textsubscript{gb} dataset using SMI-TED-Polymer models with frozen weights (\textbf{a}) or fine-tuned with CPG (\textbf{b}) and PSMILES (\textbf{c}).}\label{egb_stp_r2}
\end{figure}

\begin{figure}[H]
\centering
\includegraphics[width=1\textwidth]{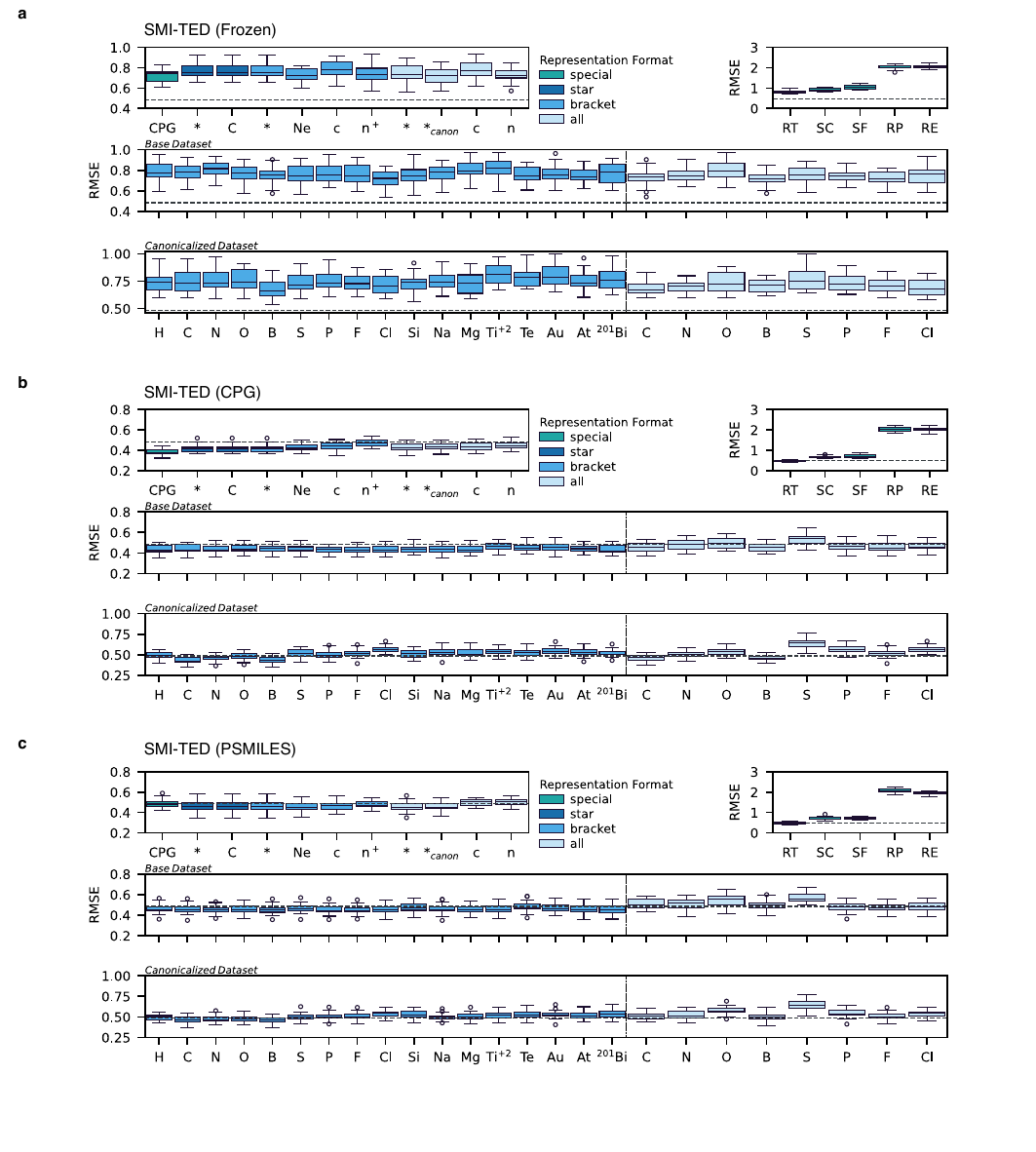}
\caption{RMSE values from repeated 5-fold cross-validation experiments with the E\textsubscript{gb} dataset using SMI-TED models with frozen weights (\textbf{a}) or fine-tuned with CPG (\textbf{b}) and PSMILES (\textbf{c}).}\label{egb_st_rmse}
\end{figure}

\begin{figure}[H]
\centering
\includegraphics[width=1\textwidth]{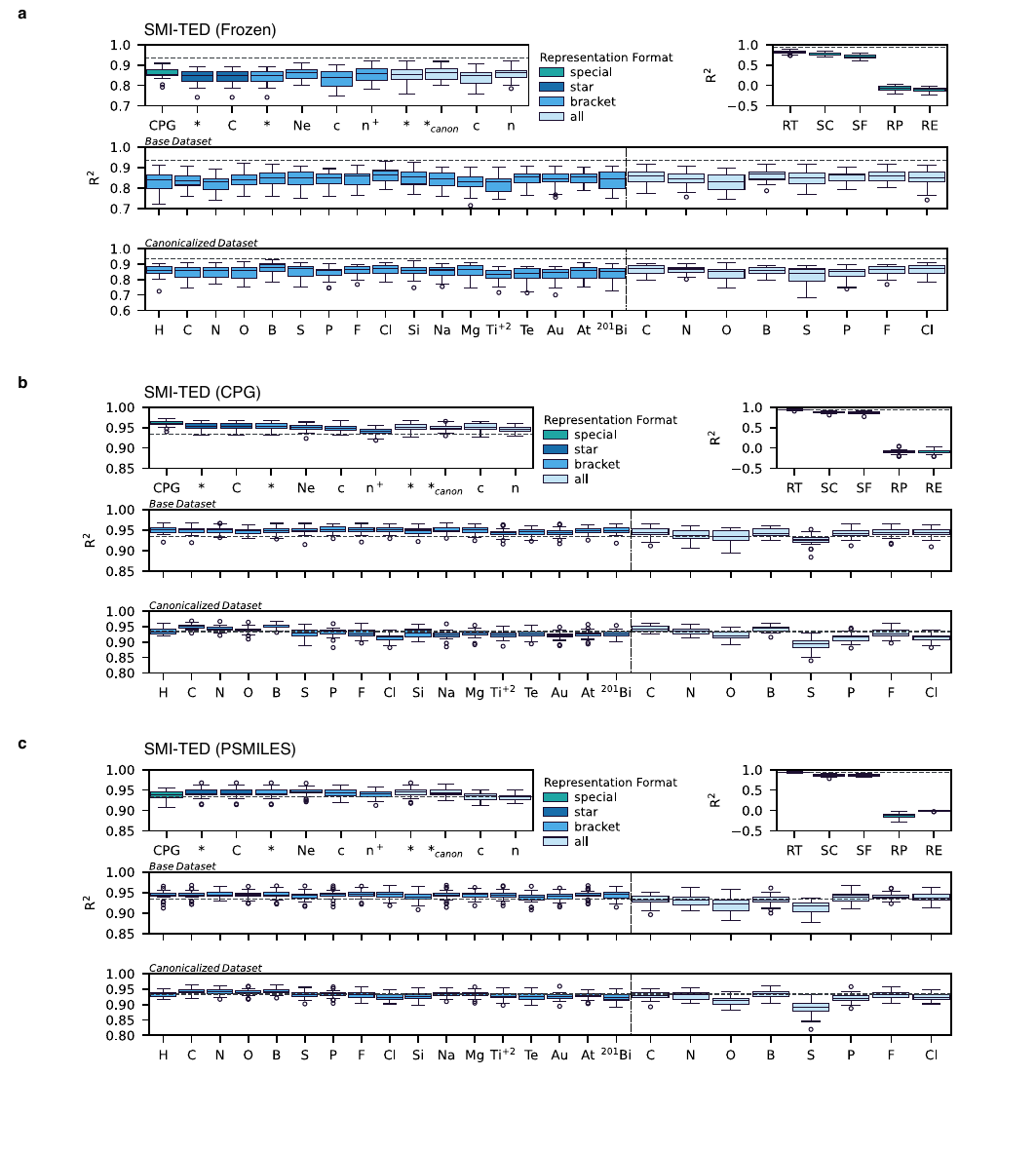}
\caption{R\textsuperscript{2} values from repeated 5-fold cross-validation experiments with the E\textsubscript{gb} dataset using SMI-TED models with frozen weights (\textbf{a}) or fine-tuned with CPG (\textbf{b}) and PSMILES (\textbf{c}).}\label{egb_st_r2}
\end{figure}

\subsubsection{Polymer Ionization Energy (E\textsubscript{i})}\label{cvEi}

\begin{figure}[H]
\centering
\includegraphics[width=1\textwidth]{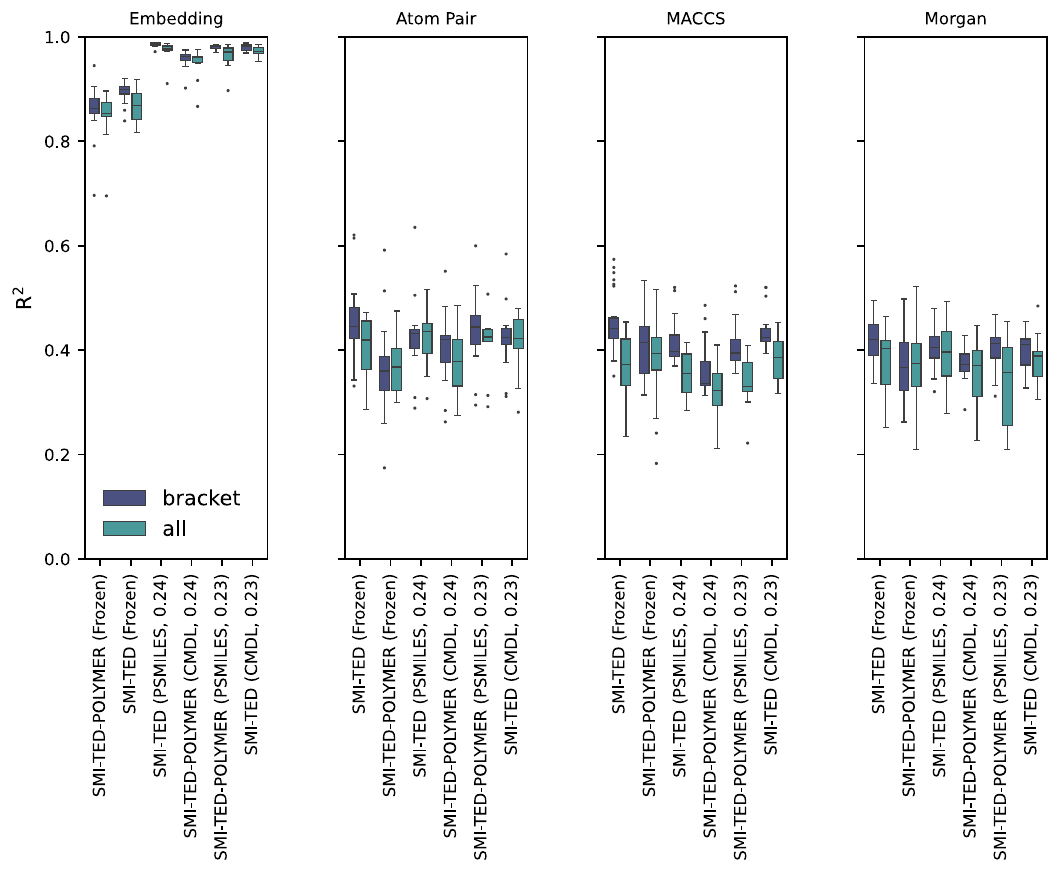}
\caption{Plots of embededing and fingerprint R\textsuperscript{2} value distributions for the E\textsubscript{i} benchmark dataset for the \textit{bracket} and \textit{all} substitution strategy.}\label{ei_knn_fp_bracket}
\end{figure}

\begin{figure}[H]
\centering
\includegraphics[width=1\textwidth]{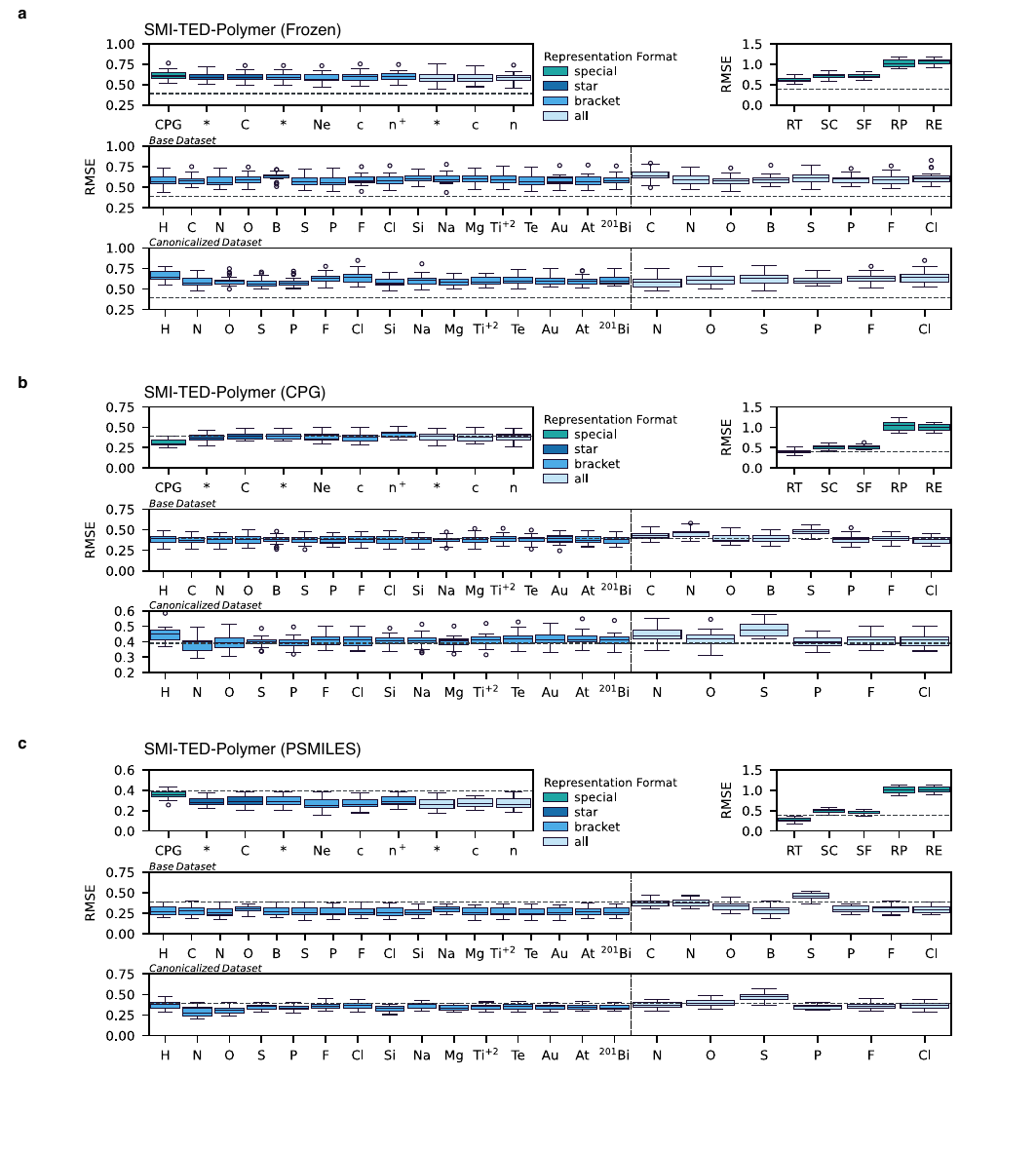}
\caption{RMSE values from repeated 5-fold cross-validation experiments with the E\textsubscript{i} dataset using SMI-TED-Polymer models with frozen weights (\textbf{a}) or fine-tuned with CPG (\textbf{b}) and PSMILES (\textbf{c}).}\label{ei_stp_rmse}
\end{figure}

\begin{figure}[H]
\centering
\includegraphics[width=1\textwidth]{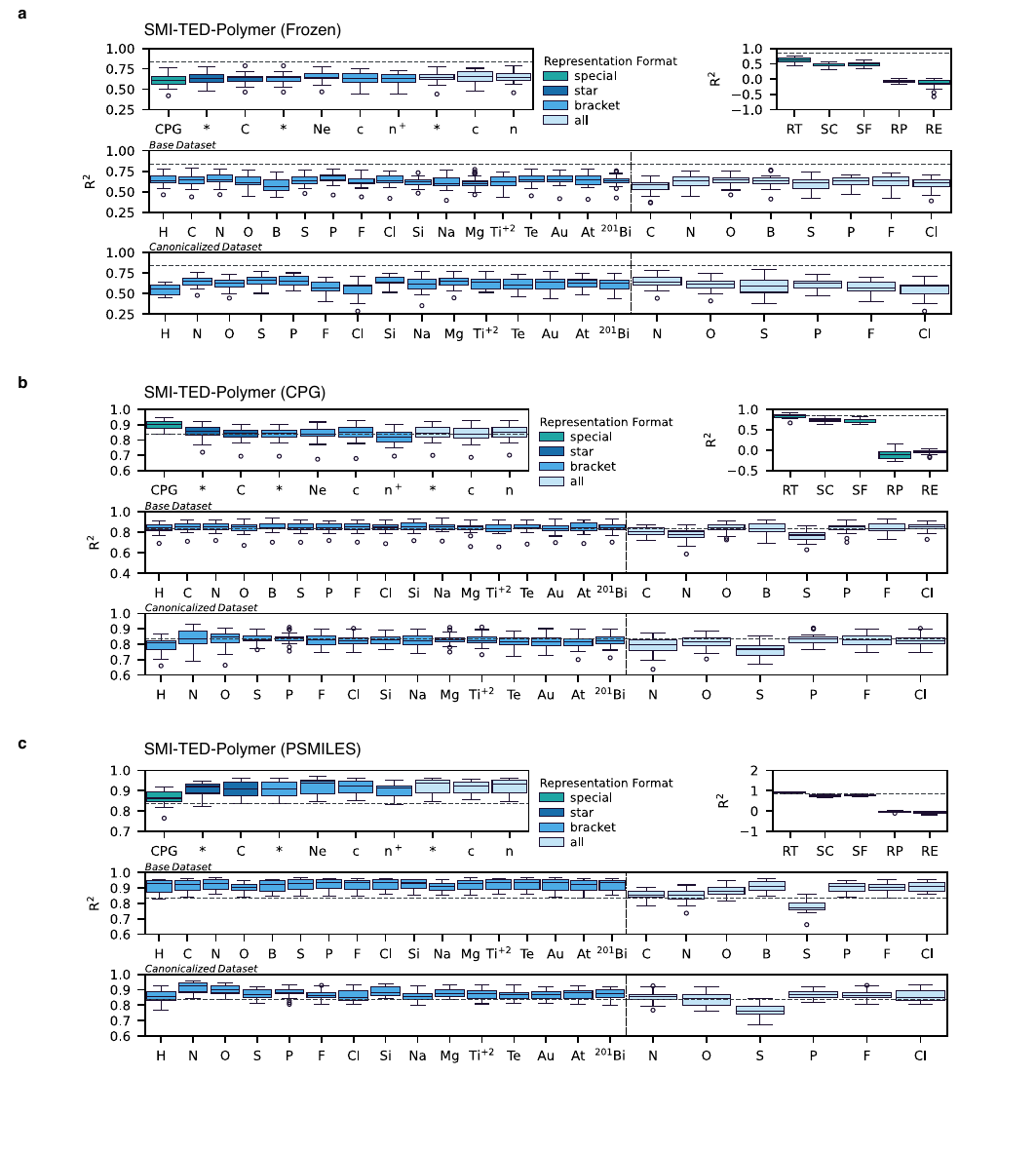}
\caption{R\textsuperscript{2} values from repeated 5-fold cross-validation experiments with the E\textsubscript{i} dataset using SMI-TED-Polymer models with frozen weights (\textbf{a}) or fine-tuned with CPG (\textbf{b}) and PSMILES (\textbf{c}).}\label{ei_stp_r2}
\end{figure}

\begin{figure}[H]
\centering
\includegraphics[width=1\textwidth]{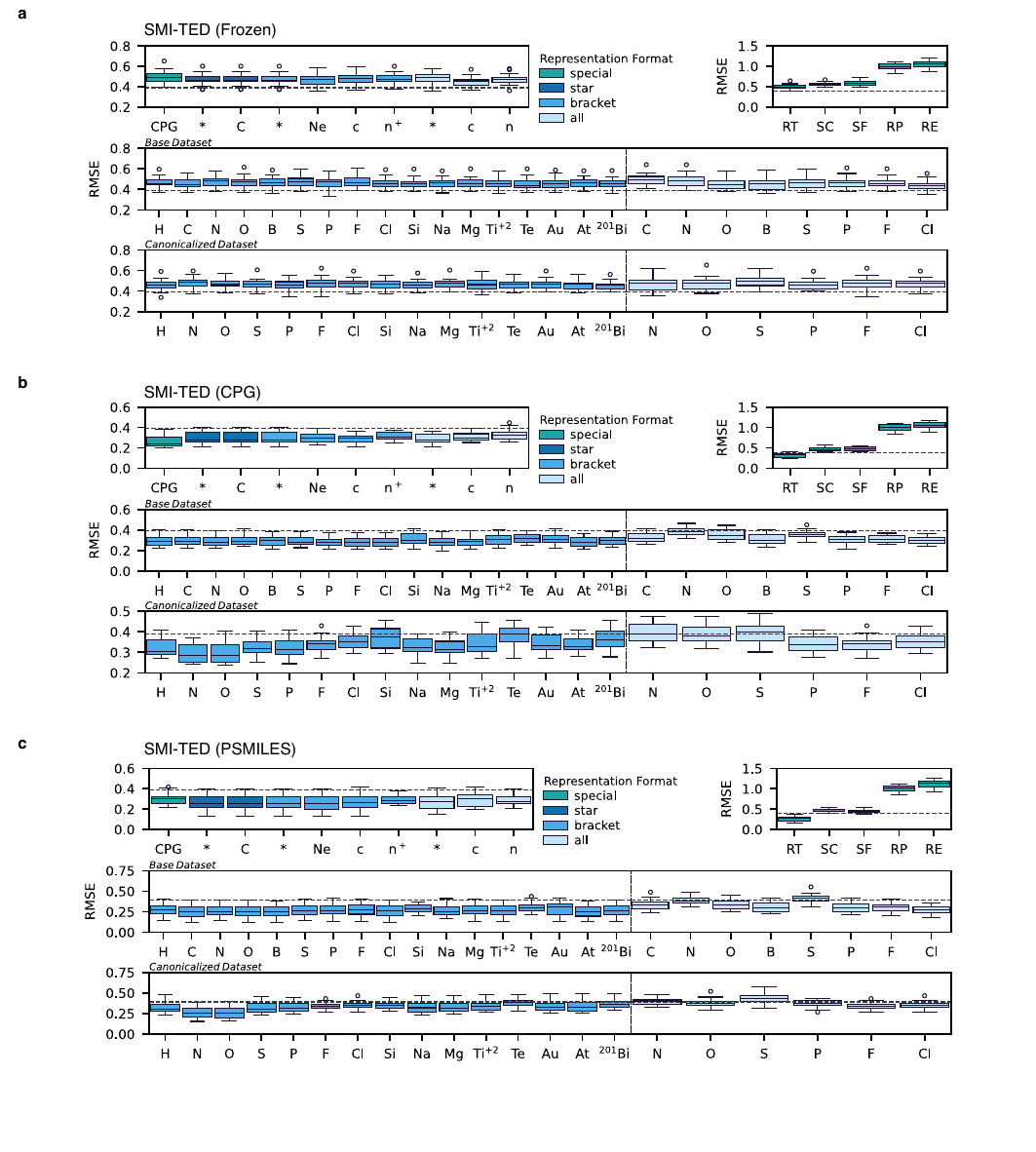}
\caption{RMSE values from repeated 5-fold cross-validation experiments with the E\textsubscript{i} dataset using SMI-TED models with frozen weights (\textbf{a}) or fine-tuned with CPG (\textbf{b}) and PSMILES (\textbf{c}).}\label{ei_st_rmse}
\end{figure}

\begin{figure}[H]
\centering
\includegraphics[width=1\textwidth]{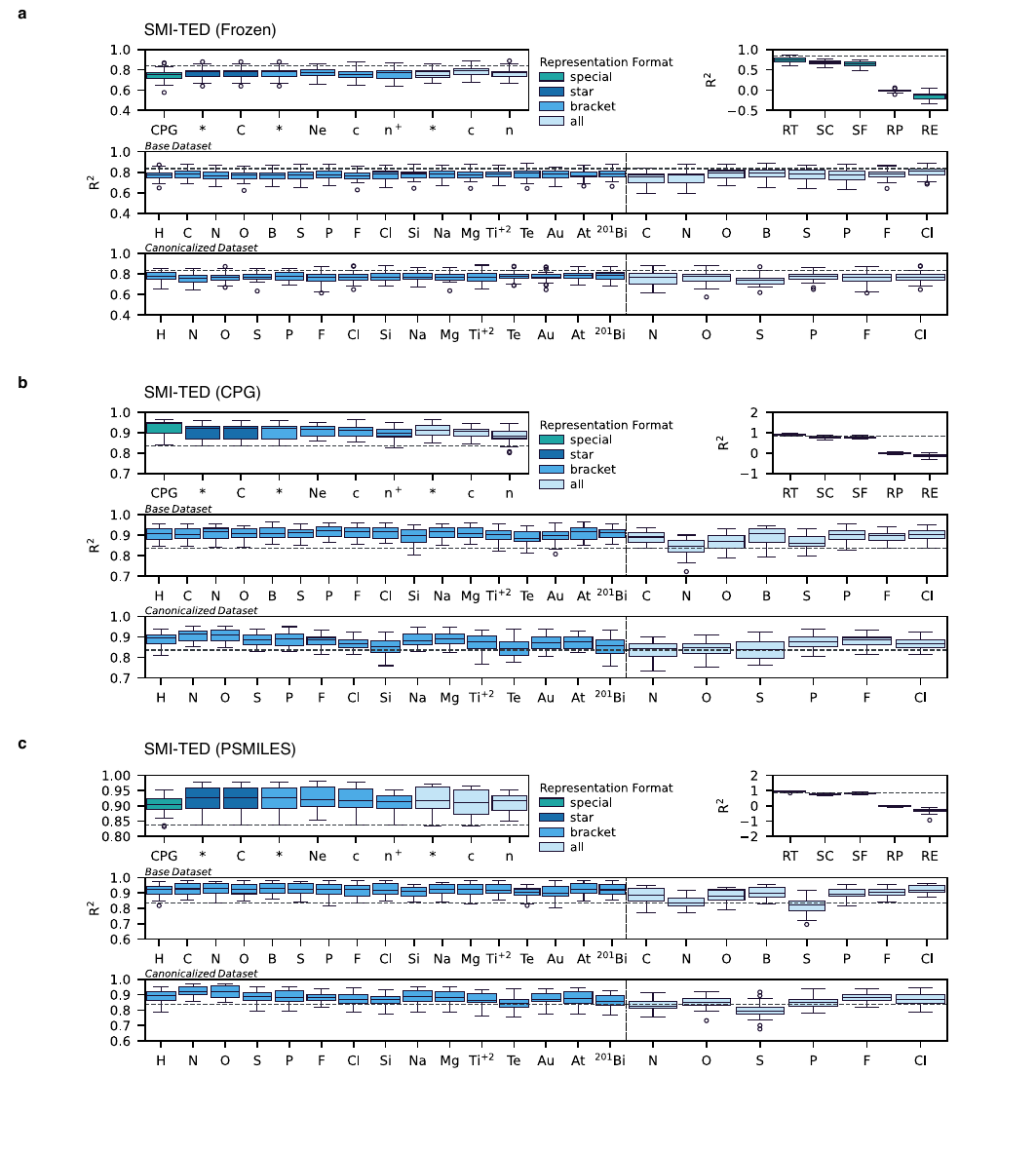}
\caption{R\textsuperscript{2} values from repeated 5-fold cross-validation experiments with the E\textsubscript{i} dataset using SMI-TED models with frozen weights (\textbf{a}) or fine-tuned with CPG (\textbf{b}) and PSMILES (\textbf{c}).}\label{ei_st_r2}
\end{figure}

\subsubsection{Polymer Dielectric Constant (EPS)}\label{cvEPS}

\begin{figure}[H]
\centering
\includegraphics[width=1\textwidth]{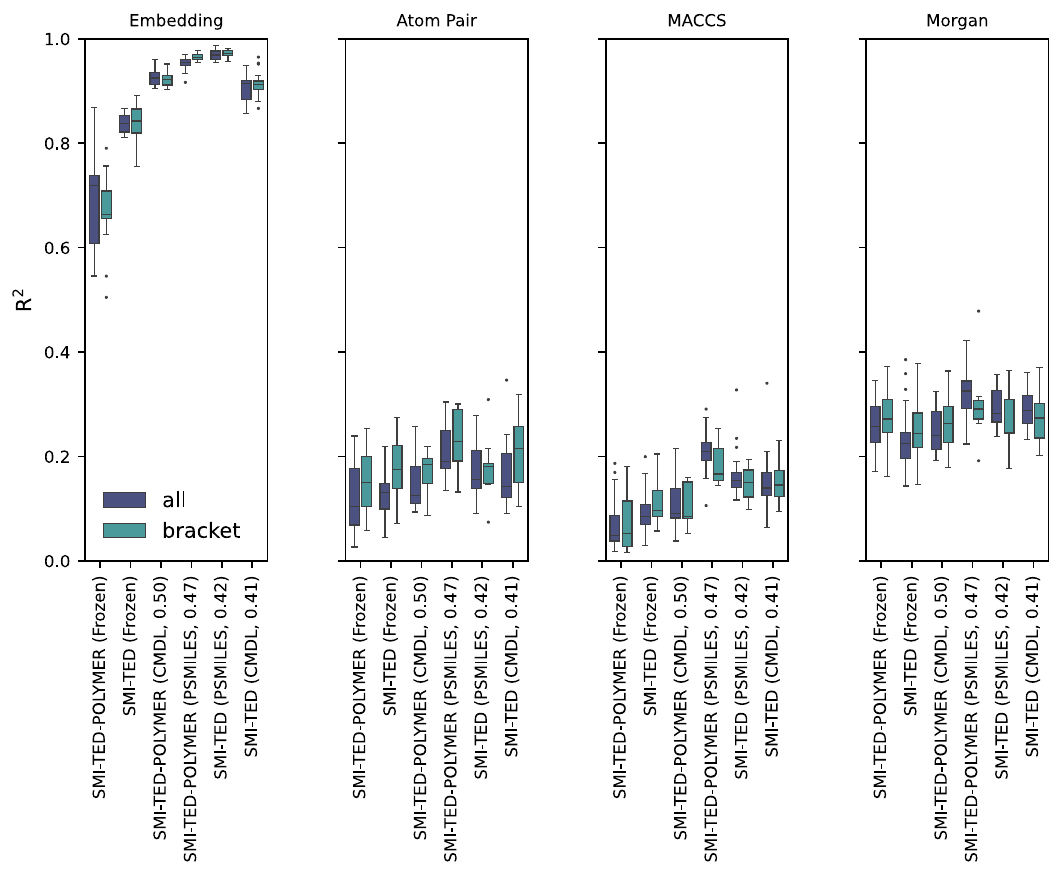}
\caption{Plots of embedding and fingerprint R\textsuperscript{2} value distributions for the EPS benchmark dataset using the \textit{bracket} and \textit{all} substitution strategy.}\label{eps_knn_fp_bracket}
\end{figure}

\begin{figure}[H]
\centering
\includegraphics[width=1\textwidth]{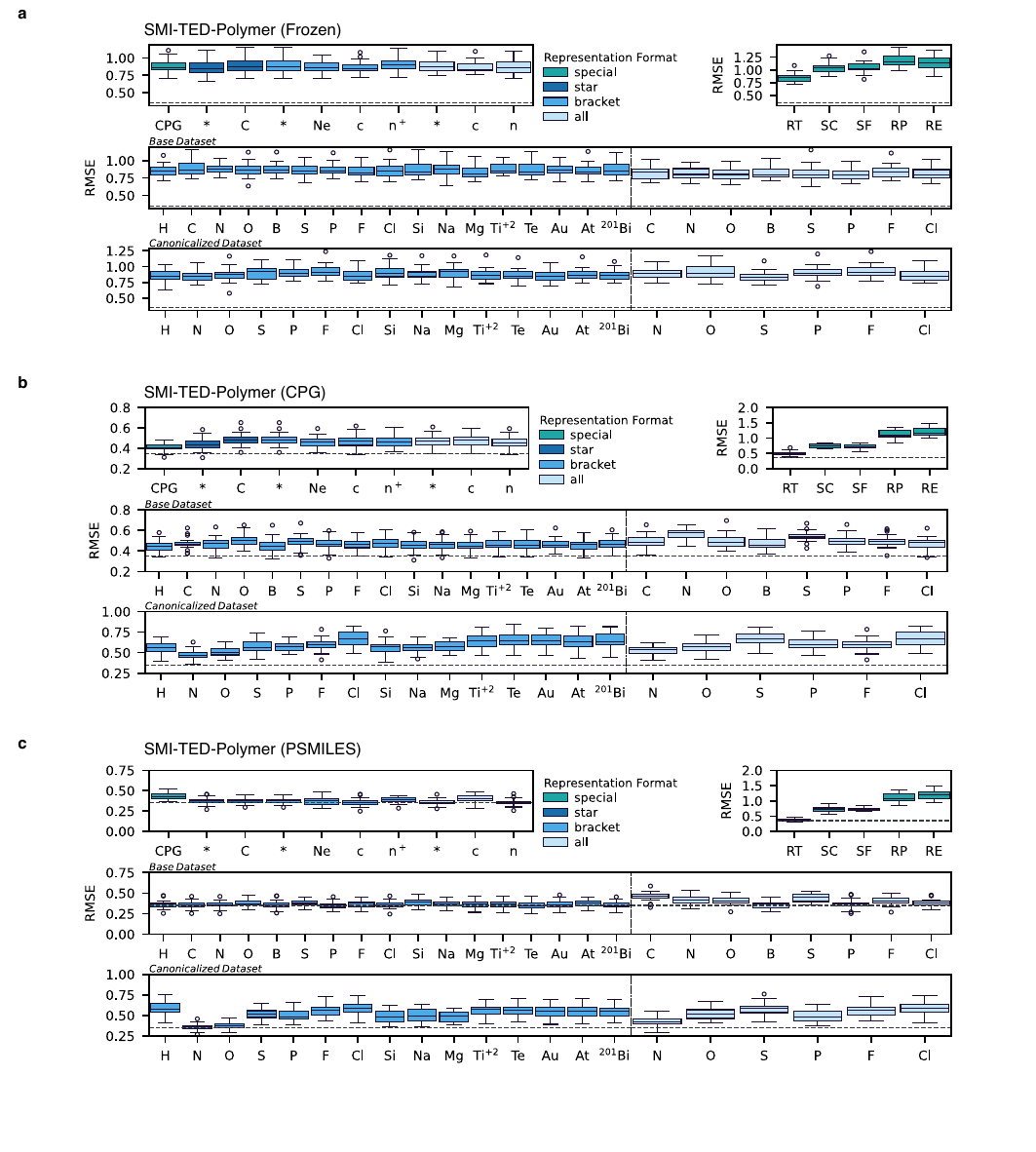}
\caption{RMSE values from repeated 5-fold cross-validation experiments with the EPS dataset using SMI-TED-Polymer models with frozen weights (\textbf{a}) or fine-tuned with CPG (\textbf{b}) and PSMILES (\textbf{c}).}\label{eps_stp_rmse}
\end{figure}

\begin{figure}[H]
\centering
\includegraphics[width=1\textwidth]{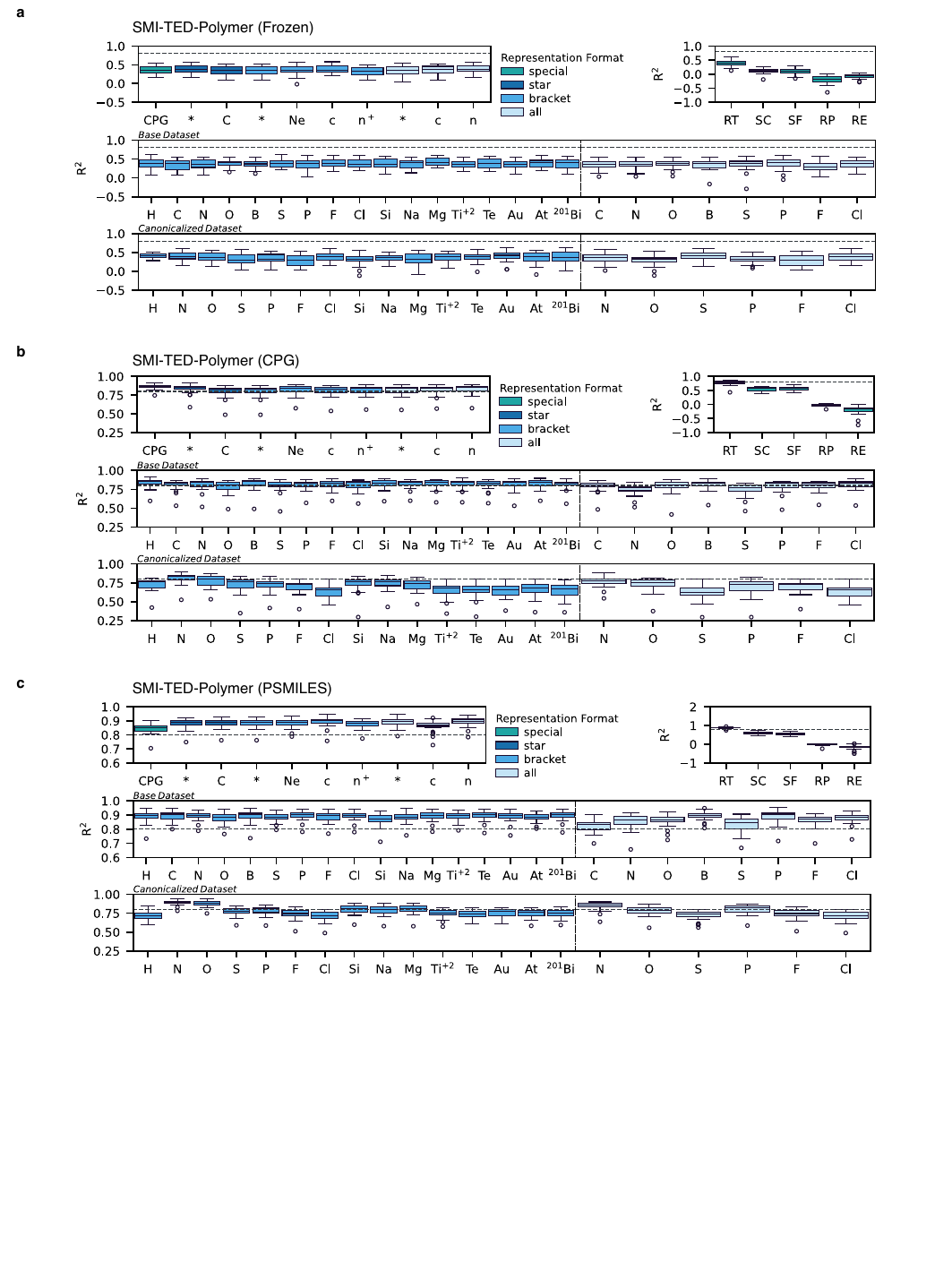}
\caption{R\textsuperscript{2} values from repeated 5-fold cross-validation experiments with the EPS dataset using SMI-TED-Polymer models with frozen weights (\textbf{a}) or fine-tuned with CPG (\textbf{b}) and PSMILES (\textbf{c}).}\label{eps_stp_r2}
\end{figure}

\begin{figure}[H]
\centering
\includegraphics[width=1\textwidth]{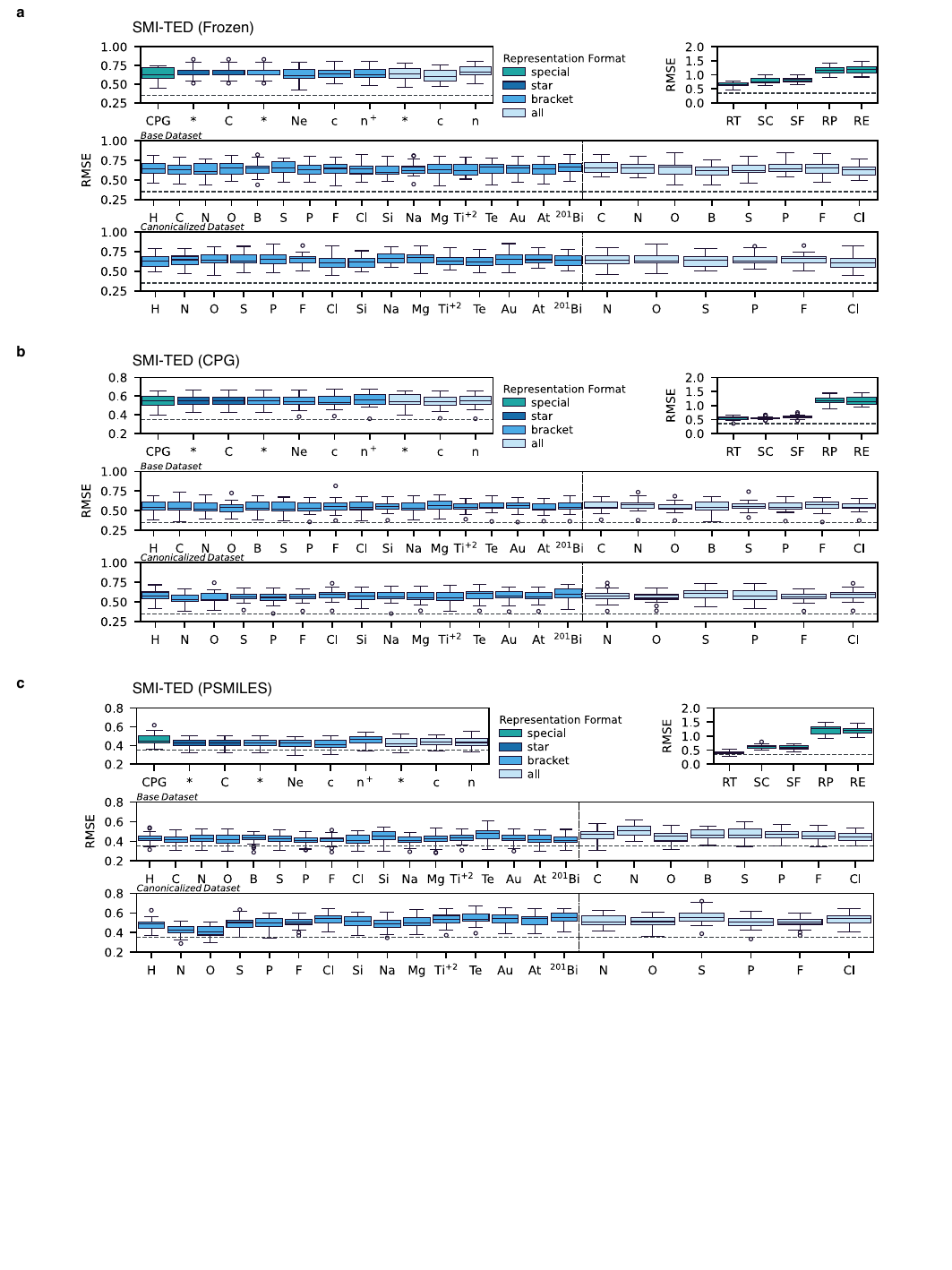}
\caption{RMSE values from repeated 5-fold cross-validation experiments with the EPS dataset using SMI-TED models with frozen weights (\textbf{a}) or fine-tuned with CPG (\textbf{b}) and PSMILES (\textbf{c}).}\label{eps_st_rmse}
\end{figure}

\begin{figure}[H]
\centering
\includegraphics[width=1\textwidth]{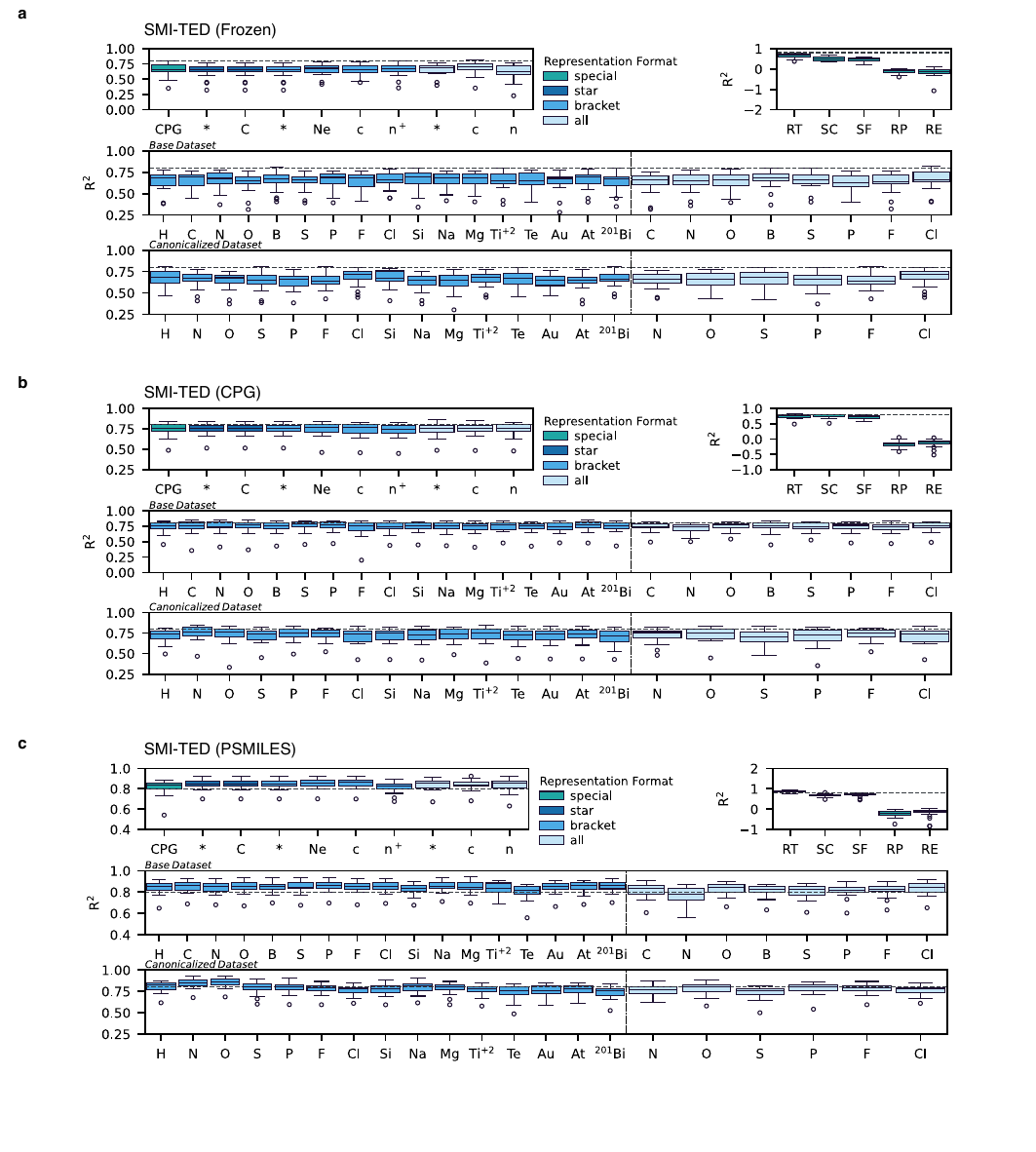}
\caption{R\textsuperscript{2} values from repeated 5-fold cross-validation experiments with the EPS dataset using SMI-TED models with frozen weights (\textbf{a}) or fine-tuned with CPG (\textbf{b}) and PSMILES (\textbf{c}).}\label{eps_st_r2}
\end{figure}

\subsubsection{Polymer Electron Affinity (E\textsubscript{ea})}\label{cvEea}

\begin{figure}[H]
\centering
\includegraphics[width=1\textwidth]{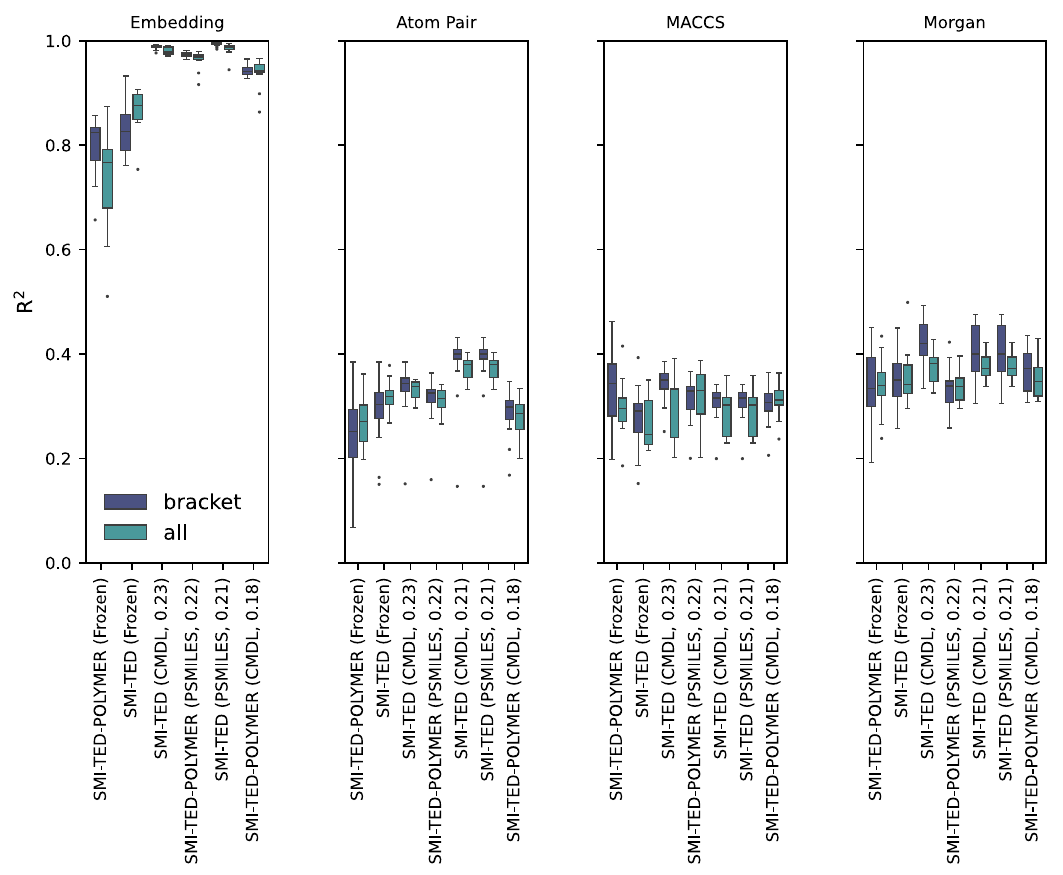}
\caption{Plots of embedding and fingerprint R\textsuperscript{2} value distributions for the E\textsubscript{ea} benchmark dataset using the \textit{bracket} and \textit{all} substitution strategy.}\label{eea_knn_fp_bracket}
\end{figure}

\begin{figure}[H]
\centering
\includegraphics[width=1\textwidth]{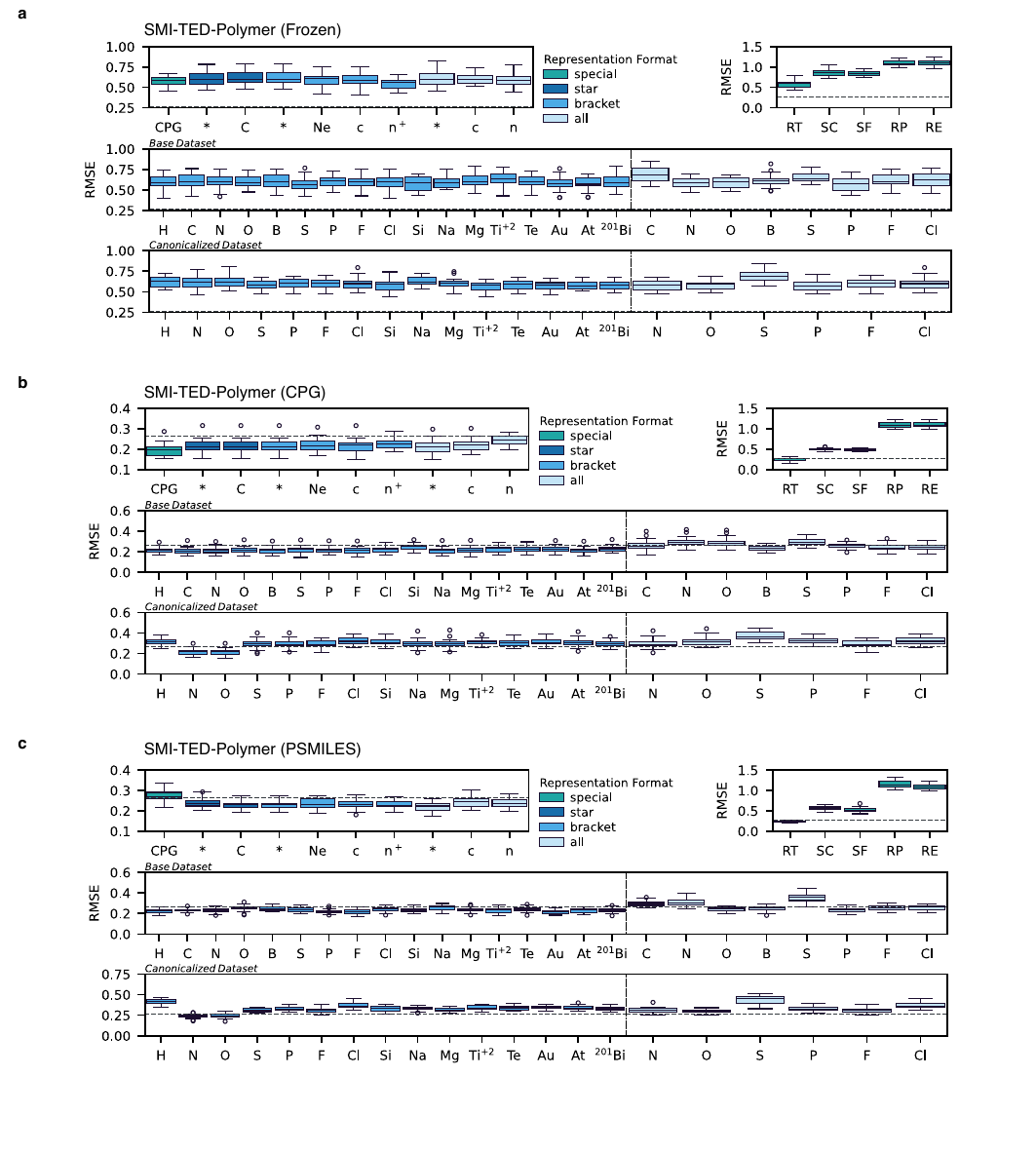}
\caption{RMSE values from repeated 5-fold cross-validation experiments with the E\textsubscript{ea} dataset using SMI-TED-Polymer models with frozen weights (\textbf{a}) or fine-tuned with CPG (\textbf{b}) and PSMILES (\textbf{c}).}\label{eea_stp_rmse}
\end{figure}

\begin{figure}[H]
\centering
\includegraphics[width=1\textwidth]{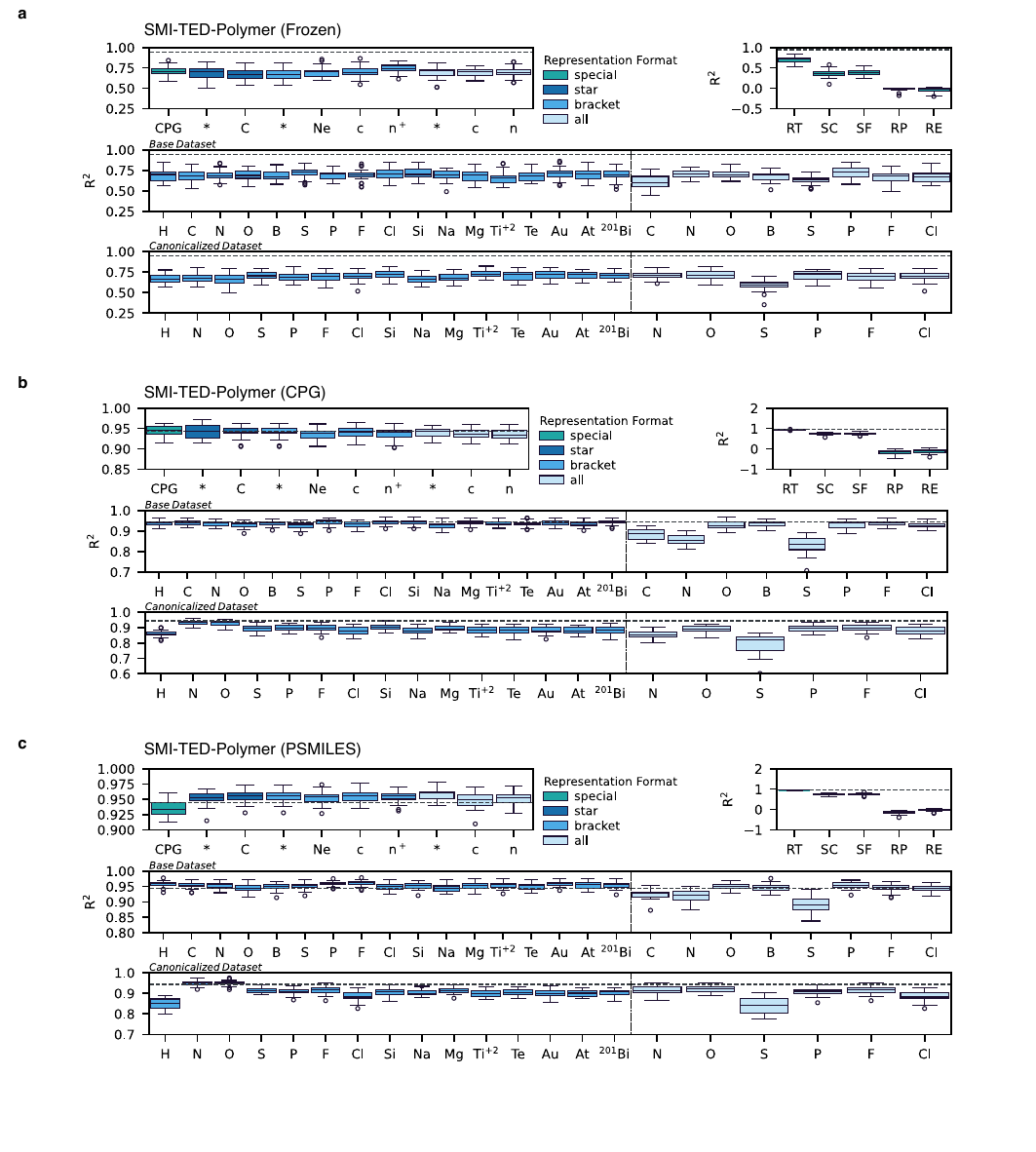}
\caption{R\textsuperscript{2} values from repeated 5-fold cross-validation experiments with the E\textsubscript{ea} dataset using SMI-TED-Polymer models with frozen weights (\textbf{a}) or fine-tuned with CPG (\textbf{b}) and PSMILES (\textbf{c}).}\label{eea_stp_r2}
\end{figure}

\begin{figure}[H]
\centering
\includegraphics[width=1\textwidth]{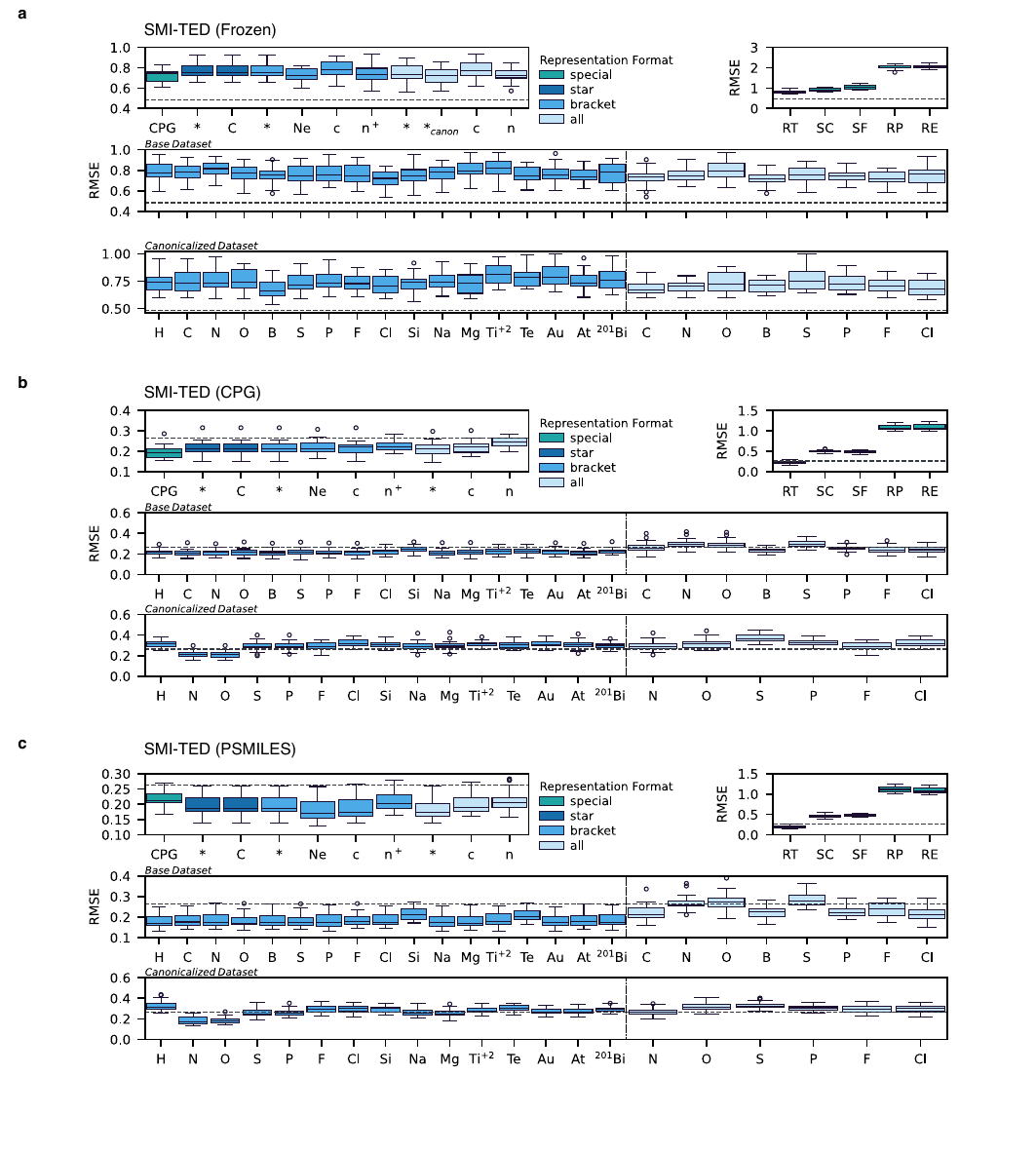}
\caption{RMSE values from repeated 5-fold cross-validation experiments with the E\textsubscript{ea} dataset using SMI-TED models with frozen weights (\textbf{a}) or fine-tuned with CPG (\textbf{b}) and PSMILES (\textbf{c}).}\label{eea_st_rmse}
\end{figure}

\begin{figure}[H]
\centering
\includegraphics[width=1\textwidth]{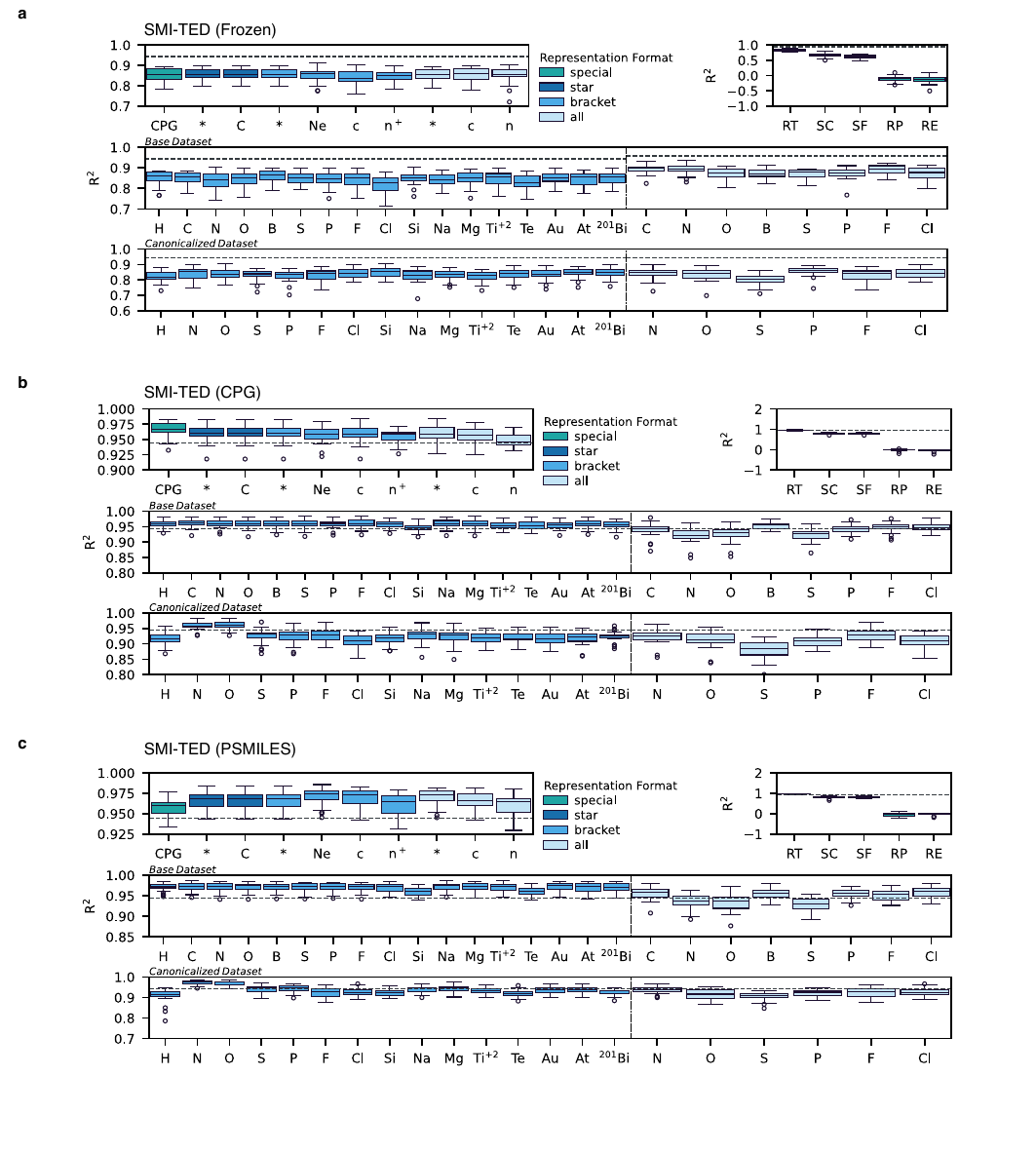}
\caption{R\textsuperscript{2} values from repeated 5-fold cross-validation experiments with the E\textsubscript{ea} dataset using SMI-TED models with frozen weights (\textbf{a}) or fine-tuned with CPG (\textbf{b}) and PSMILES (\textbf{c}).}\label{eea_st_r2}
\end{figure}

\subsubsection{Polymer Atomization Energy (E\textsubscript{at})}\label{cvEat}

\begin{figure}[H]
\centering
\includegraphics[width=1\textwidth]{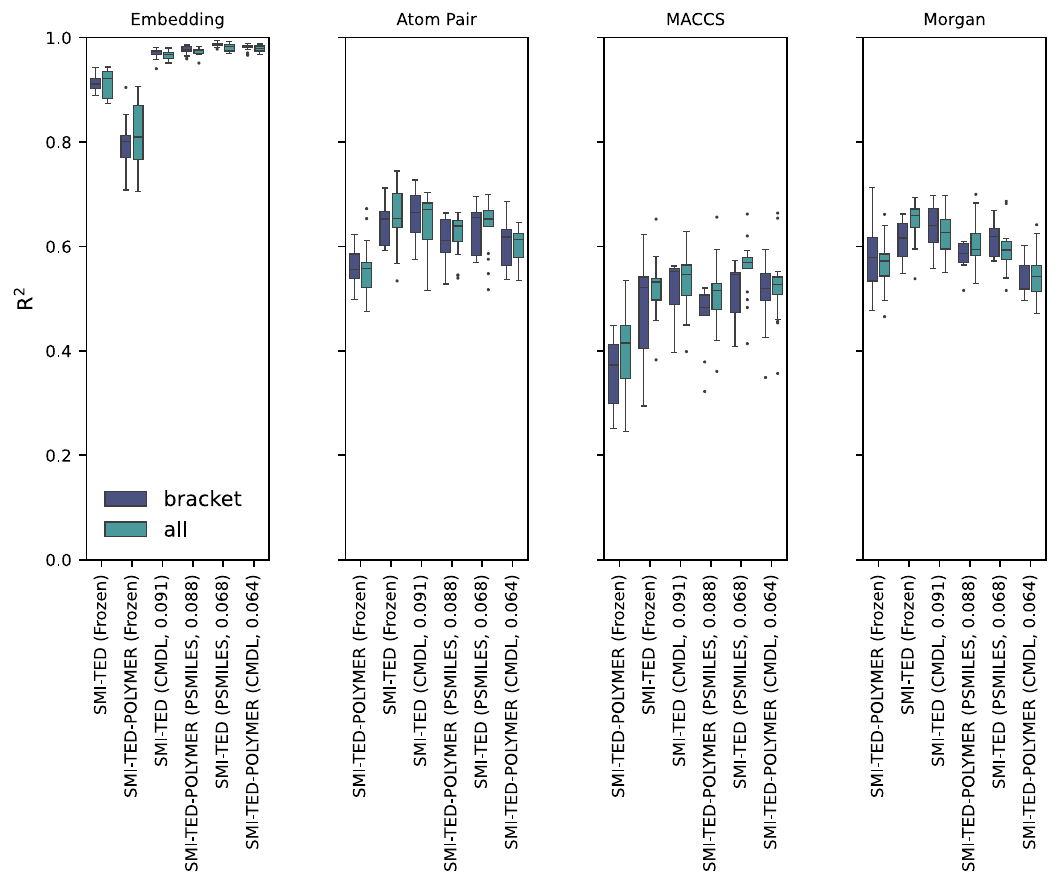}
\caption{Plots of embededing and fingerprint R\textsuperscript{2} value distributions for the E\textsubscript{at} benchmark dataset for the \textit{bracket} and \textit{all} substitution strategy.}\label{eat_knn_fp_all}
\end{figure}

\begin{figure}[H]
\centering
\includegraphics[width=1\textwidth]{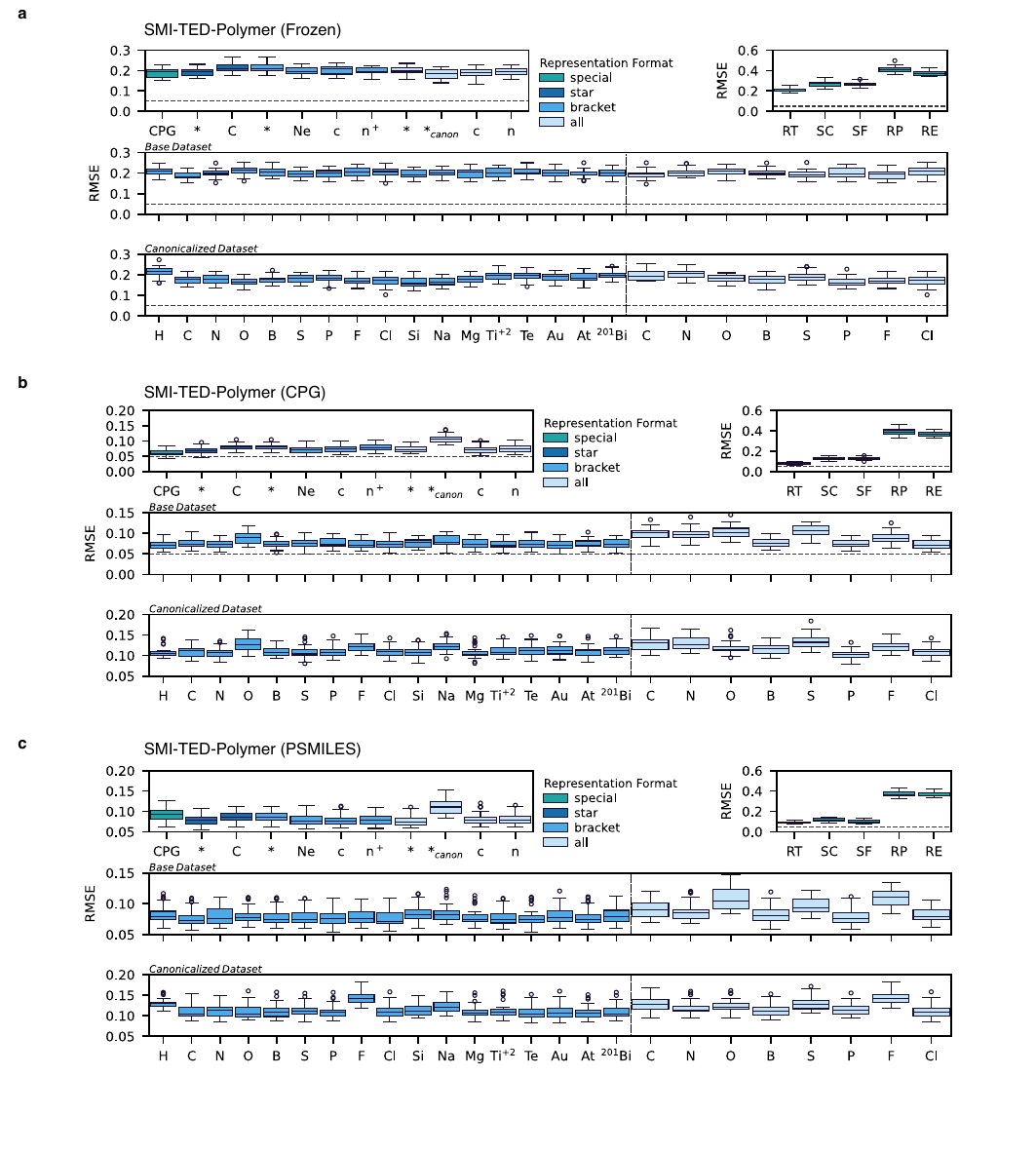}
\caption{RMSE values from repeated 5-fold cross-validation experiments with the E\textsubscript{at} dataset using SMI-TED-Polymer models with frozen weights (\textbf{a}) or fine-tuned with CPG (\textbf{b}) and PSMILES (\textbf{c}).}\label{eat_stp_rmse}
\end{figure}

\begin{figure}[H]
\centering
\includegraphics[width=1\textwidth]{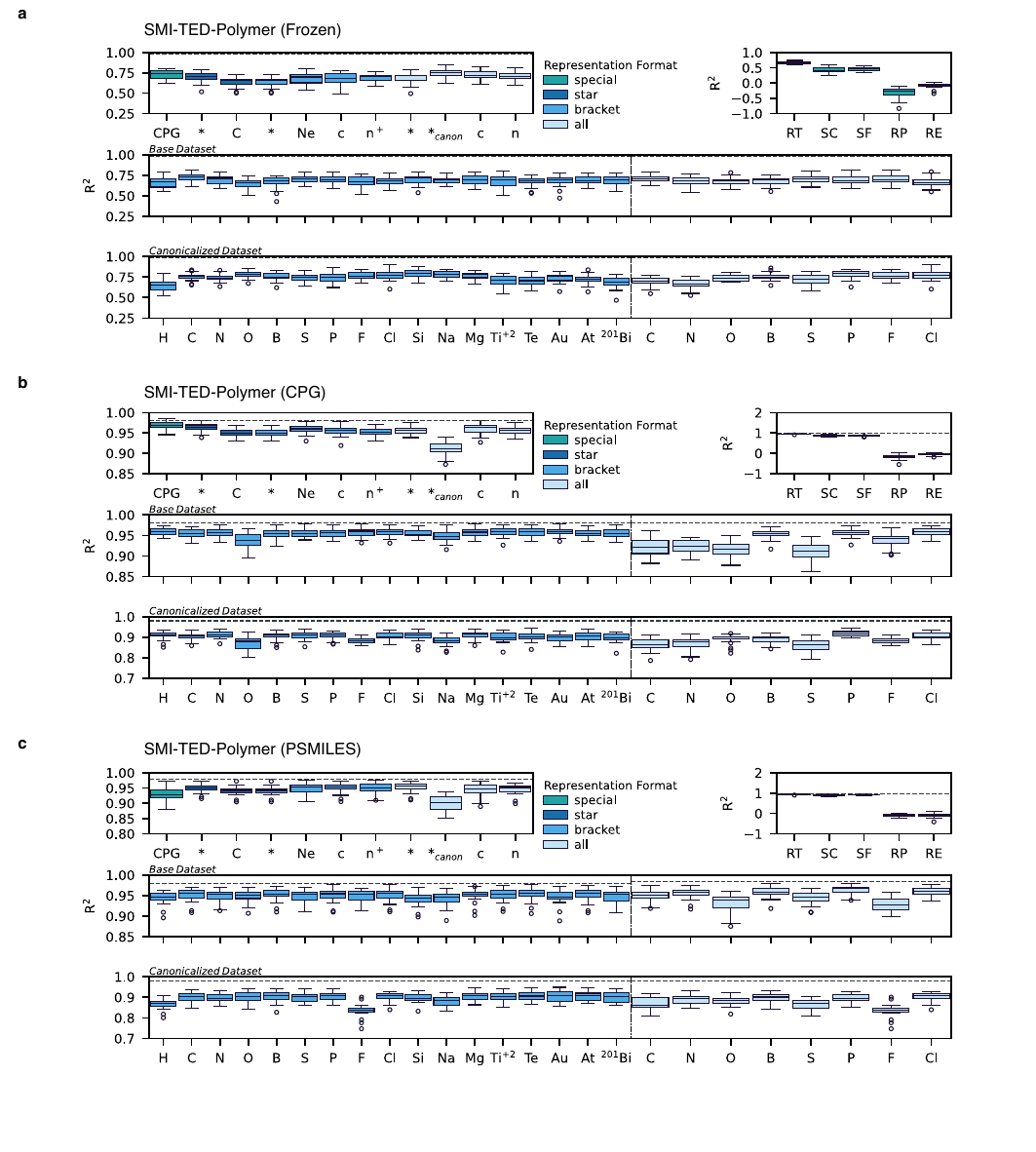}
\caption{R\textsuperscript{2} values from repeated 5-fold cross-validation experiments with the E\textsubscript{at} dataset using SMI-TED-Polymer models with frozen weights (\textbf{a}) or fine-tuned with CPG (\textbf{b}) and PSMILES (\textbf{c}).}\label{eat_stp_r2}
\end{figure}

\begin{figure}[H]
\centering
\includegraphics[width=1\textwidth]{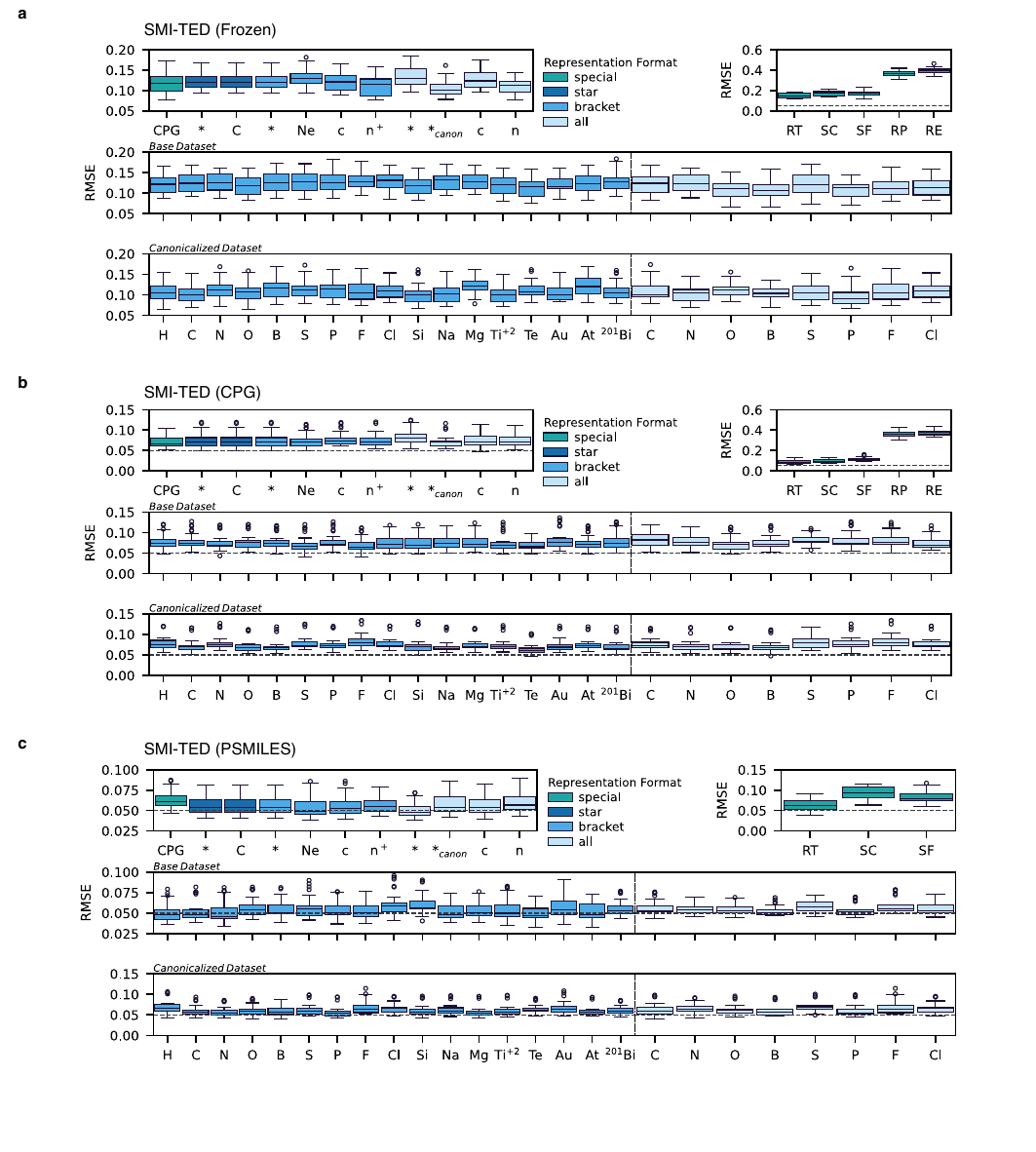}
\caption{RMSE values from repeated 5-fold cross-validation experiments with the E\textsubscript{at} dataset using SMI-TED models with frozen weights (\textbf{a}) or fine-tuned with CPG (\textbf{b}) and PSMILES (\textbf{c}).}\label{eat_st_rmse}
\end{figure}

\begin{figure}[H]
\centering
\includegraphics[width=1\textwidth]{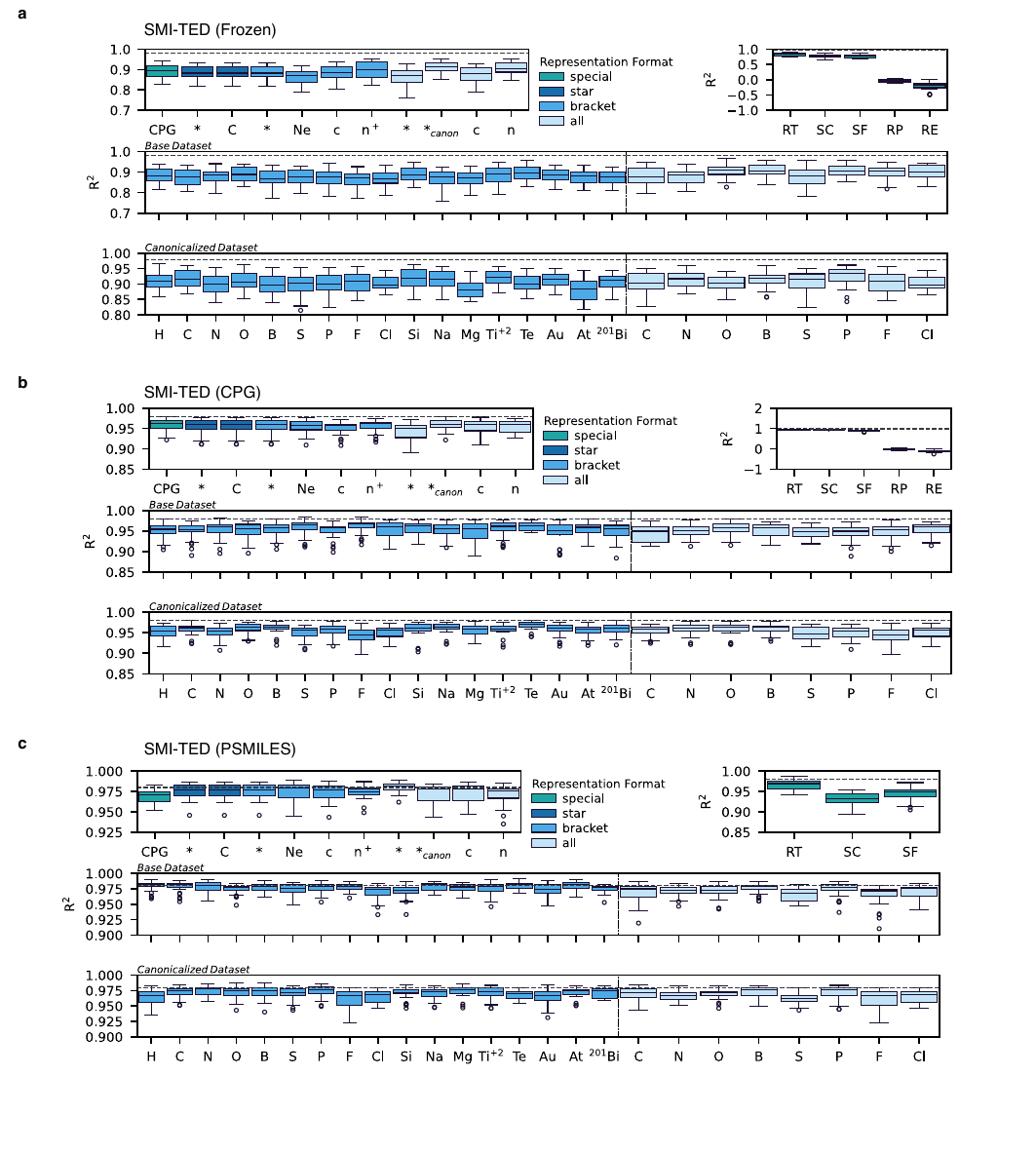}
\caption{R\textsuperscript{2} values from repeated 5-fold cross-validation experiments with the E\textsubscript{at} dataset using SMI-TED models with frozen weights (\textbf{a}) or fine-tuned with CPG (\textbf{b}) and PSMILES (\textbf{c}).}\label{eat_st_r2}
\end{figure}

\subsubsection{Refractivity Index (N\textsubscript{c})}\label{cvNc}

\begin{figure}[H]
\centering
\includegraphics[width=1\textwidth]{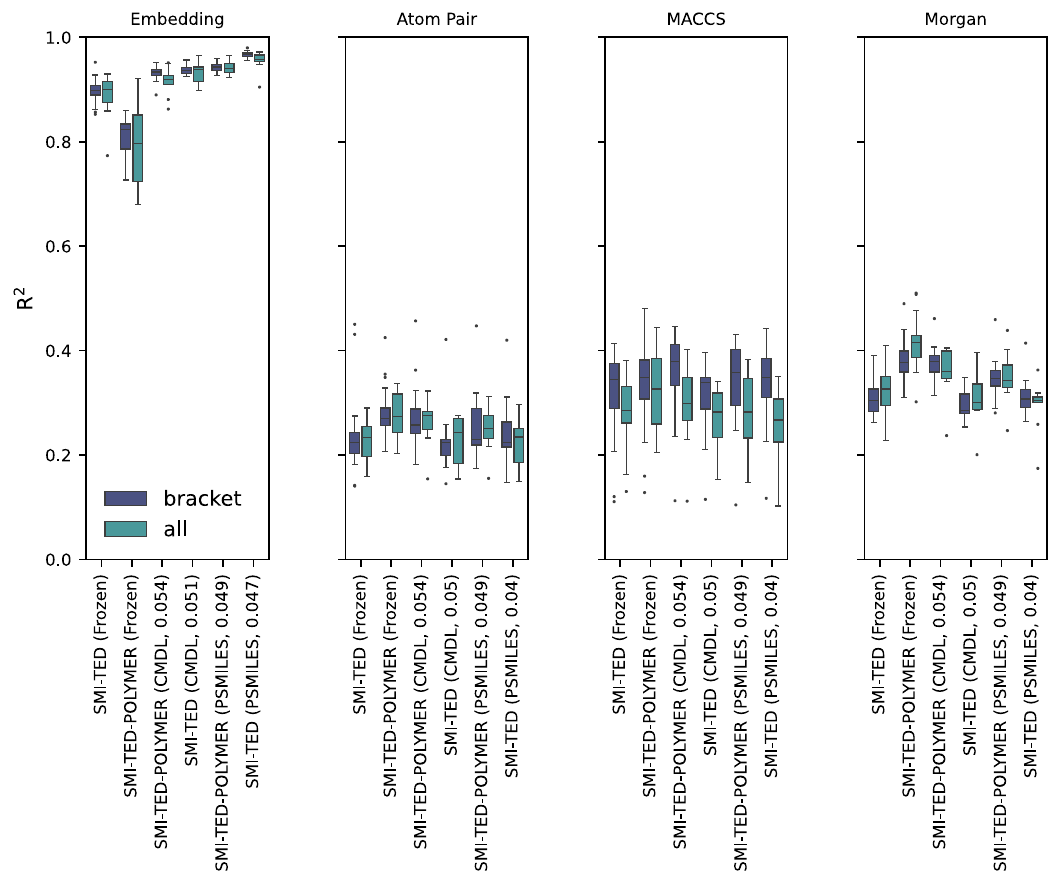}
\caption{Plots of embedding and fingerprint R\textsuperscript{2} value distributions for the N\textsubscript{c} benchmark dataset using the \textit{bracket} and \textit{all} substitution strategy.}\label{nc_ami_knn_embed}
\end{figure}

\begin{figure}[H]
\centering
\includegraphics[width=1\textwidth]{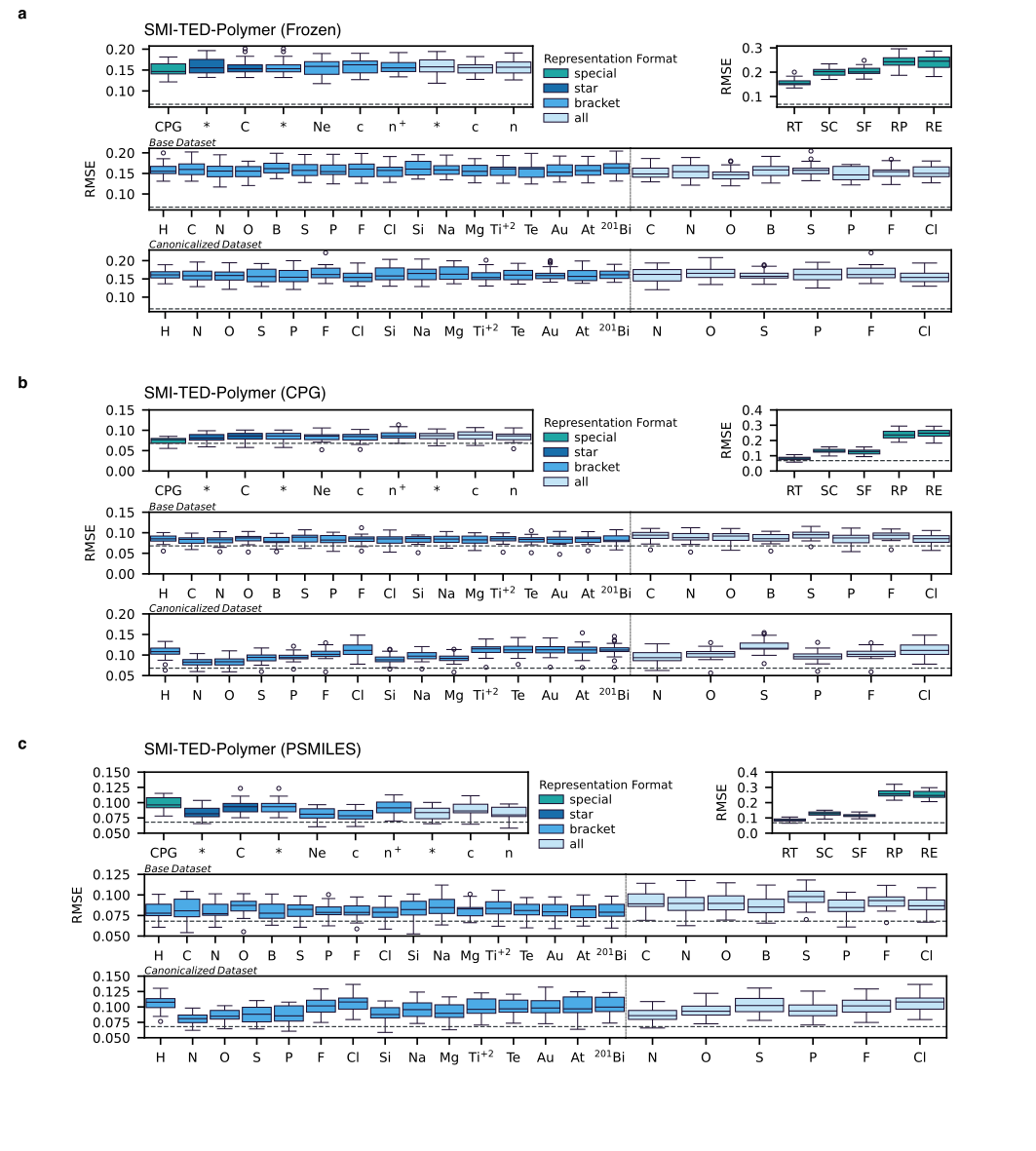}
\caption{RMSE values from repeated 5-fold cross-validation experiments with the N\textsubscript{c} dataset using SMI-TED-Polymer models with frozen weights (\textbf{a}) or fine-tuned with CPG (\textbf{b}) and PSMILES (\textbf{c}).}\label{nc_stp_rmse}
\end{figure}

\begin{figure}[H]
\centering
\includegraphics[width=1\textwidth]{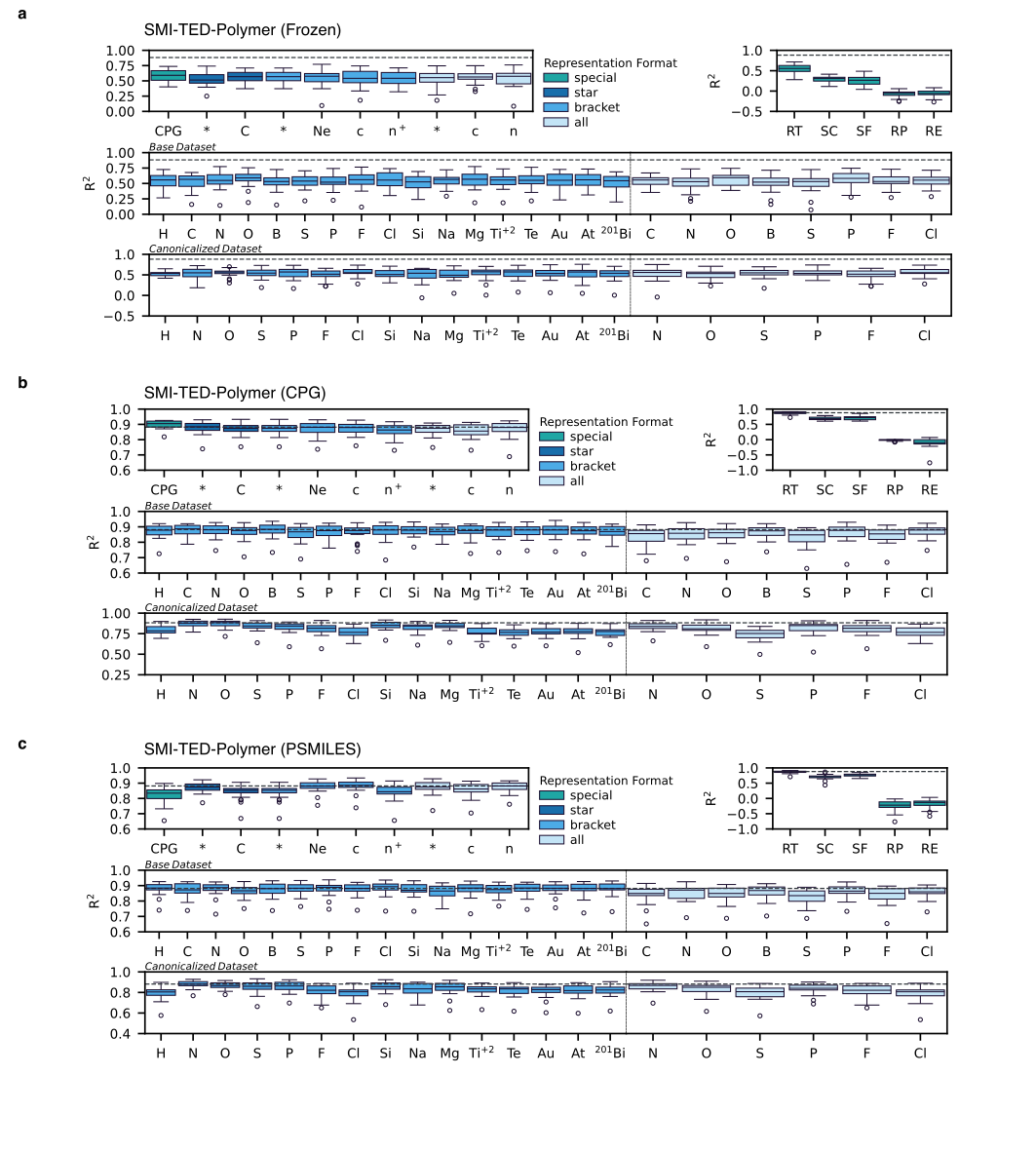}
\caption{R\textsuperscript{2} values from repeated 5-fold cross-validation experiments with the N\textsubscript{c} dataset using SMI-TED-Polymer models with frozen weights (\textbf{a}) or fine-tuned with CPG (\textbf{b}) and PSMILES (\textbf{c}).}\label{nc_stp_r2}
\end{figure}

\begin{figure}[H]
\centering
\includegraphics[width=1\textwidth]{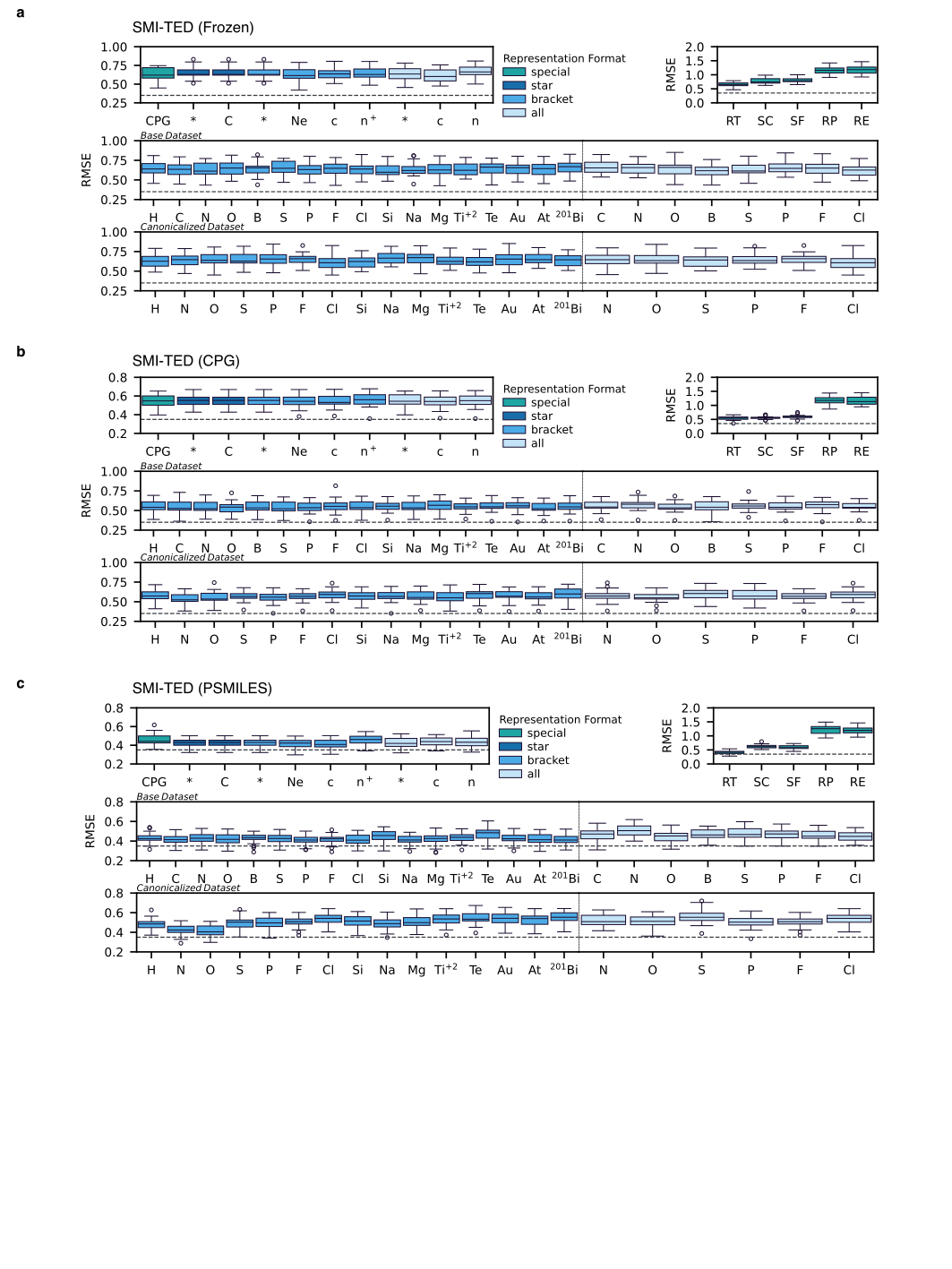}
\caption{RMSE values from repeated 5-fold cross-validation experiments with the N\textsubscript{c} dataset using SMI-TED models with frozen weights (\textbf{a}) or fine-tuned with CPG (\textbf{b}) and PSMILES (\textbf{c}).}\label{nc_st_rmse}
\end{figure}

\begin{figure}[H]
\centering
\includegraphics[width=1\textwidth]{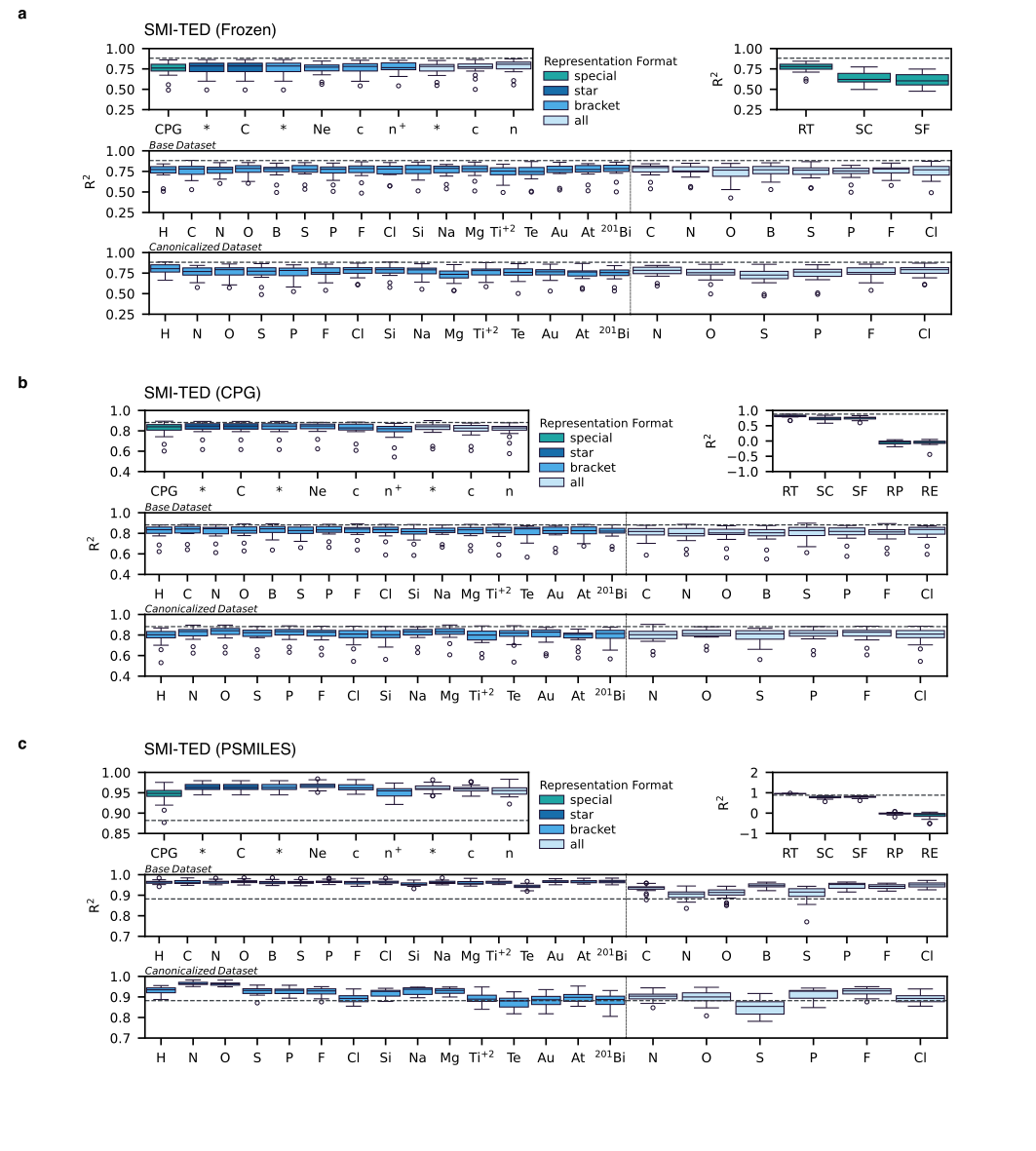}
\caption{R\textsuperscript{2} values from repeated 5-fold cross-validation experiments with the N\textsubscript{c} dataset using SMI-TED models with frozen weights (\textbf{a}) or fine-tuned with CPG (\textbf{b}) and PSMILES (\textbf{c}).}\label{nc_st_r2}
\end{figure}

\subsubsection{Crystallization Tendency (X\textsubscript{c})}\label{cvXc}

\begin{figure}[H]
\centering
\includegraphics[width=1\textwidth]{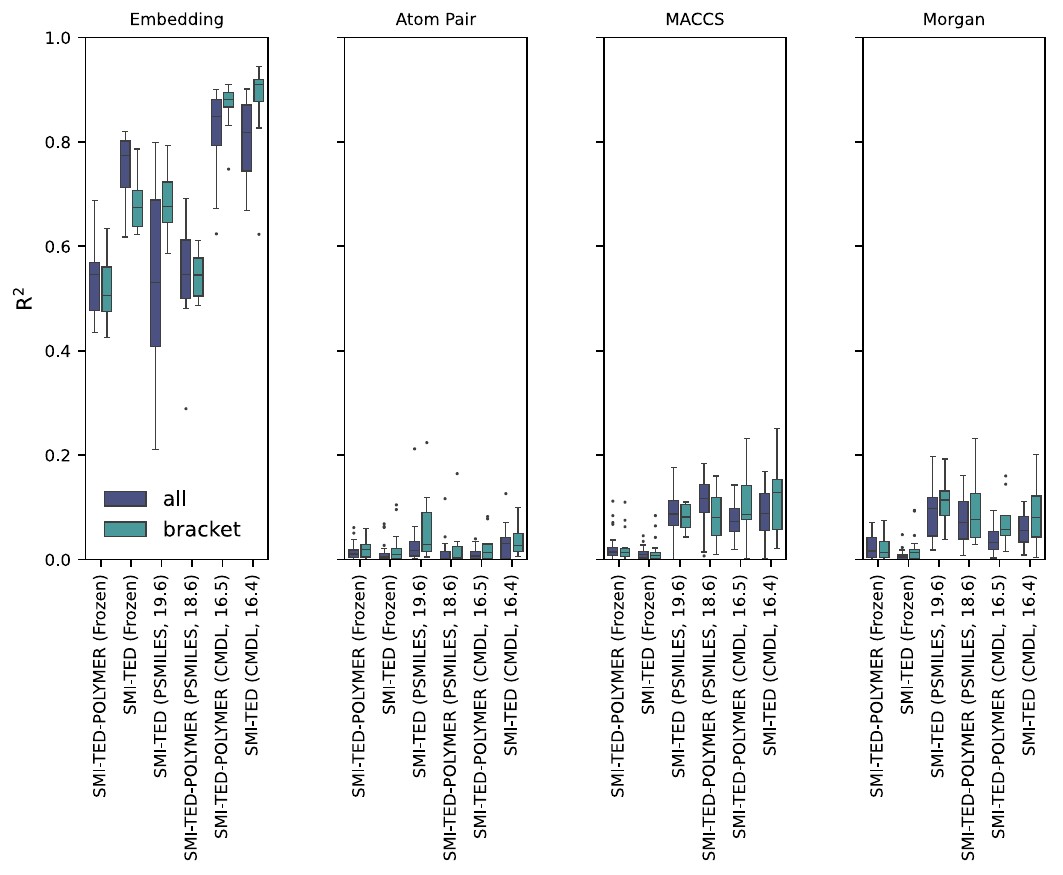}
\caption{Plots of embedding and fingerprint R\textsuperscript{2} value distributions for the X\textsubscript{c} benchmark dataset using the \textit{bracket} and \textit{all} substitution strategy.}\label{xc_knn_fp_bracket}
\end{figure}

\begin{figure}[H]
\centering
\includegraphics[width=1\textwidth]{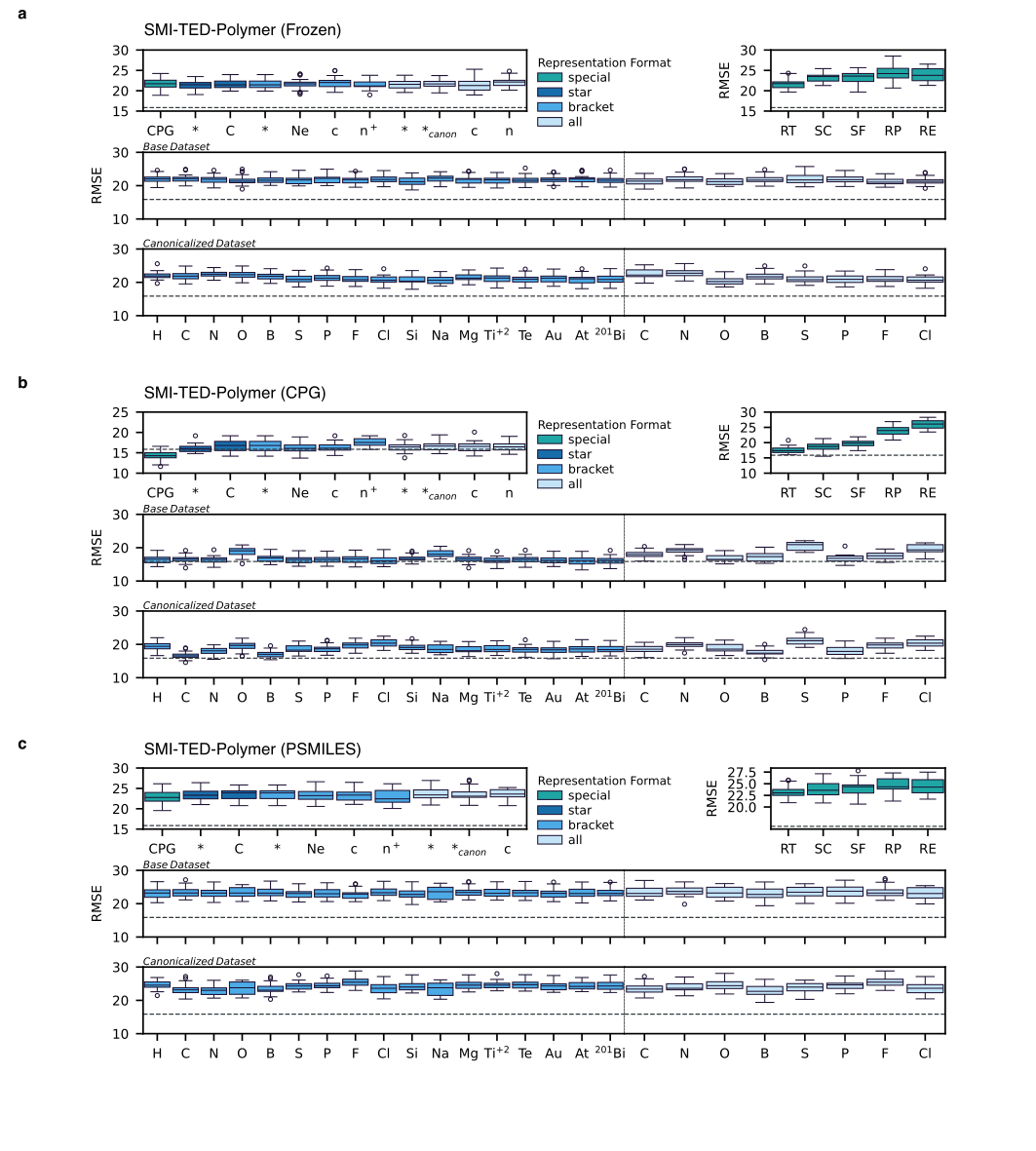}
\caption{RMSE values from repeated 5-fold cross-validation experiments with the X\textsubscript{c} dataset using SMI-TED-Polymer models with frozen weights (\textbf{a}) or fine-tuned with CPG (\textbf{b}) and PSMILES (\textbf{c}).}\label{xc_stp_rmse}
\end{figure}

\begin{figure}[H]
\centering
\includegraphics[width=1\textwidth]{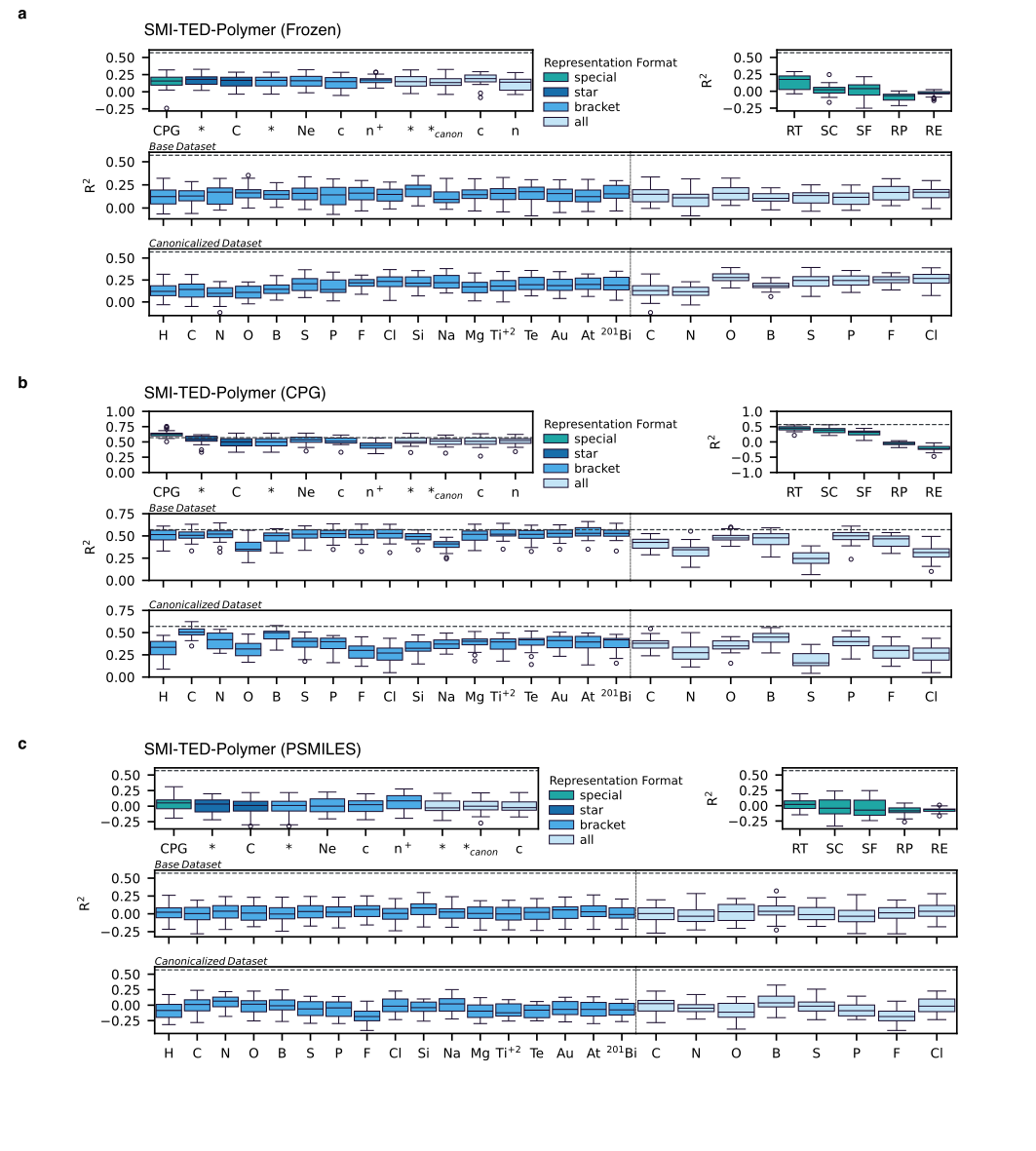}
\caption{R\textsuperscript{2} values from repeated 5-fold cross-validation experiments with the X\textsubscript{c} dataset using SMI-TED-Polymer models with frozen weights (\textbf{a}) or fine-tuned with CPG (\textbf{b}) and PSMILES (\textbf{c}).}\label{xc_stp_r2}
\end{figure}

\begin{figure}[H]
\centering
\includegraphics[width=1\textwidth]{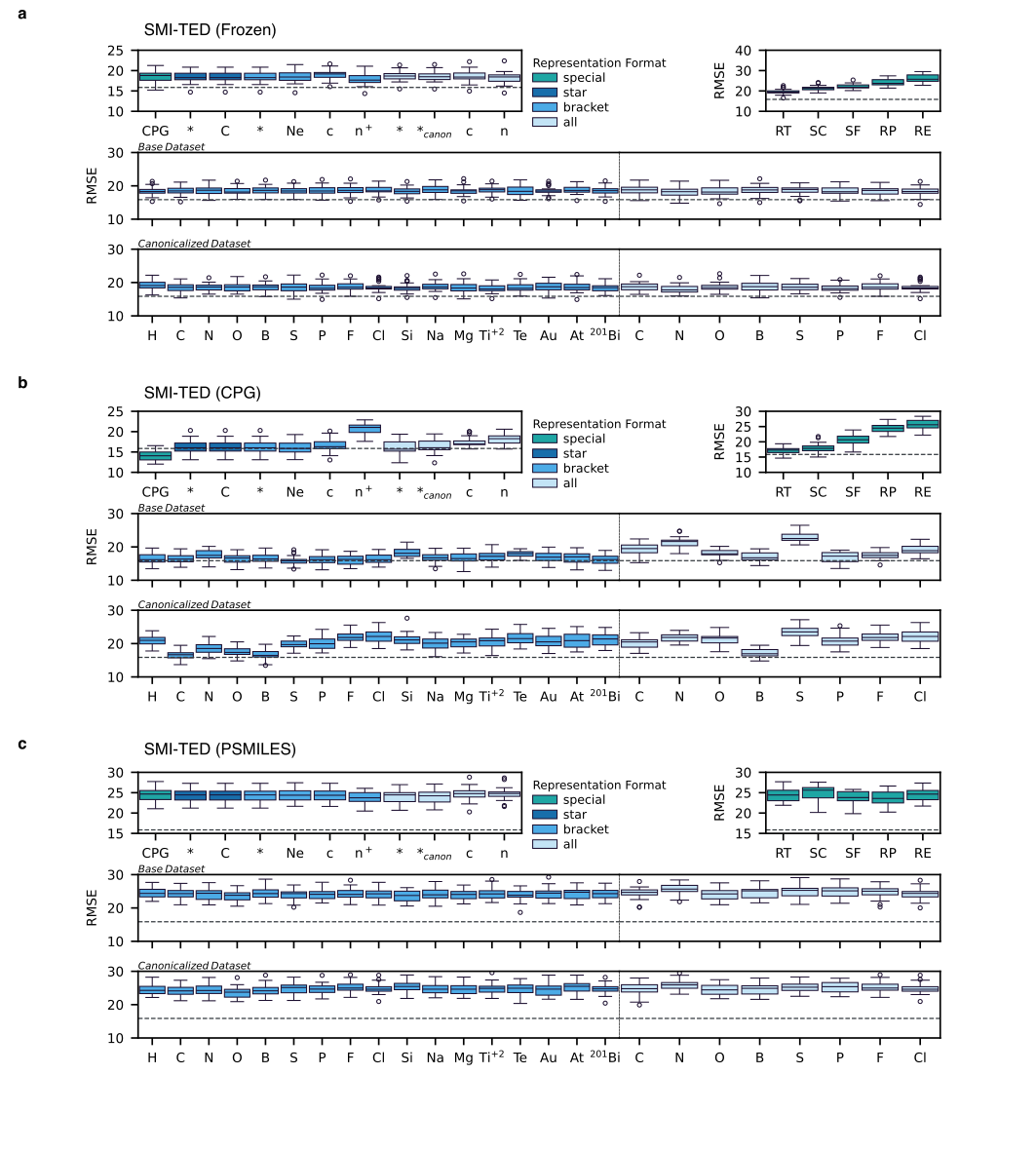}
\caption{RMSE values from repeated 5-fold cross-validation experiments with the X\textsubscript{c} dataset using SMI-TED models with frozen weights (\textbf{a}) or fine-tuned with CPG (\textbf{b}) and PSMILES (\textbf{c}).}\label{xc_st_rmse}
\end{figure}

\begin{figure}[H]
\centering
\includegraphics[width=1\textwidth]{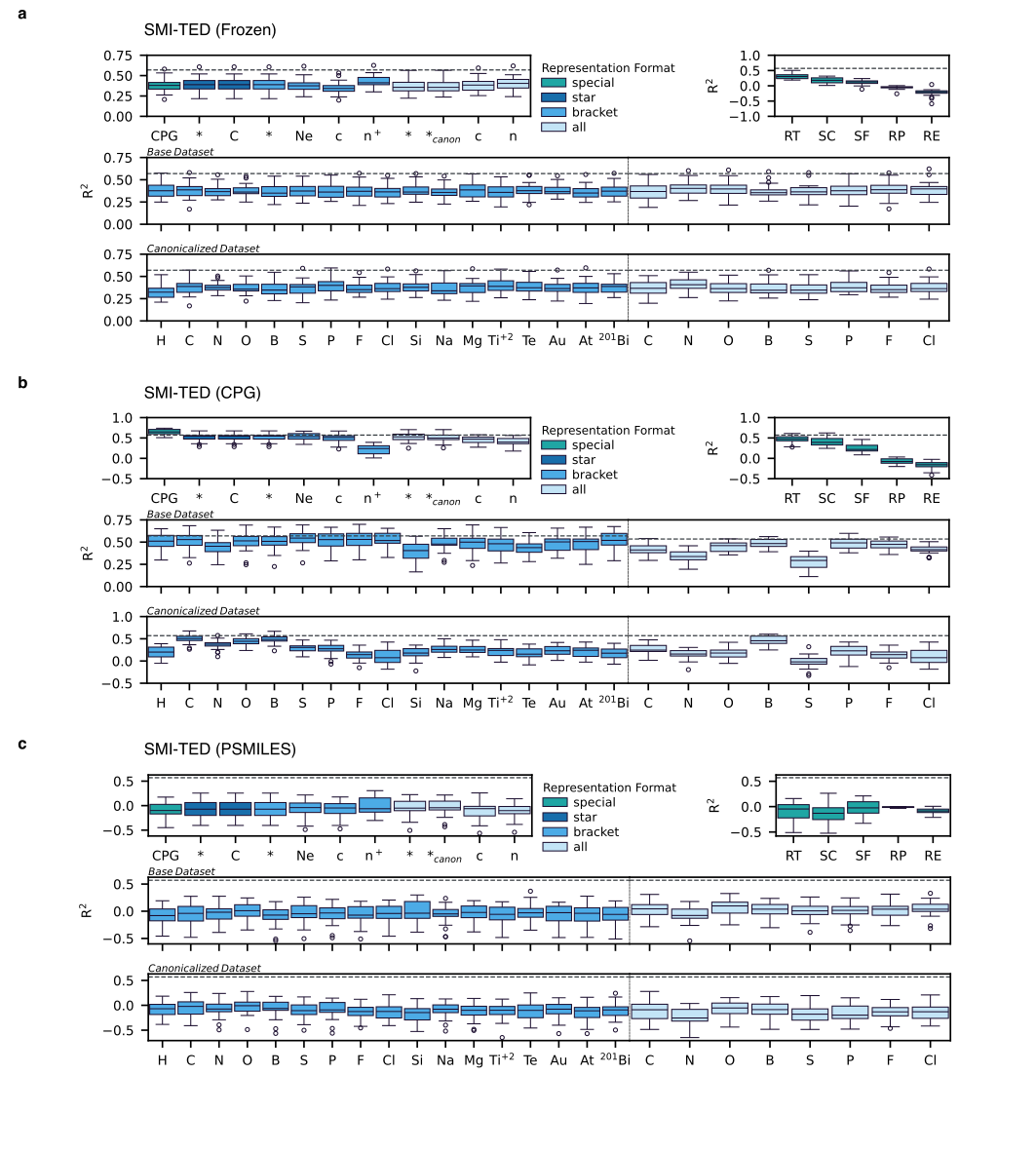}
\caption{R\textsuperscript{2} values from repeated 5-fold cross-validation experiments with the X\textsubscript{c} dataset using SMI-TED models with frozen weights (\textbf{a}) or fine-tuned with CPG (\textbf{b}) and PSMILES (\textbf{c}).}\label{xc_st_r2}
\end{figure}

\subsubsection{Glass Transition Temperature (T\textsubscript{g}-II)}\label{cvAcsTg}

\begin{figure}[H]
\centering
\includegraphics[width=1\textwidth]{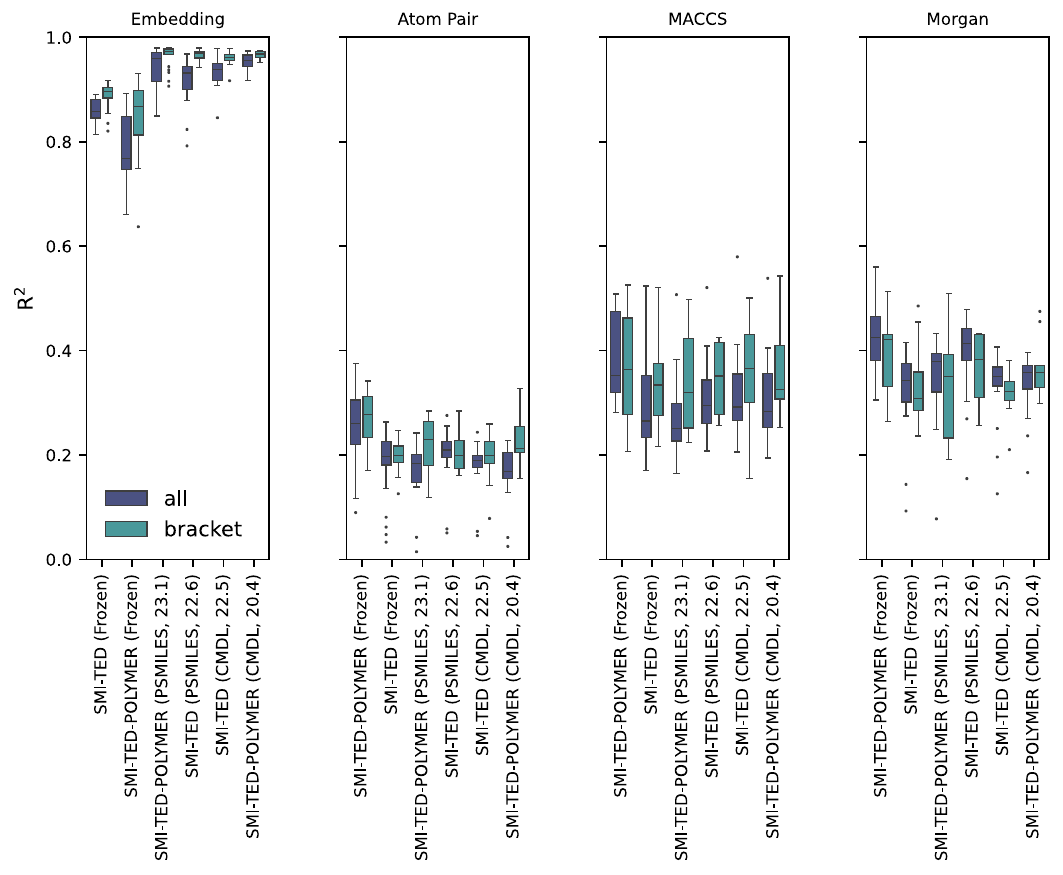}
\caption{Plots of embedding and fingerprint R\textsuperscript{2} value distributions for the T\textsubscript{g}-II benchmark dataset using the \textit{bracket} and \textit{all} substitution strategy.}\label{acs_ami_knn_embed}
\end{figure}

\begin{figure}[H]
\centering
\includegraphics[width=1\textwidth]{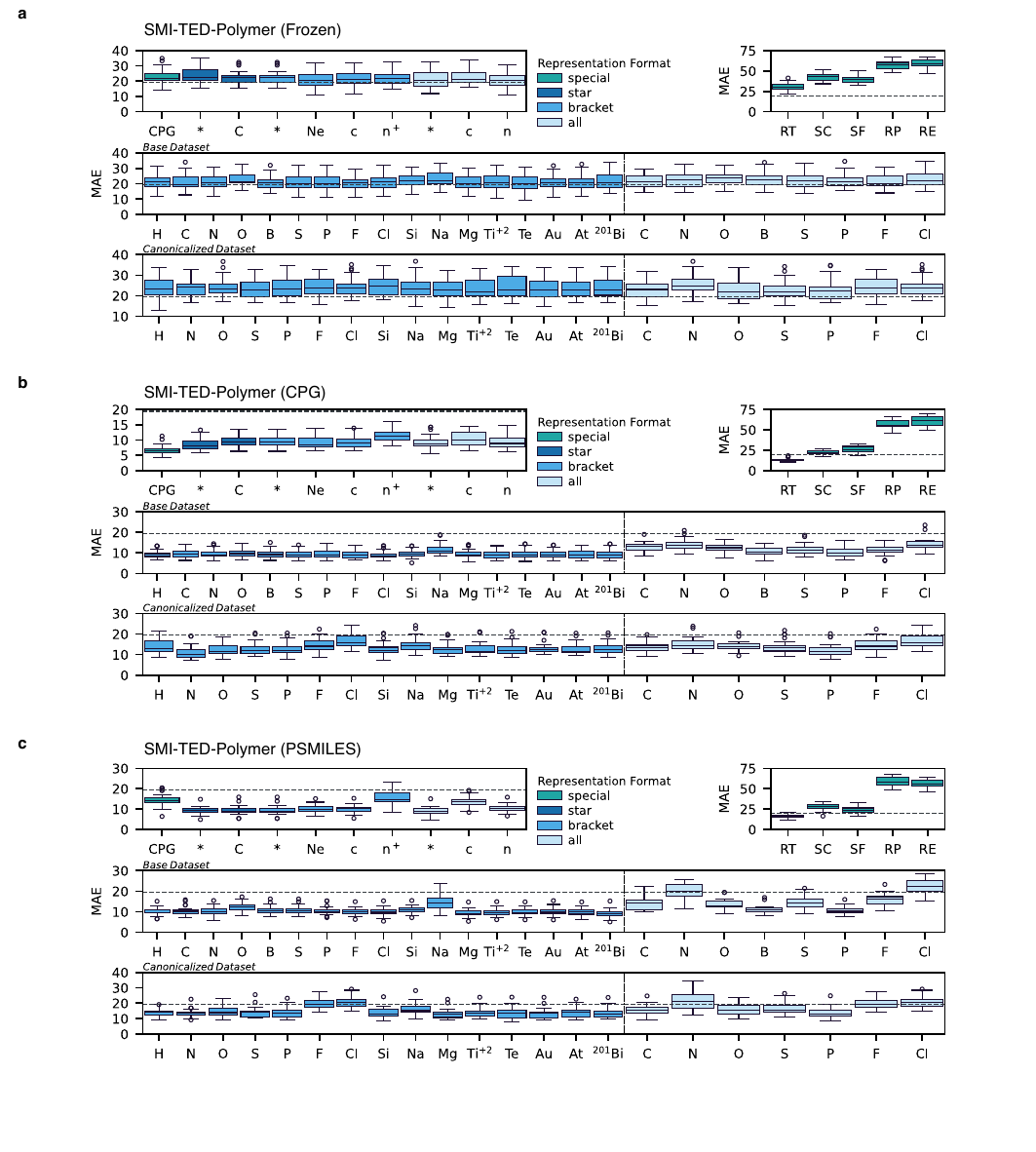}
\caption{MAE values from repeated 5-fold cross-validation experiments with the T\textsubscript{g}-II dataset using SMI-TED-Polymer models with frozen weights (\textbf{a}) or fine-tuned with CPG (\textbf{b}) and PSMILES (\textbf{c}).}\label{acs_ami_stp_mae}
\end{figure}

\begin{figure}[H]
\centering
\includegraphics[width=1\textwidth]{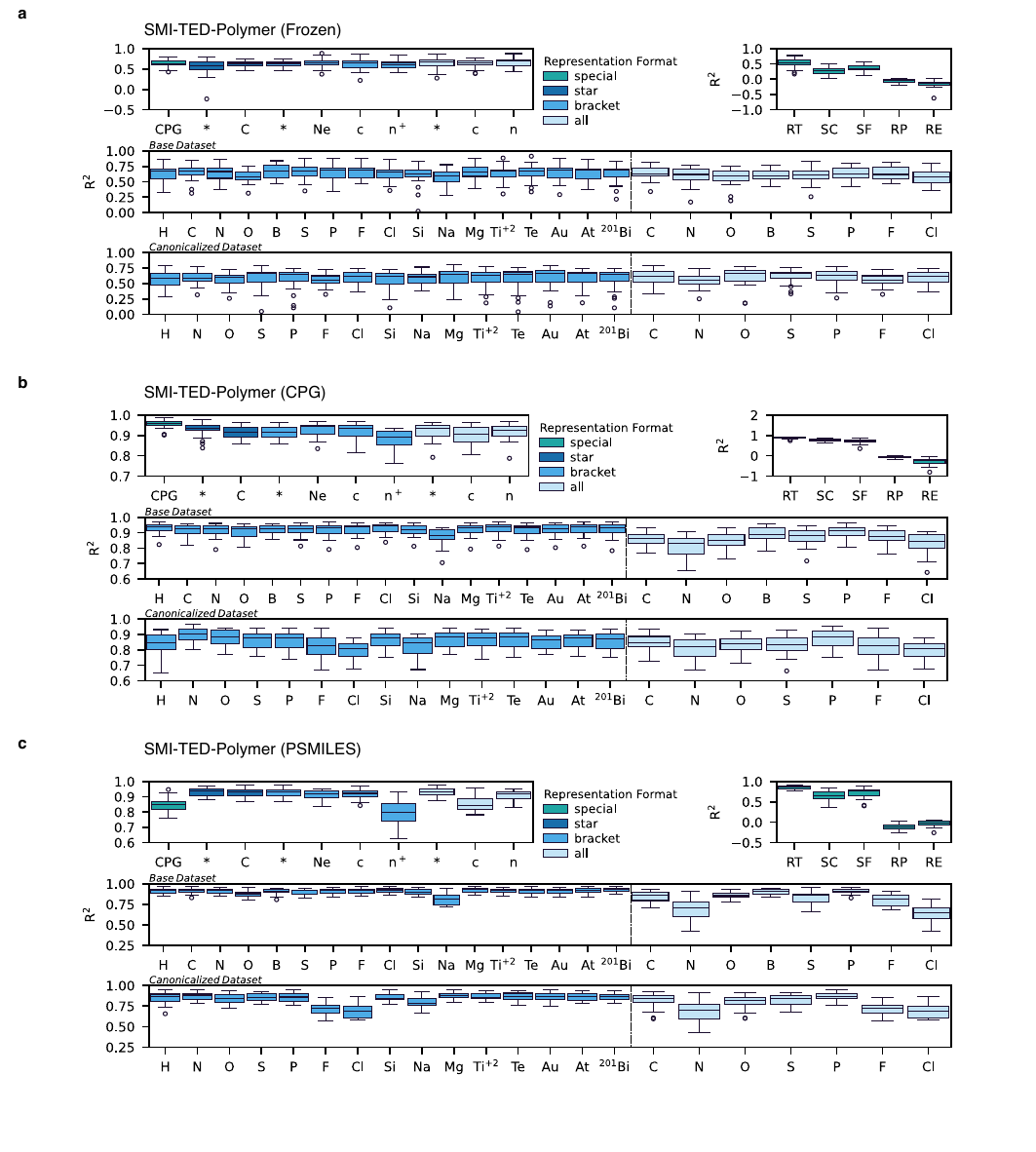}
\caption{R\textsuperscript{2} values from repeated 5-fold cross-validation experiments with the T\textsubscript{g}-II dataset using SMI-TED-Polymer models with frozen weights (\textbf{a}) or fine-tuned with CPG (\textbf{b}) and PSMILES (\textbf{c}).}\label{acs_ami_stp_r2}
\end{figure}

\begin{figure}[H]
\centering
\includegraphics[width=1\textwidth]{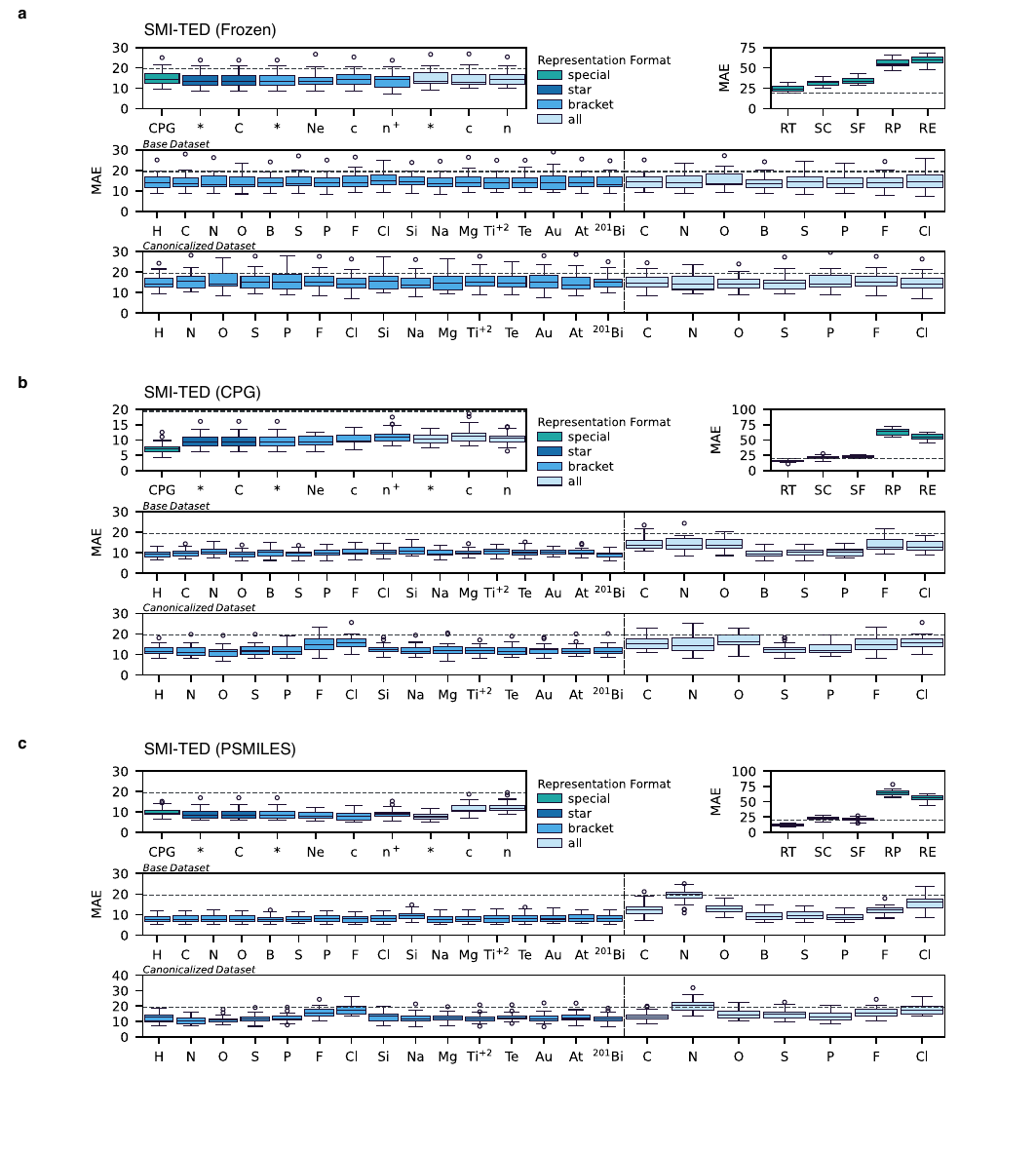}
\caption{MAE values from repeated 5-fold cross-validation experiments with the T\textsubscript{g}-II dataset using SMI-TED models with frozen weights (\textbf{a}) or fine-tuned with CPG (\textbf{b}) and PSMILES (\textbf{c}).}\label{acs_ami_st_mae}
\end{figure}

\begin{figure}[H]
\centering
\includegraphics[width=1\textwidth]{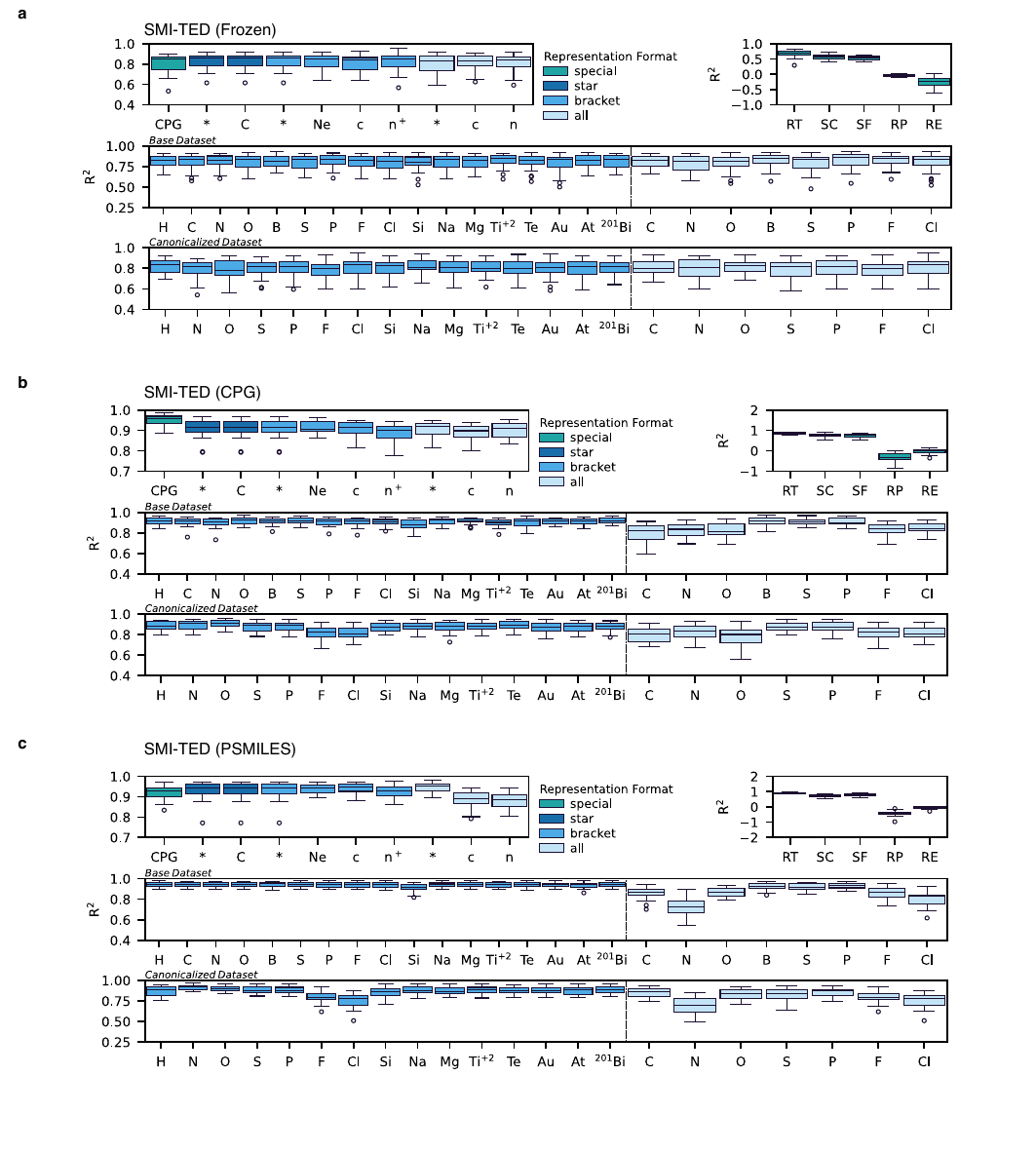}
\caption{R\textsuperscript{2} values from repeated 5-fold cross-validation experiments with the T\textsubscript{g}-II dataset using SMI-TED models with frozen weights (\textbf{a}) or fine-tuned with CPG (\textbf{b}) and PSMILES (\textbf{c}).}\label{acs_ami_st_r2}
\end{figure}

\subsubsection{CO\textsubscript{2} Gas Permeability}\label{cv_netl_co2}

\begin{figure}[H]
\centering
\includegraphics[width=1\textwidth]{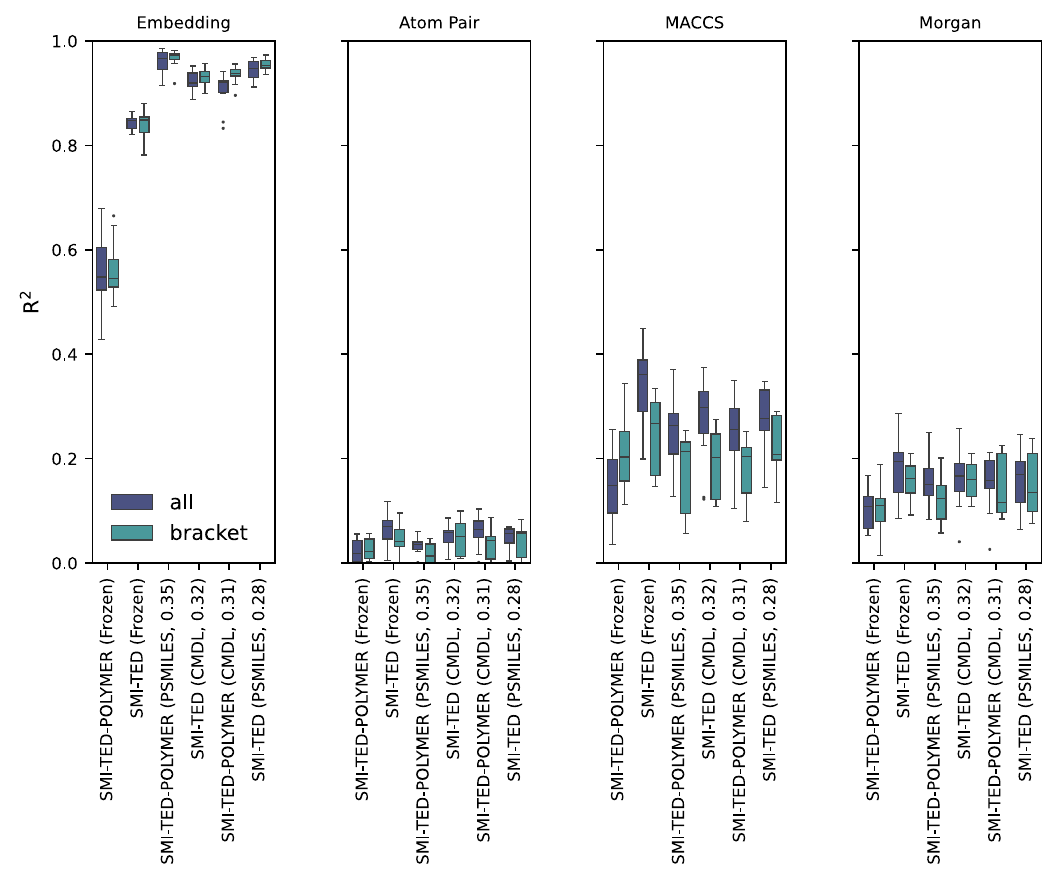}
\caption{Plots of embedding and fingerprint R\textsuperscript{2} value distributions for the CO\textsubscript{2} permeability benchmark dataset using the \textit{bracket} and \textit{all} substitution strategy.}\label{netl_co2_knn_fp_all}
\end{figure}

\begin{figure}[H]
\centering
\includegraphics[width=1\textwidth]{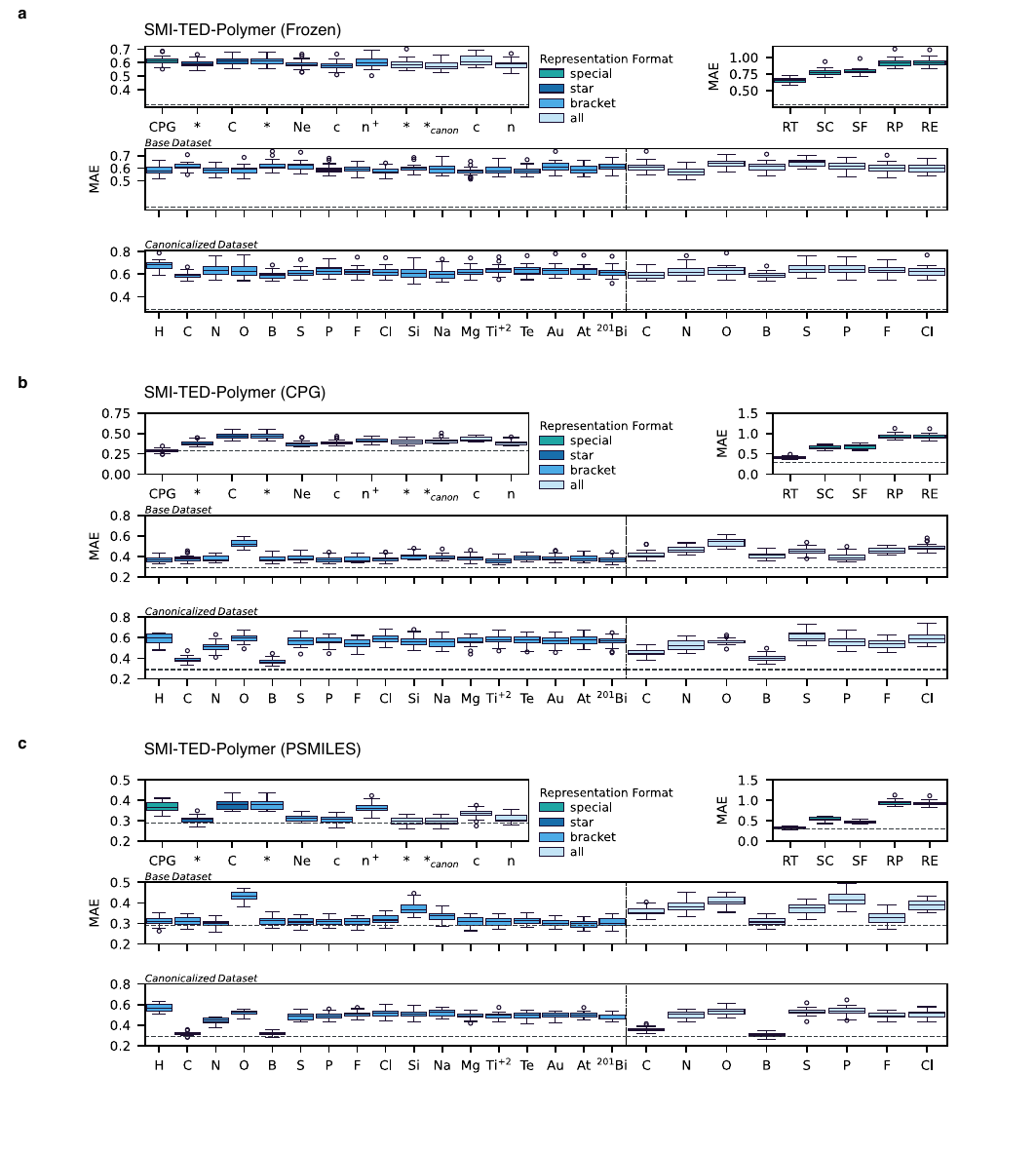}
\caption{MAE values from repeated 5-fold cross-validation experiments with the CO\textsubscript{2} Gas Permeability dataset using SMI-TED-Polymer models with frozen weights (\textbf{a}) or fine-tuned with CPG (\textbf{b}) and PSMILES (\textbf{c}).}\label{netl_co2_stp_mae}
\end{figure}

\begin{figure}[H]
\centering
\includegraphics[width=1\textwidth]{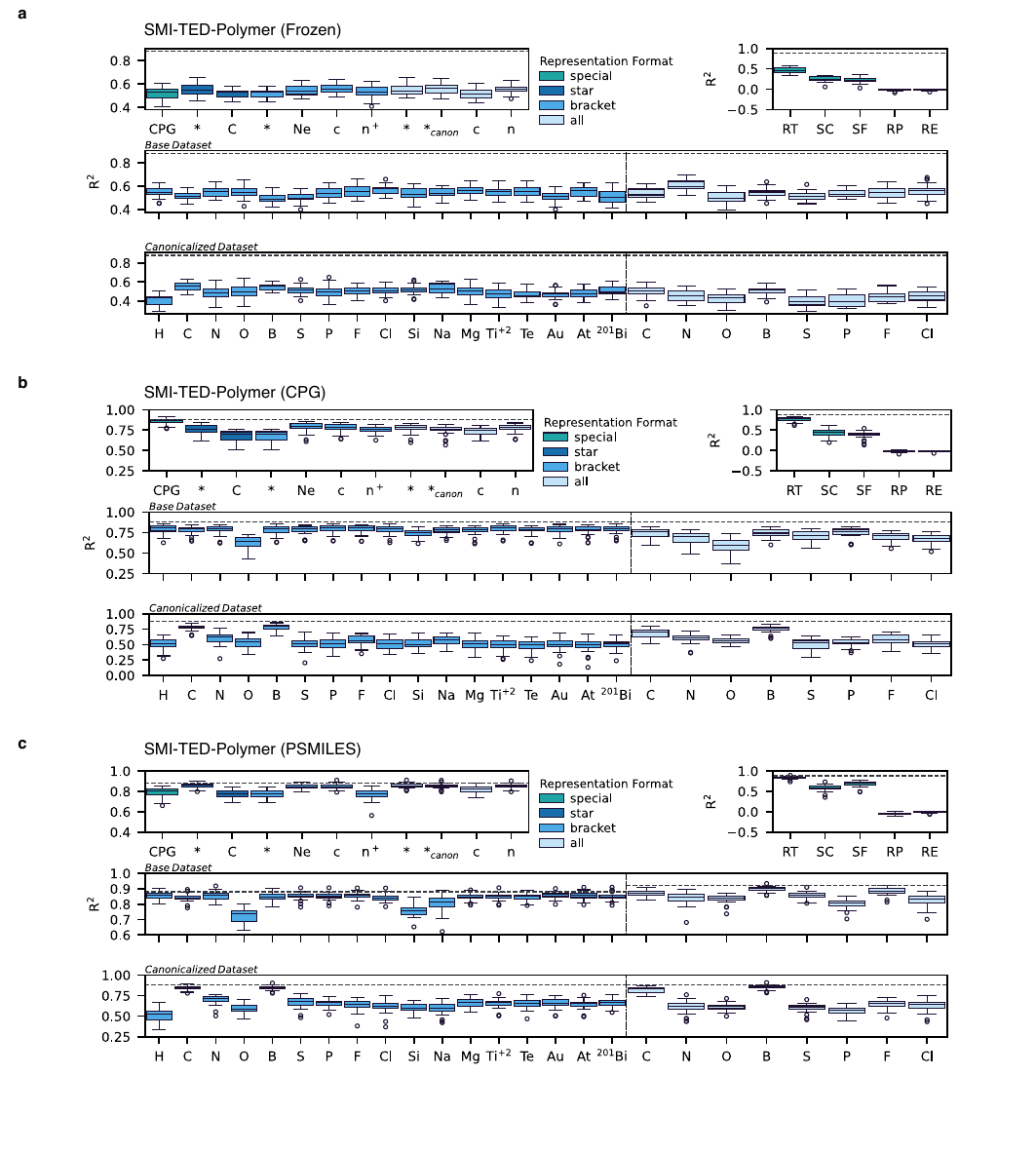}
\caption{R\textsuperscript{2} values from repeated 5-fold cross-validation experiments with the CO\textsubscript{2} Gas Permeability dataset using SMI-TED-Polymer models with frozen weights (\textbf{a}) or fine-tuned with CPG (\textbf{b}) and PSMILES (\textbf{c}).}\label{netl_co2_stp_r2}
\end{figure}

\begin{figure}[H]
\centering
\includegraphics[width=1\textwidth]{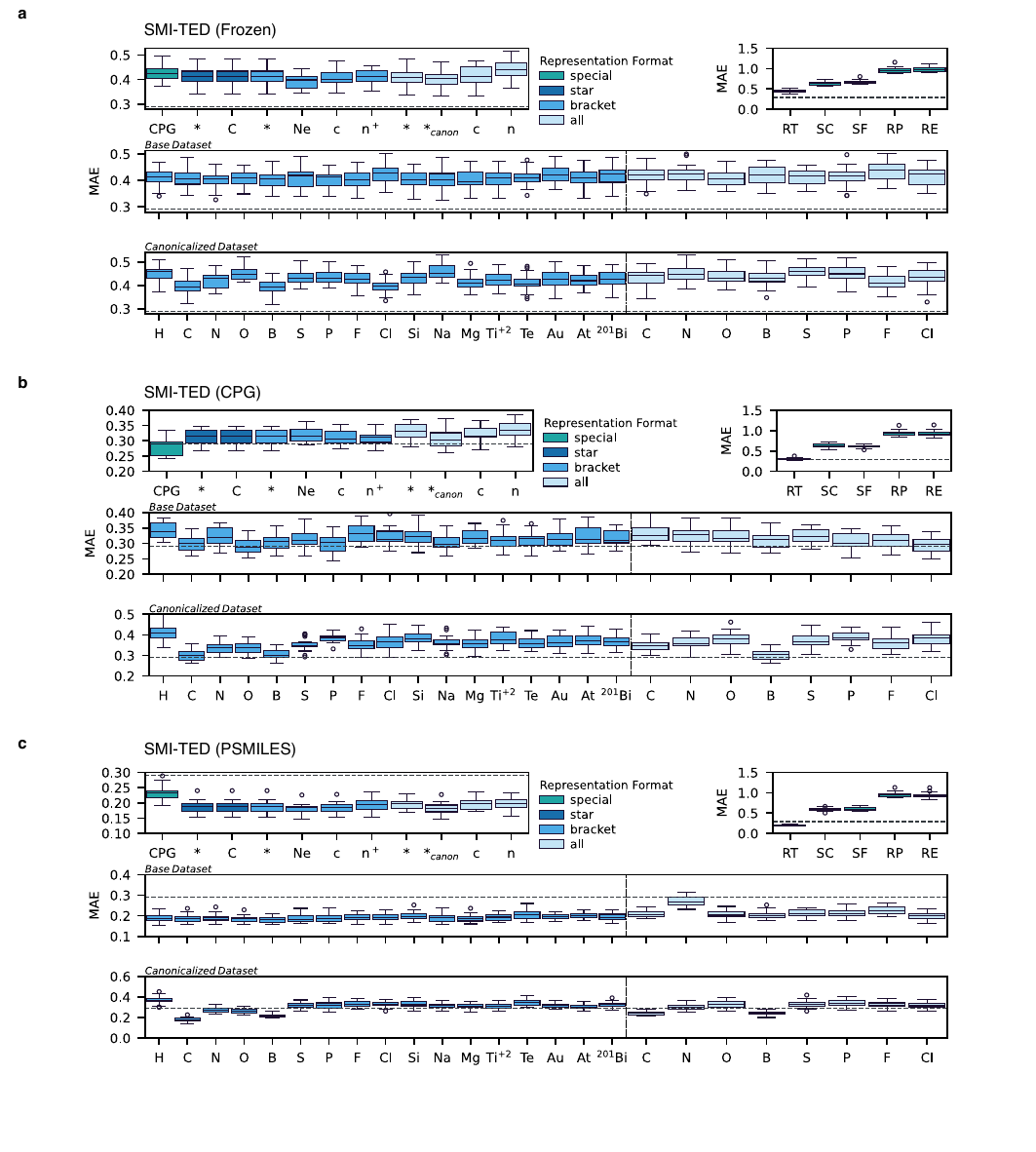}
\caption{MAE values from repeated 5-fold cross-validation experiments with the CO\textsubscript{2} Gas Permeability dataset using SMI-TED models with frozen weights (\textbf{a}) or fine-tuned with CPG (\textbf{b}) and PSMILES (\textbf{c}).}\label{netl_co2_st_mae}
\end{figure}

\begin{figure}[H]
\centering
\includegraphics[width=1\textwidth]{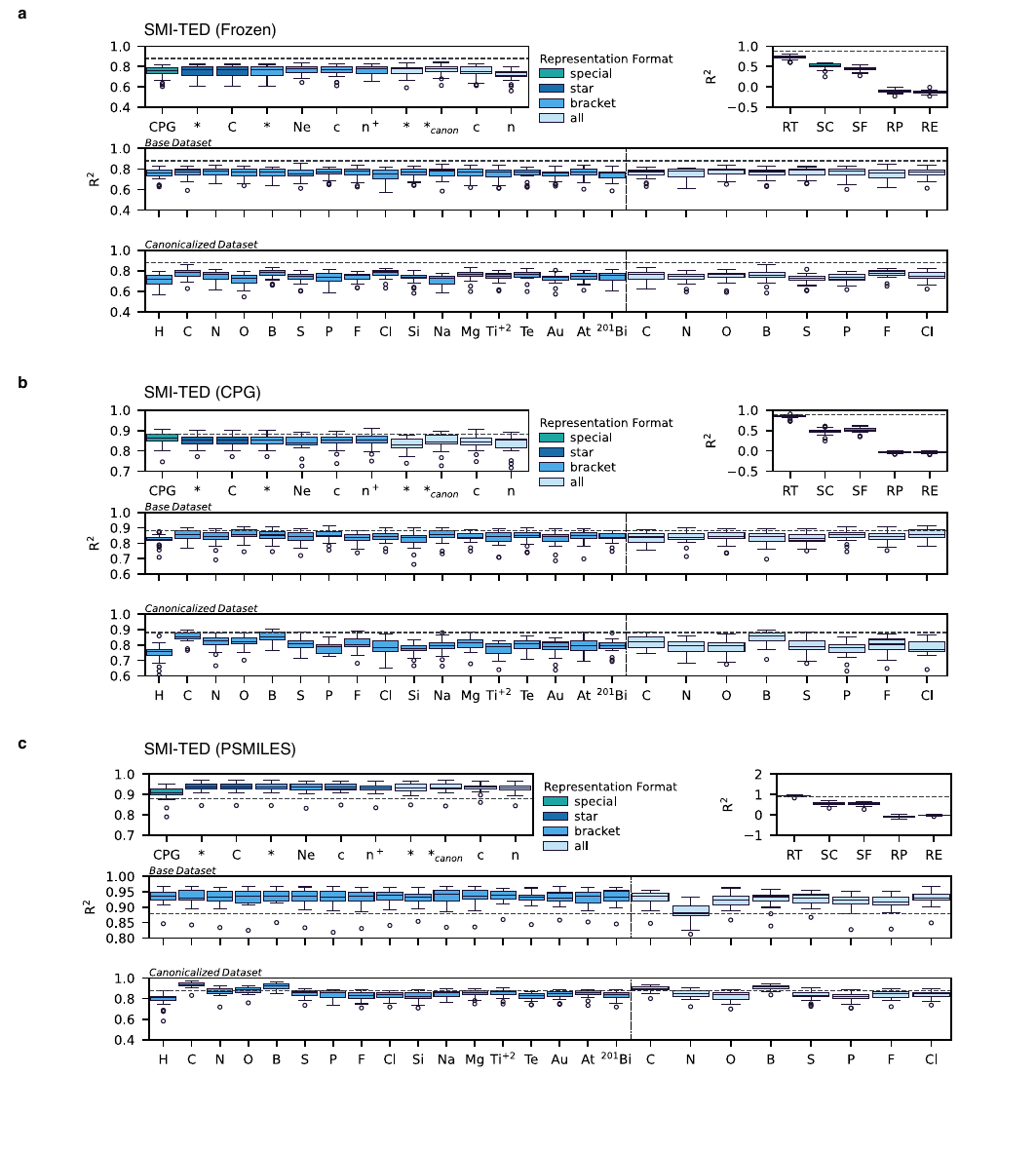}
\caption{R\textsuperscript{2} values from repeated 5-fold cross-validation experiments with the CO\textsubscript{2} Gas Permeability dataset using SMI-TED models with frozen weights (\textbf{a}) or fine-tuned with CPG (\textbf{b}) and PSMILES (\textbf{c}).}\label{netl_co2_st_r2}
\end{figure}

\subsubsection{N\textsubscript{2} Gas Permeability}\label{cv_netl_n2}

\begin{figure}[H]
\centering
\includegraphics[width=1\textwidth]{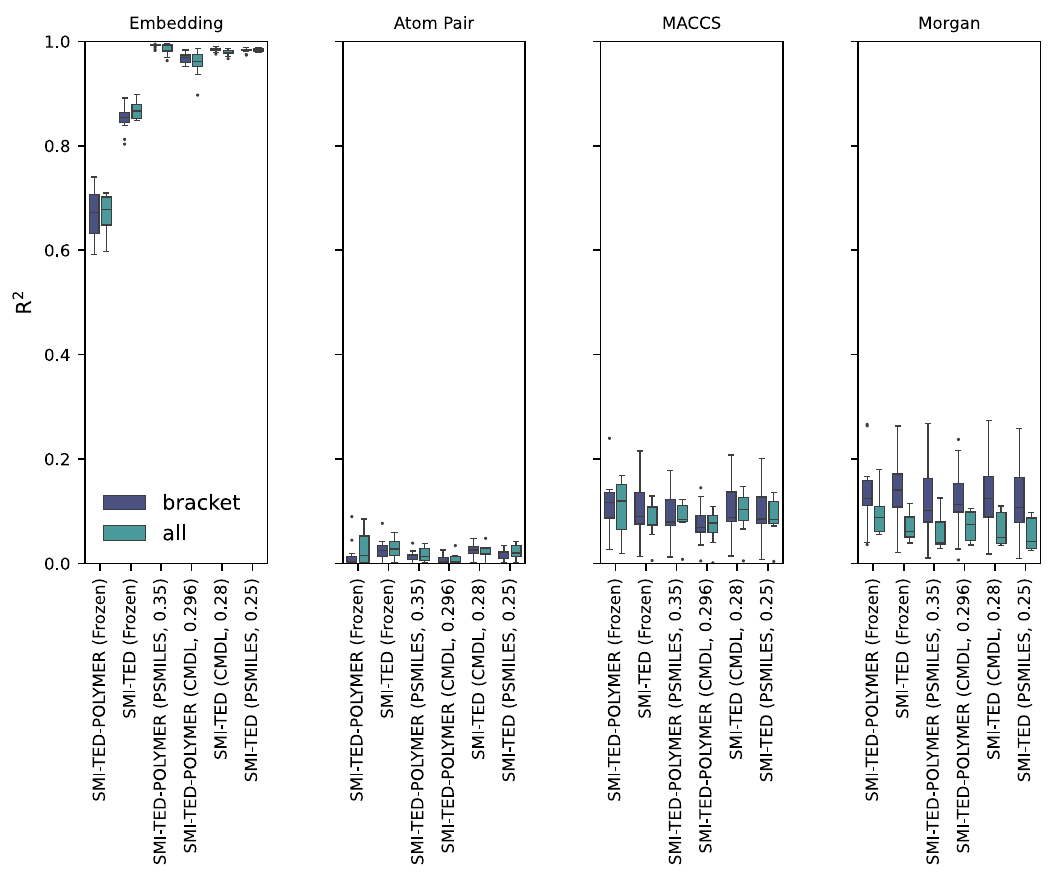}
\caption{Plots of embedding and fingerprint R\textsuperscript{2} value distributions for the N\textsubscript{2} permeability benchmark dataset using the \textit{bracket} and \textit{all} substitution strategy.}\label{netl_n2_knn_fp_all}
\end{figure}

\begin{figure}[H]
\centering
\includegraphics[width=1\textwidth]{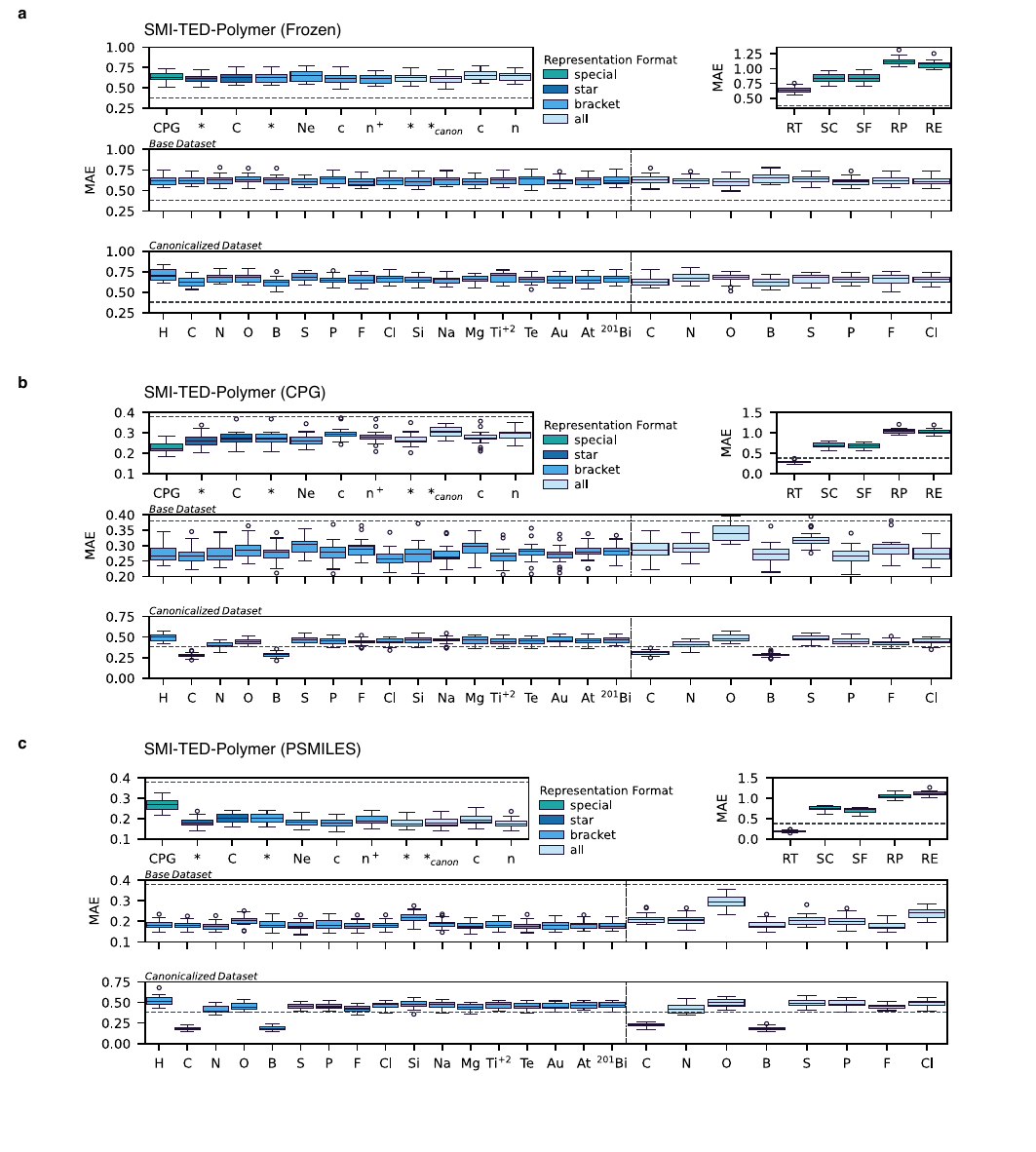}
\caption{MAE values from repeated 5-fold cross-validation experiments with the N\textsubscript{2} Gas Permeability dataset using SMI-TED-Polymer models with frozen weights (\textbf{a}) or fine-tuned with CPG (\textbf{b}) and PSMILES (\textbf{c}).}\label{netl_n2_stp_mae}
\end{figure}

\begin{figure}[H]
\centering
\includegraphics[width=1\textwidth]{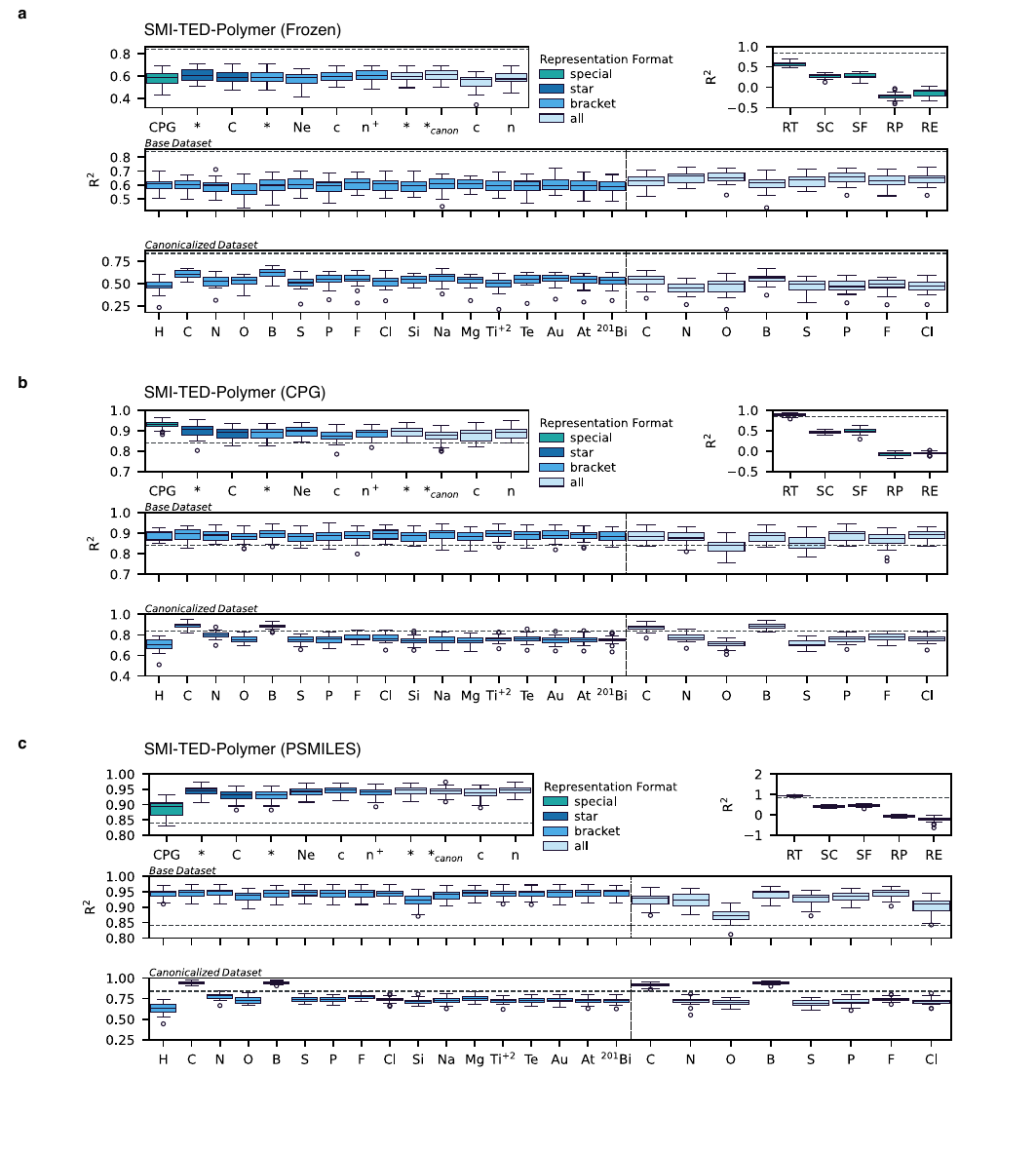}
\caption{R\textsuperscript{2} values from repeated 5-fold cross-validation experiments with the N\textsubscript{2} Gas Permeability dataset using SMI-TED-Polymer models with frozen weights (\textbf{a}) or fine-tuned with CPG (\textbf{b}) and PSMILES (\textbf{c}).}\label{netl_n2_stp_r2}
\end{figure}

\begin{figure}[H]
\centering
\includegraphics[width=1\textwidth]{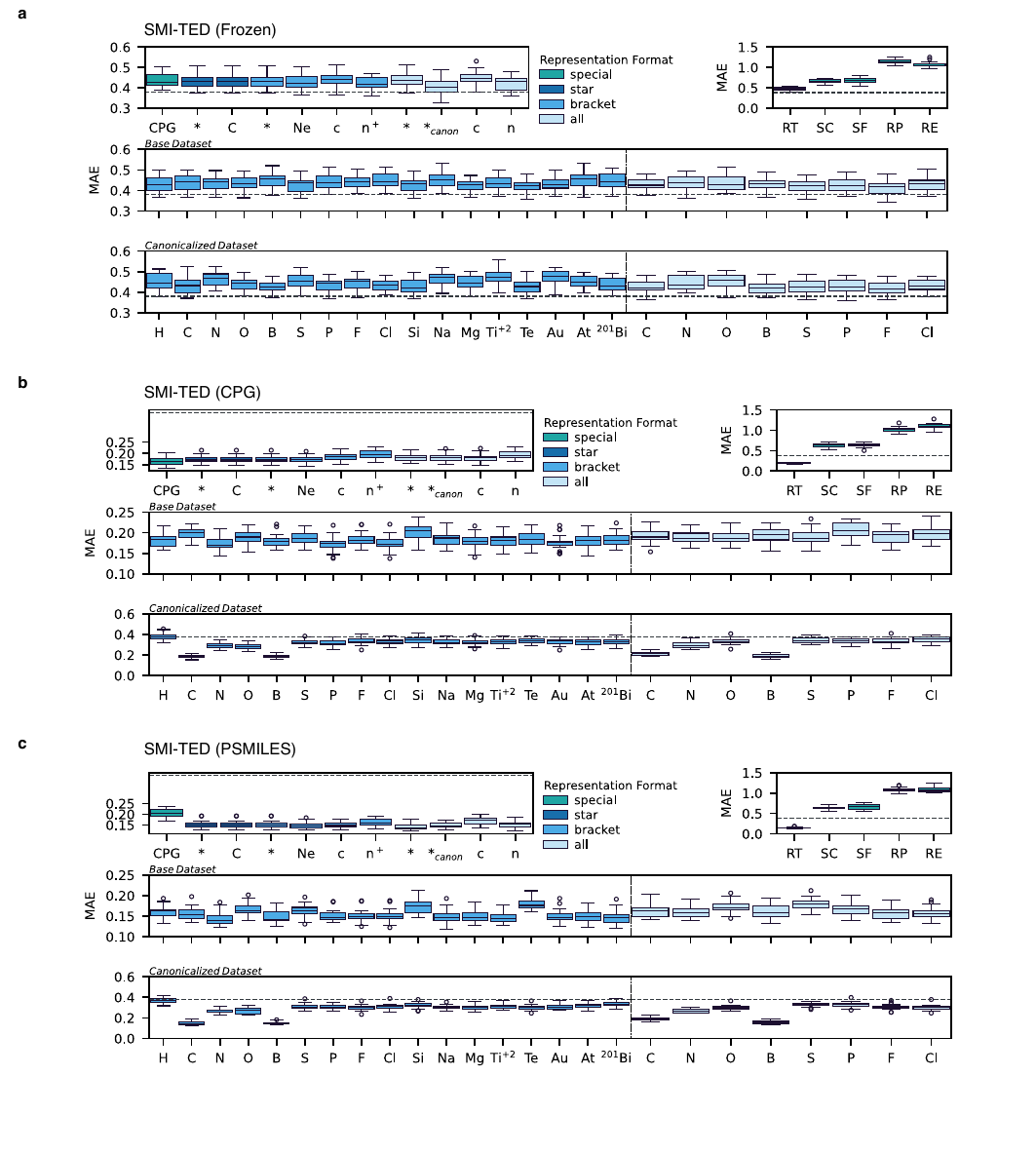}
\caption{MAE values from repeated 5-fold cross-validation experiments with the N\textsubscript{2} Gas Permeability dataset using SMI-TED models with frozen weights (\textbf{a}) or fine-tuned with CPG (\textbf{b}) and PSMILES (\textbf{c}).}\label{netl_n2_st_mae}
\end{figure}

\begin{figure}[H]
\centering
\includegraphics[width=1\textwidth]{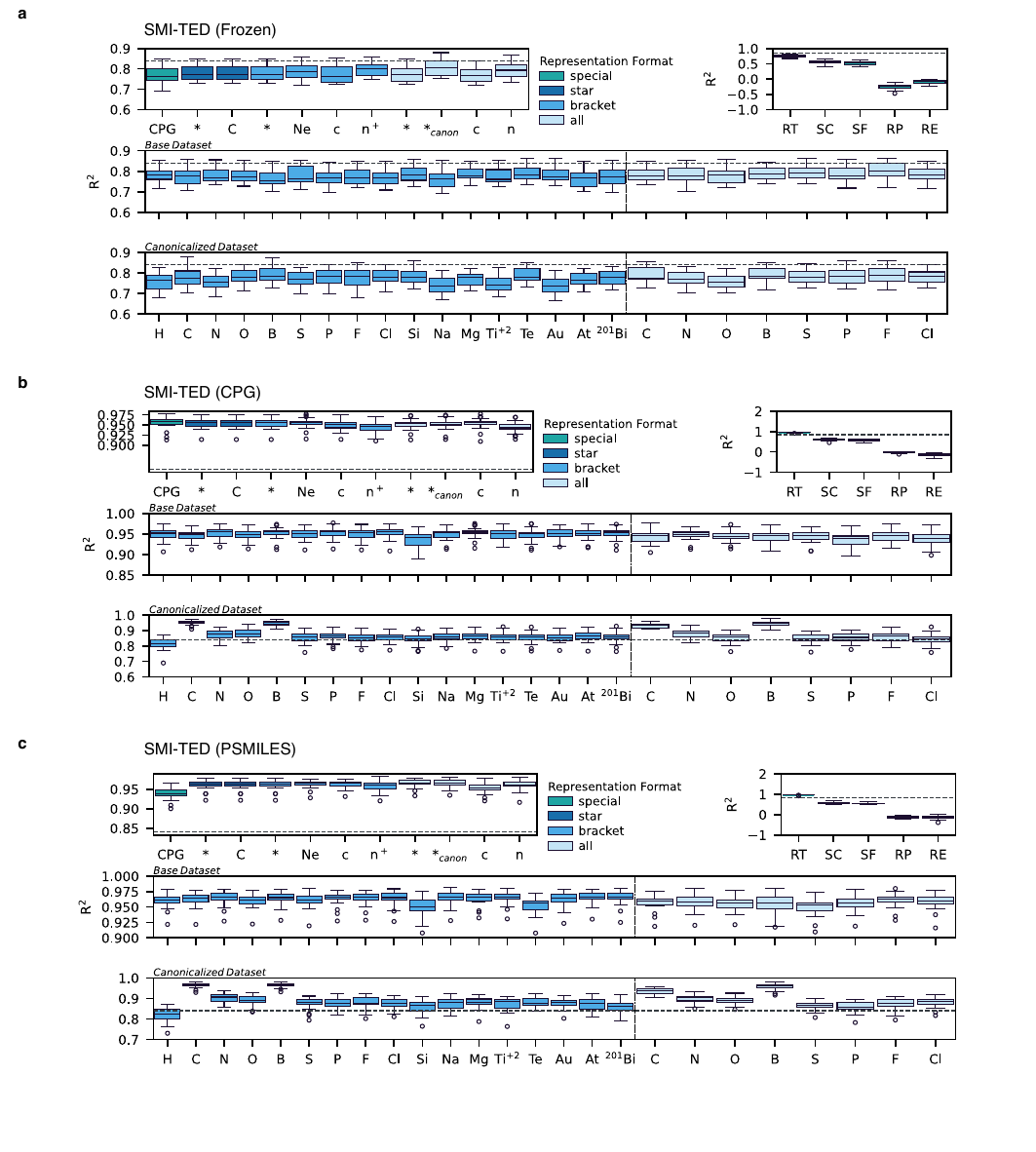}
\caption{R\textsuperscript{2} values from repeated 5-fold cross-validation experiments with the N\textsubscript{2} Gas Permeability dataset using SMI-TED models with frozen weights (\textbf{a}) or fine-tuned with CPG (\textbf{b}) and PSMILES (\textbf{c}).}\label{netl_n2_st_r2}
\end{figure}

\subsubsection{CH\textsubscript{4} Gas Permeability}\label{cv_netl_ch4}

\begin{figure}[H]
\centering
\includegraphics[width=1\textwidth]{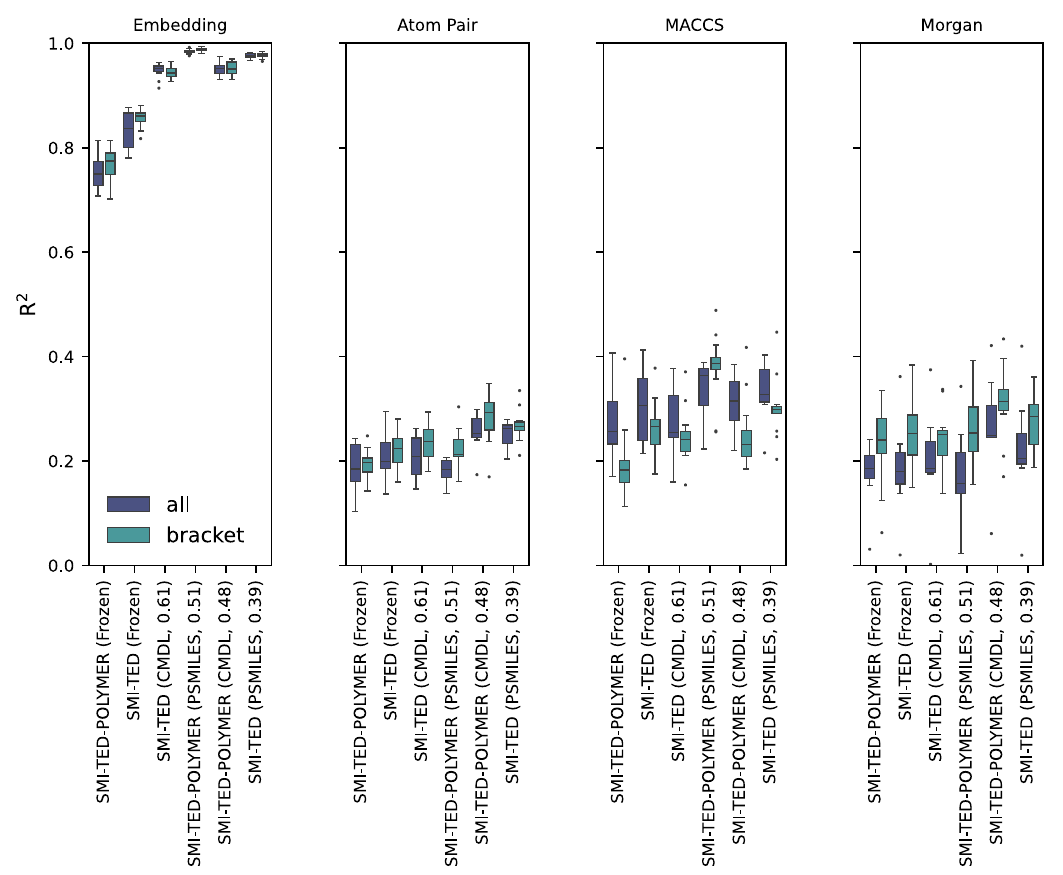}
\caption{Plots of embedding and fingerprint R\textsuperscript{2} value distributions for the CH\textsubscript{4} permeability benchmark dataset using the \textit{bracket} and \textit{all} substitution strategy.}\label{netl_ch4_knn_fp_all}
\end{figure}

\begin{figure}[H]
\centering
\includegraphics[width=1\textwidth]{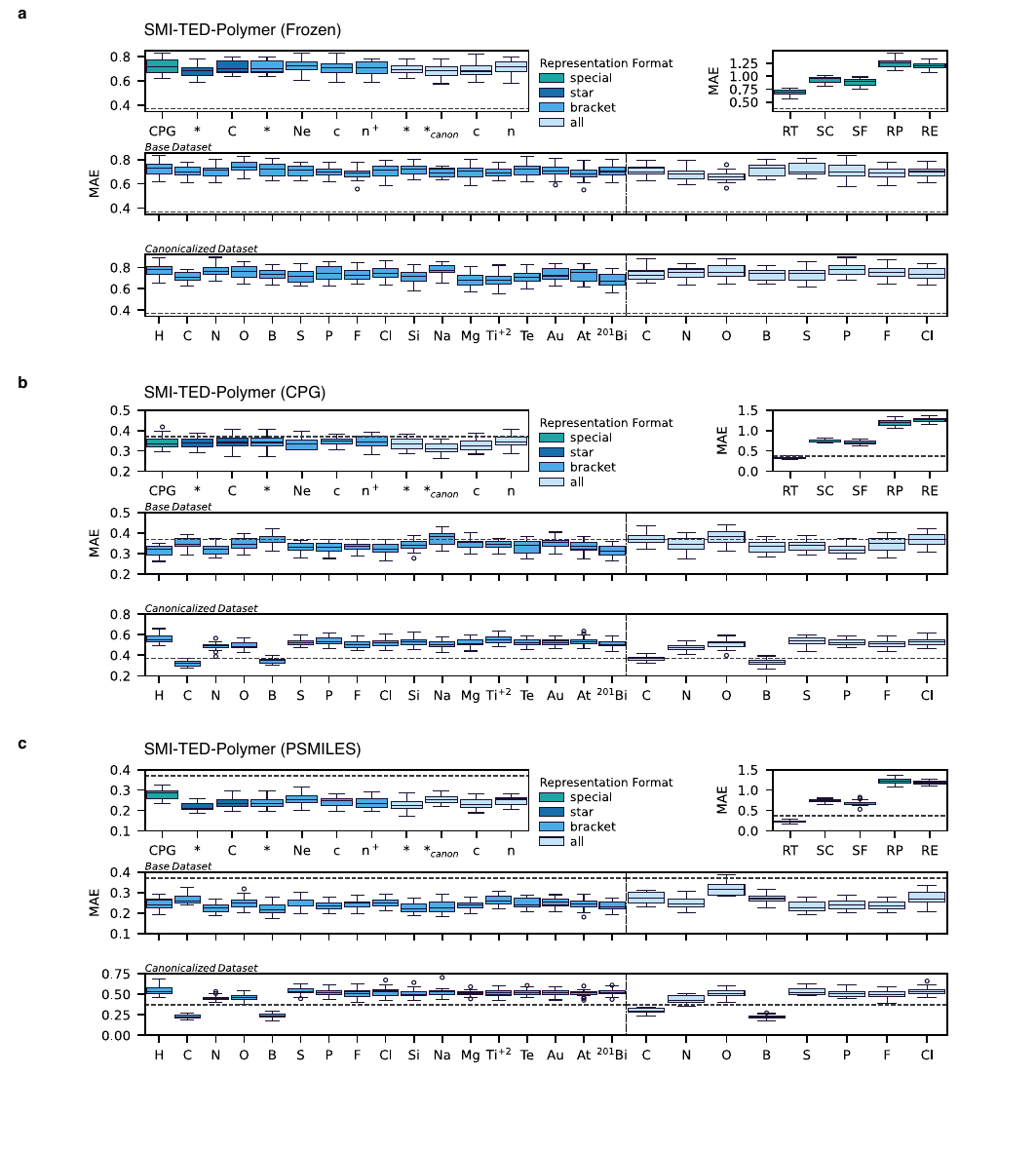}
\caption{MAE values from repeated 5-fold cross-validation experiments with the CH\textsubscript{4} Gas Permeability dataset using SMI-TED-Polymer models with frozen weights (\textbf{a}) or fine-tuned with CPG (\textbf{b}) and PSMILES (\textbf{c}).}\label{netl_ch4_stp_mae}
\end{figure}

\begin{figure}[H]
\centering
\includegraphics[width=1\textwidth]{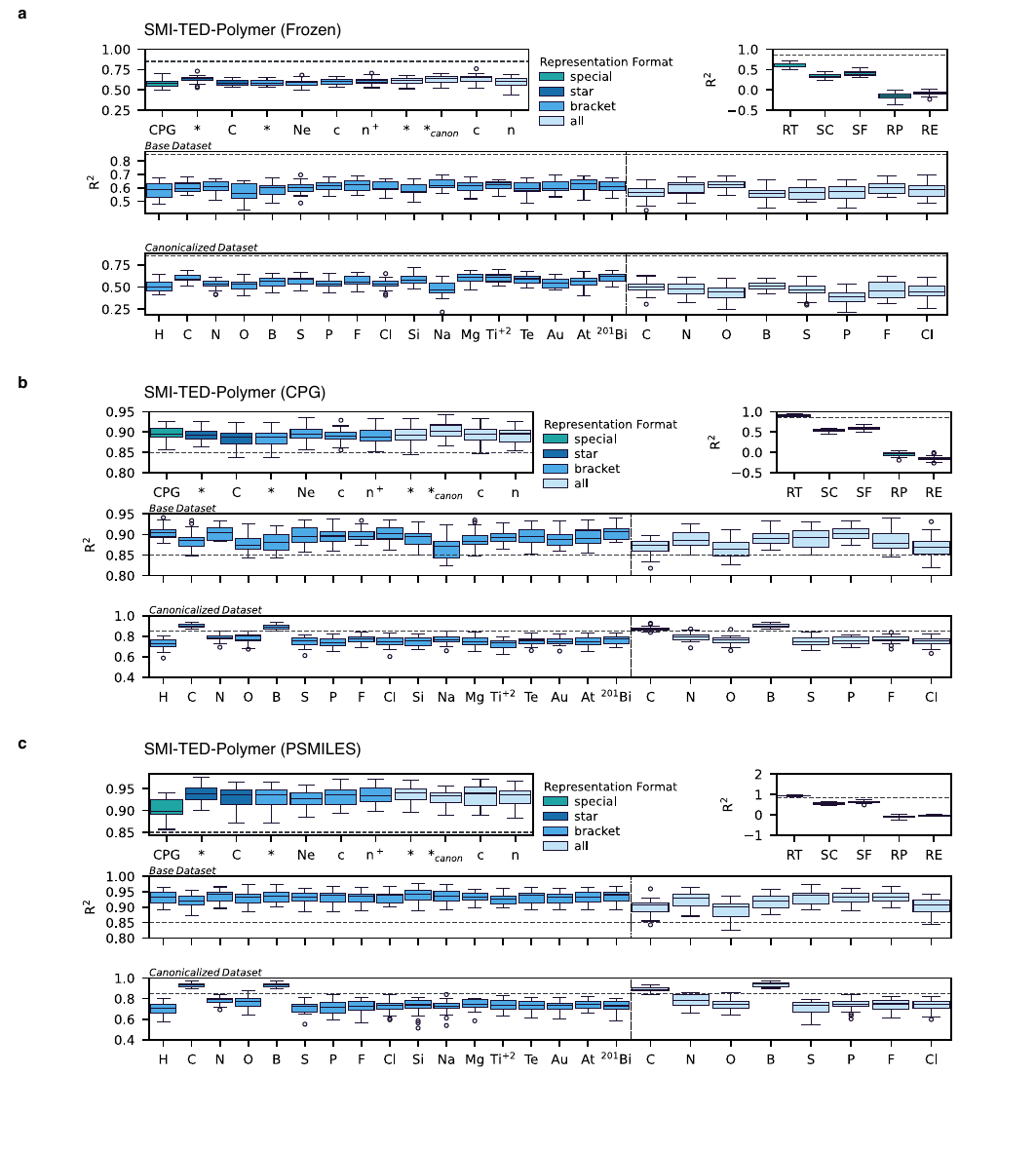}
\caption{R\textsuperscript{2} values from repeated 5-fold cross-validation experiments with the CH\textsubscript{4} Gas Permeability dataset using SMI-TED-Polymer models with frozen weights (\textbf{a}) or fine-tuned with CPG (\textbf{b}) and PSMILES (\textbf{c}).}\label{netl_ch4_stp_r2}
\end{figure}

\begin{figure}[H]
\centering
\includegraphics[width=1\textwidth]{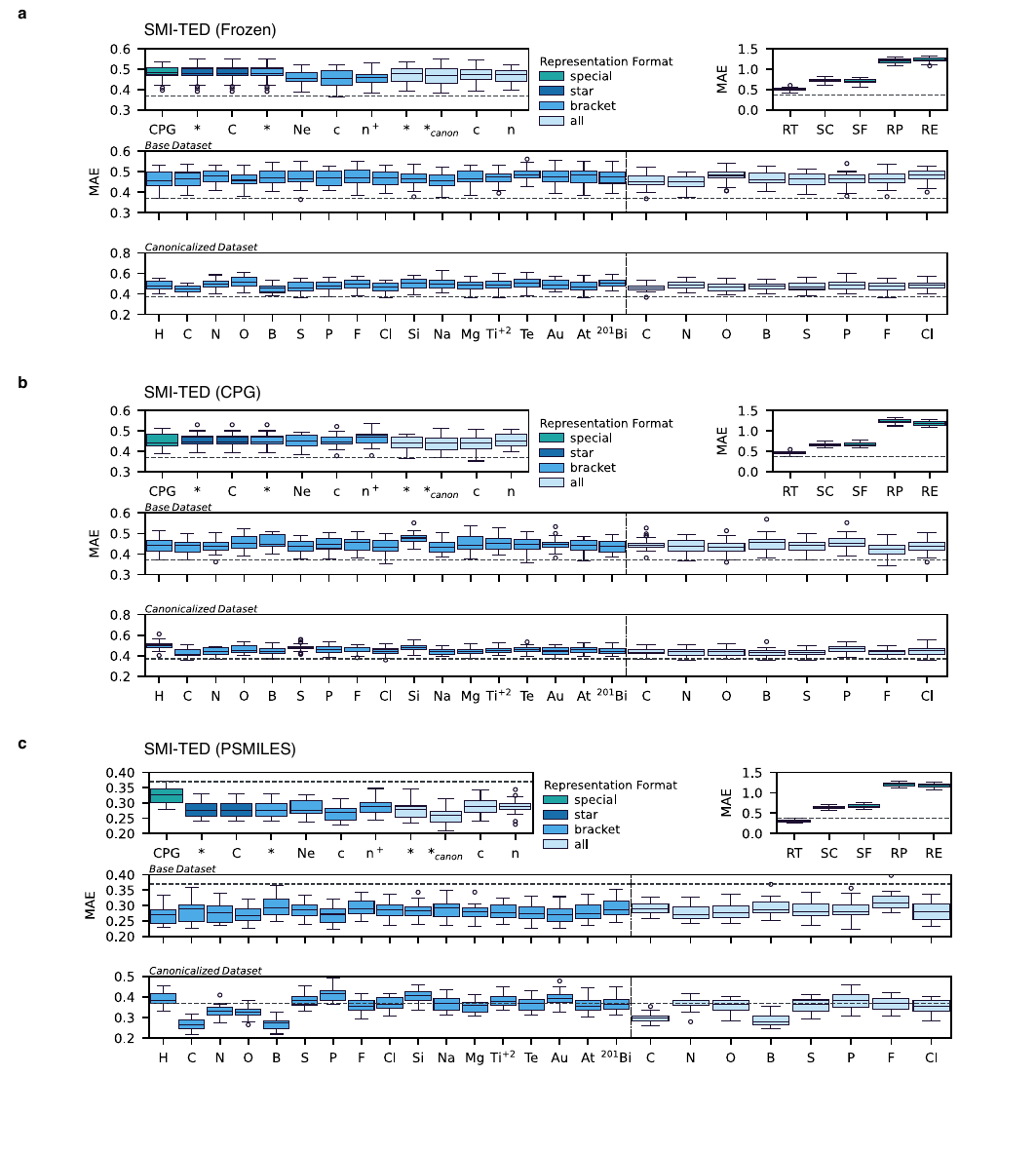}
\caption{MAE values from repeated 5-fold cross-validation experiments with the CH\textsubscript{4} Gas Permeability dataset using SMI-TED models with frozen weights (\textbf{a}) or fine-tuned with CPG (\textbf{b}) and PSMILES (\textbf{c}).}\label{netl_ch4_st_mae}
\end{figure}

\begin{figure}[H]
\centering
\includegraphics[width=1\textwidth]{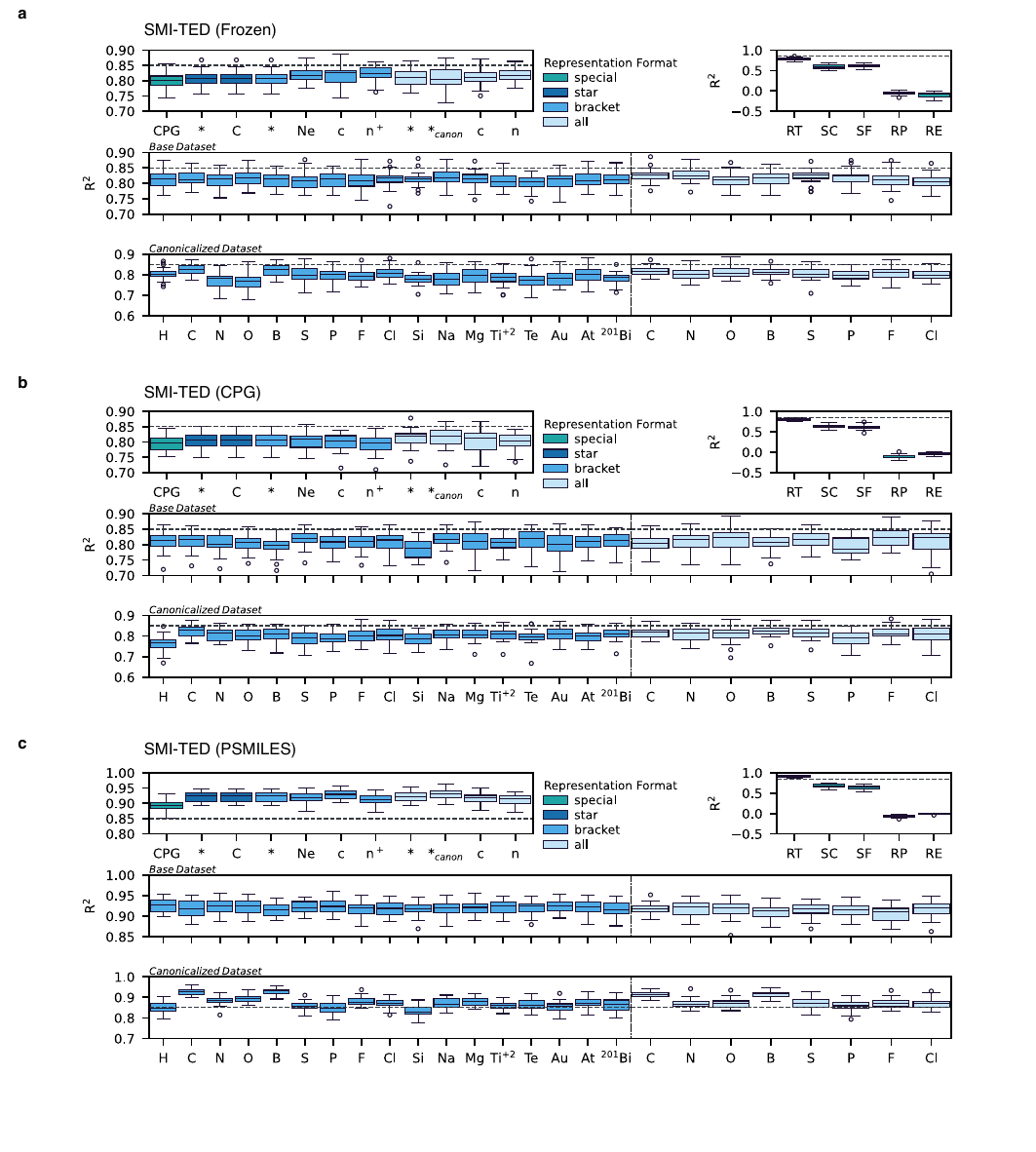}
\caption{R\textsuperscript{2} values from repeated 5-fold cross-validation experiments with the CH\textsubscript{4} Gas Permeability dataset using SMI-TED models with frozen weights (\textbf{a}) or fine-tuned with CPG (\textbf{b}) and PSMILES (\textbf{c}).}\label{netl_ch4_st_r2}
\end{figure}

\subsubsection{CO\textsubscript{2}:CH\textsubscript{4} Gas Selectivity}\label{cv_netl_co2_ch4}

\begin{figure}[H]
\centering
\includegraphics[width=1\textwidth]{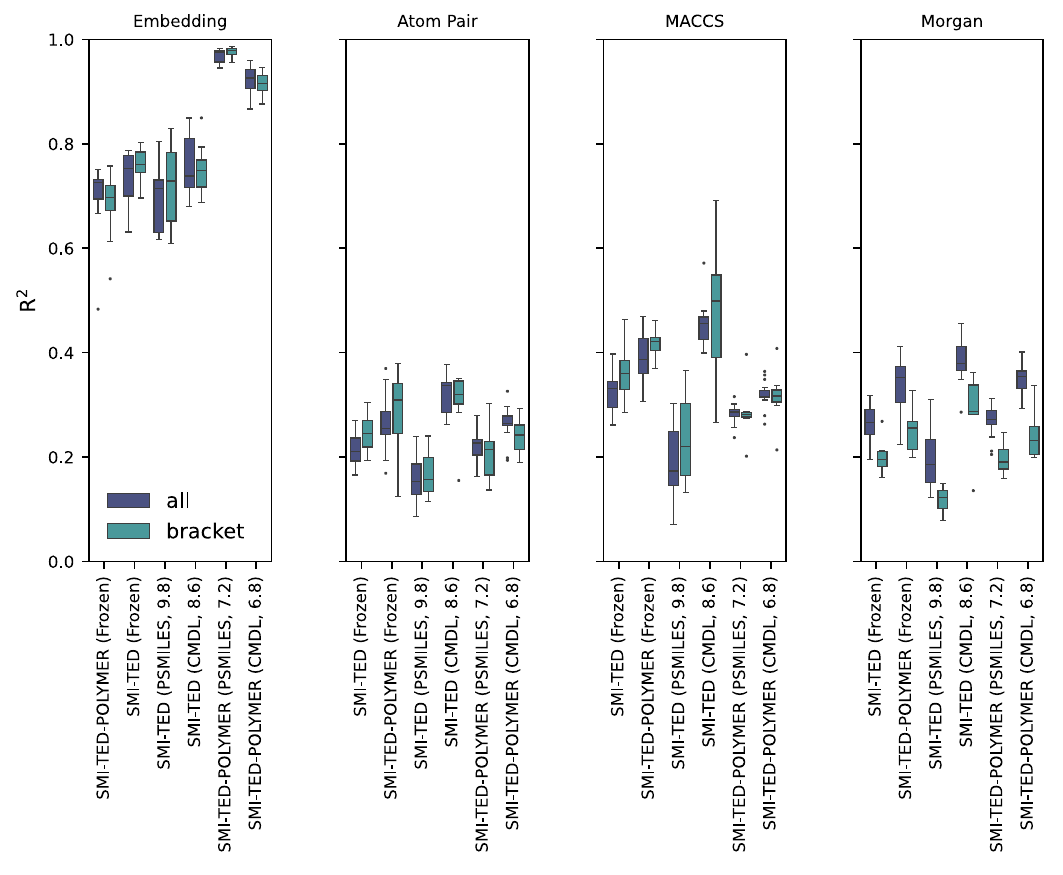}
\caption{Plots of embedding and fingerprint R\textsuperscript{2} value distributions for the CO\textsubscript{2}:CH\textsubscript{4} selectivity benchmark dataset using the \textit{bracket} and \textit{all} substitution strategy.}\label{netl_co2_ch4_knn_fp_all}
\end{figure}

\begin{figure}[H]
\centering
\includegraphics[width=1\textwidth]{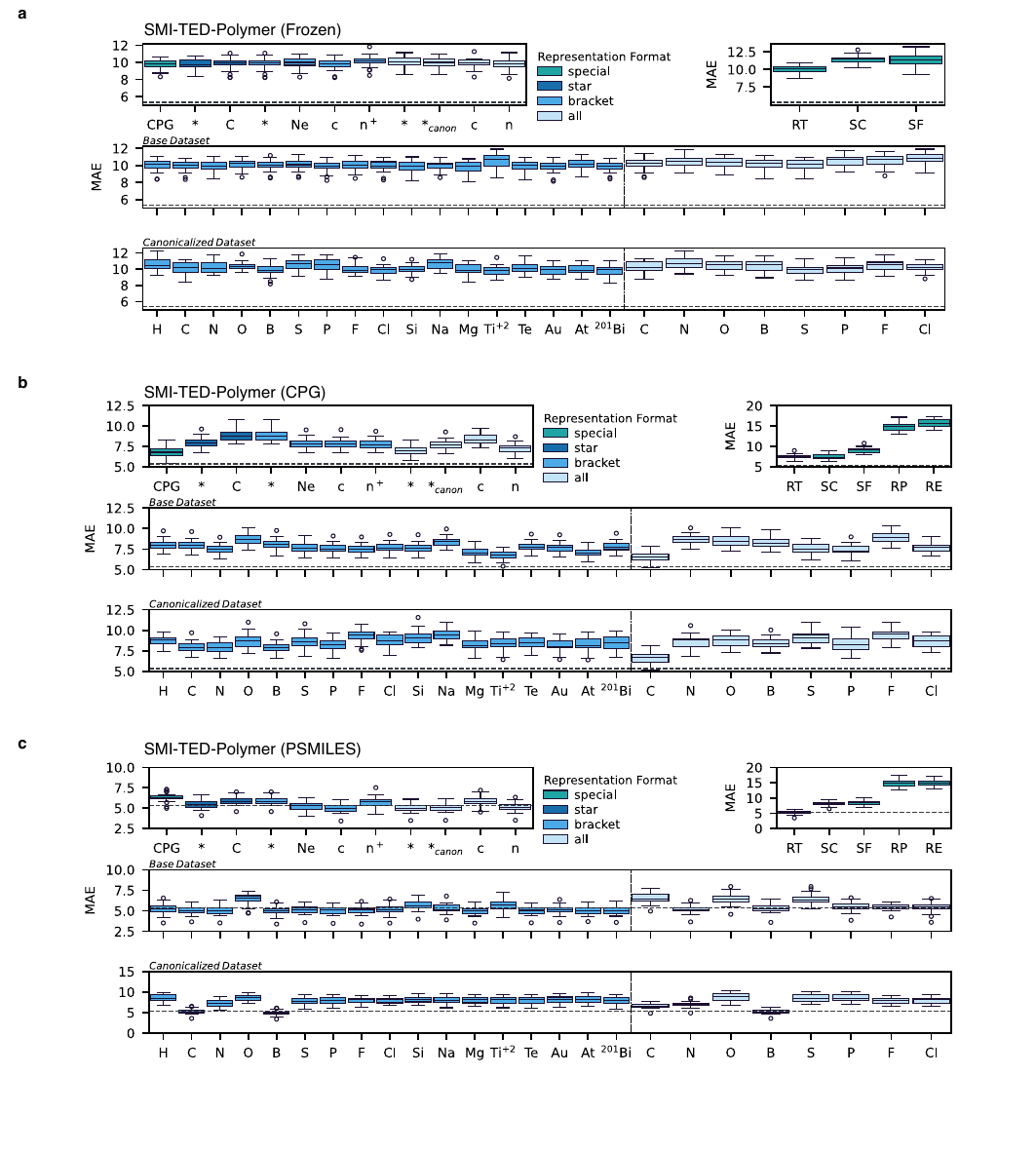}
\caption{MAE values from repeated 5-fold cross-validation experiments with the CO\textsubscript{2}:CH\textsubscript{4} Gas Selectivity dataset using SMI-TED-Polymer models with frozen weights (\textbf{a}) or fine-tuned with CPG (\textbf{b}) and PSMILES (\textbf{c}).}\label{netl_co2_ch4_stp_mae}
\end{figure}

\begin{figure}[H]
\centering
\includegraphics[width=1\textwidth]{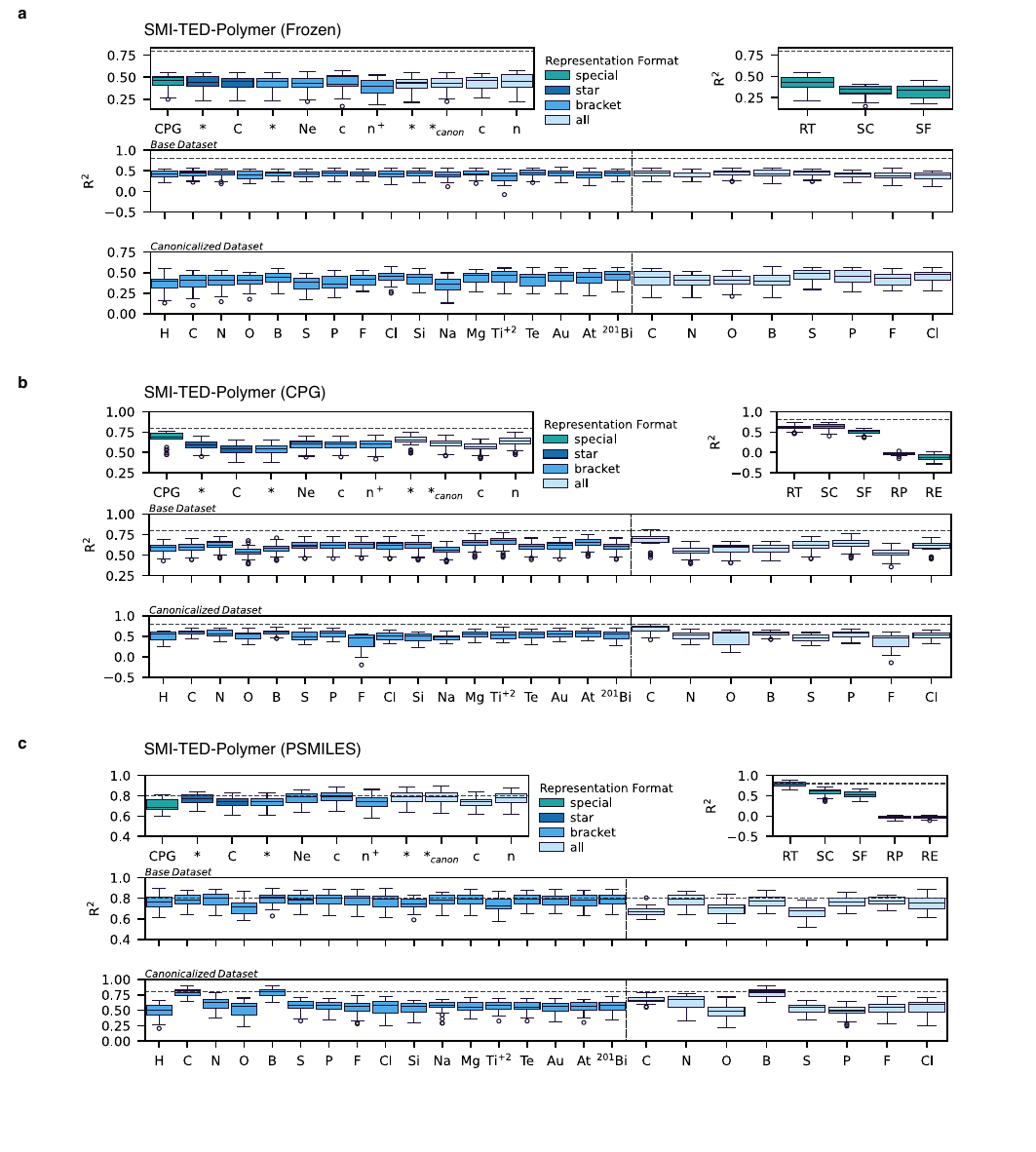}
\caption{R\textsuperscript{2} values from repeated 5-fold cross-validation experiments with the CO\textsubscript{2}:CH\textsubscript{4} Gas Selectivity dataset using SMI-TED-Polymer models with frozen weights (\textbf{a}) or fine-tuned with CPG (\textbf{b}) and PSMILES (\textbf{c}).}\label{netl_co2_ch4_stp_r2}
\end{figure}

\begin{figure}[H]
\centering
\includegraphics[width=1\textwidth]{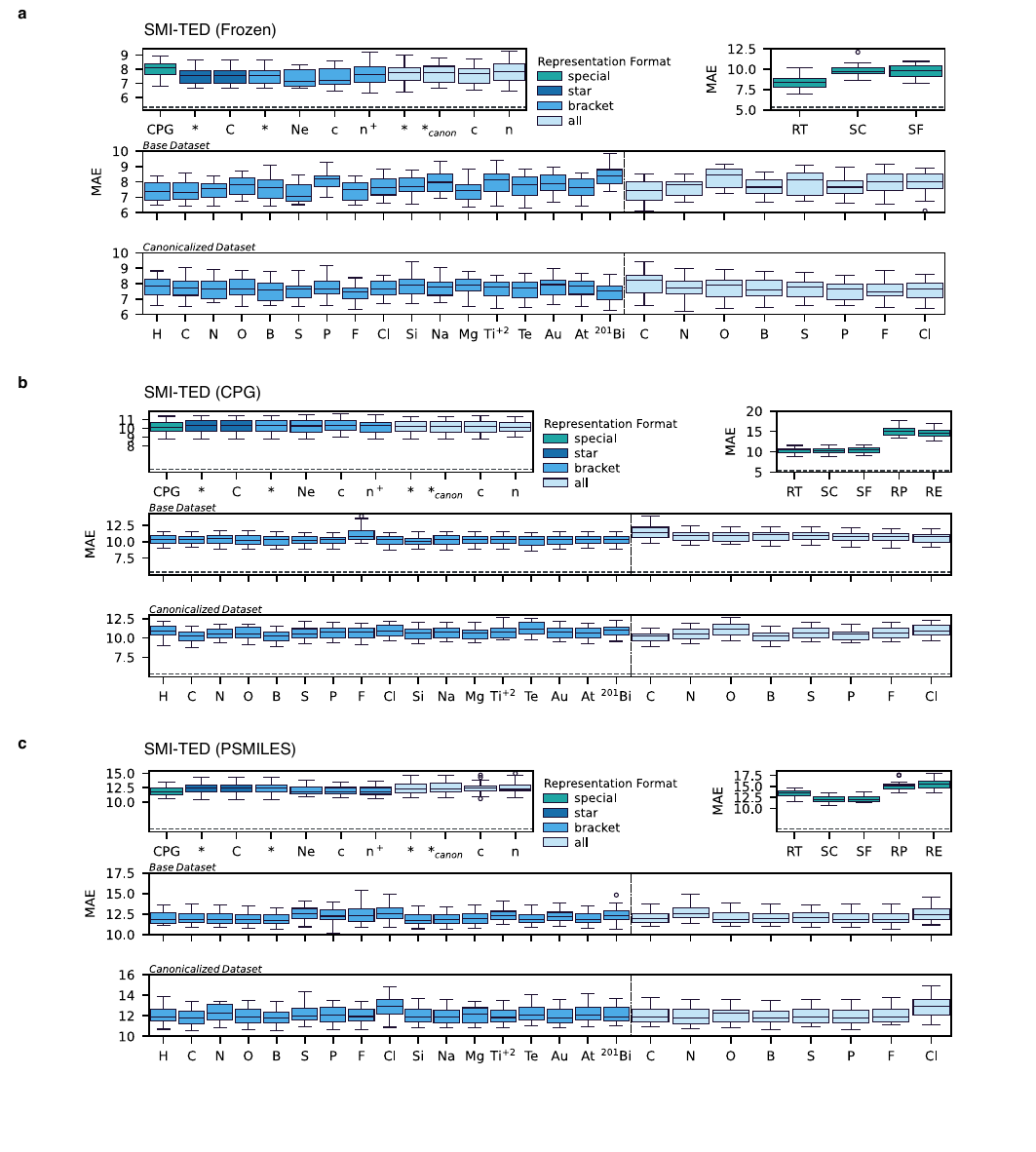}
\caption{MAE values from repeated 5-fold cross-validation experiments with the CO\textsubscript{2}:CH\textsubscript{4} Gas Selectivity dataset using SMI-TED models with frozen weights (\textbf{a}) or fine-tuned with CPG (\textbf{b}) and PSMILES (\textbf{c}).}\label{netl_co2_ch4_st_mae}
\end{figure}

\begin{figure}[H]
\centering
\includegraphics[width=1\textwidth]{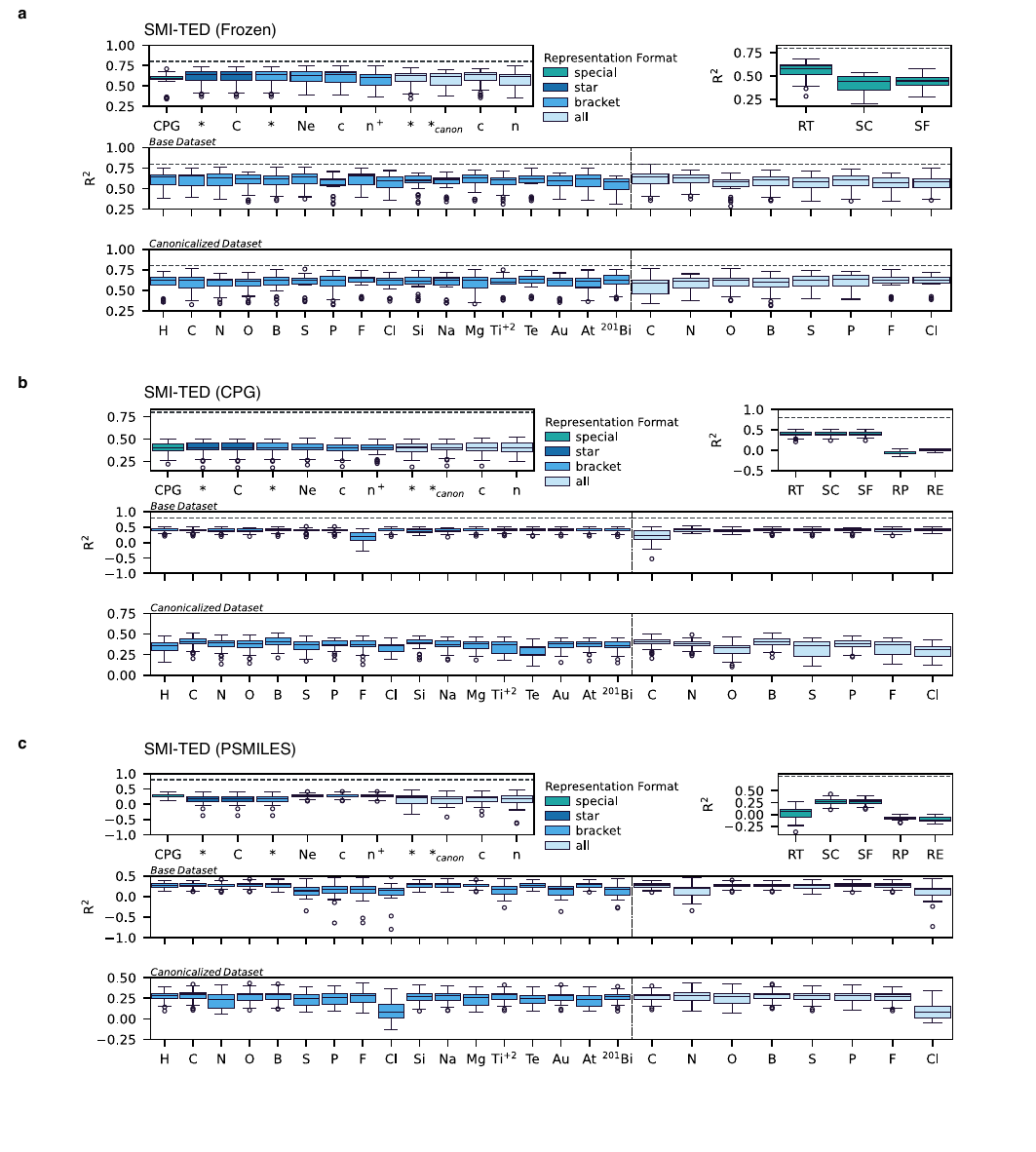}
\caption{R\textsuperscript{2} values from repeated 5-fold cross-validation experiments with the CO\textsubscript{2}:CH\textsubscript{4} Gas Selectivity dataset using SMI-TED models with frozen weights (\textbf{a}) or fine-tuned with CPG (\textbf{b}) and PSMILES (\textbf{c}).}\label{netl_co2_ch4_st_r2}
\end{figure}

\subsubsection{CO\textsubscript{2}:N\textsubscript{2} Gas Selectivity}\label{cv_netl_co2_n2}

\begin{figure}[H]
\centering
\includegraphics[width=1\textwidth]{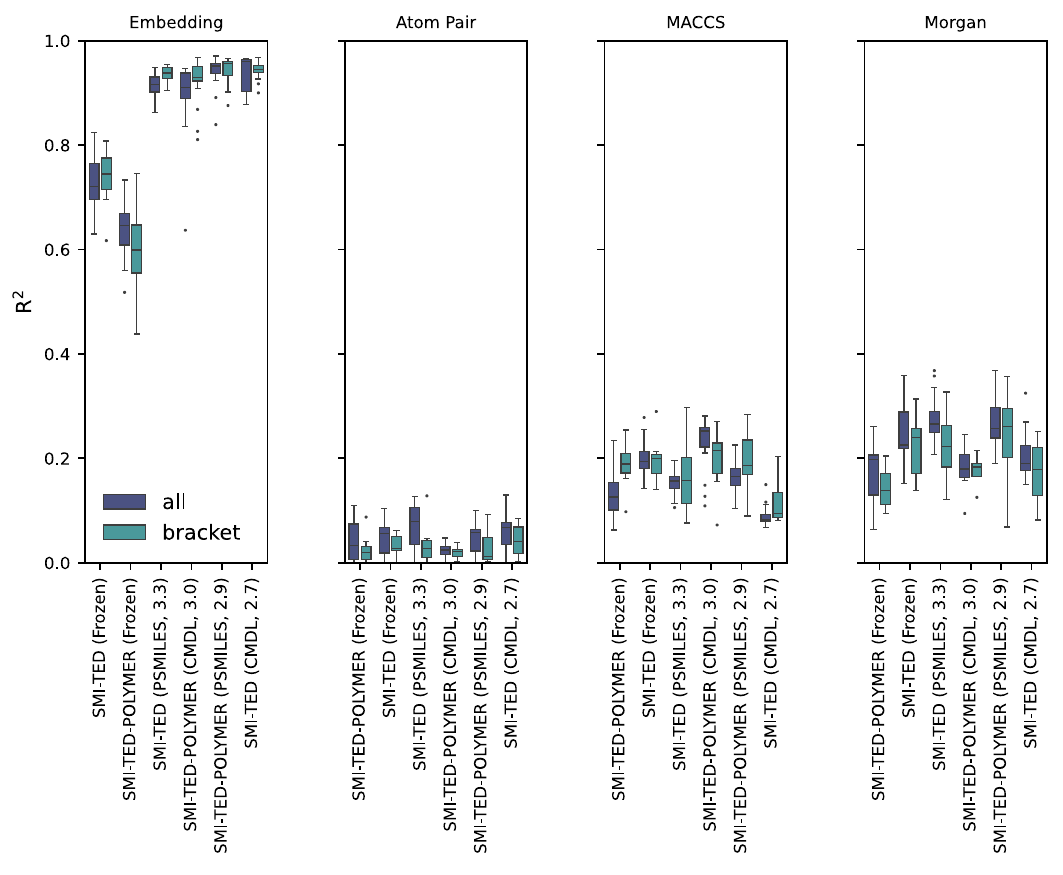}
\caption{Plots of embedding and fingerprint R\textsuperscript{2} value distributions for the CO\textsubscript{2}:N\textsubscript{2} selectivity benchmark dataset using the \textit{bracket} and \textit{all} substitution strategy.}\label{netl_co2_n2_knn_fp_all}
\end{figure}

\begin{figure}[H]
\centering
\includegraphics[width=1\textwidth]{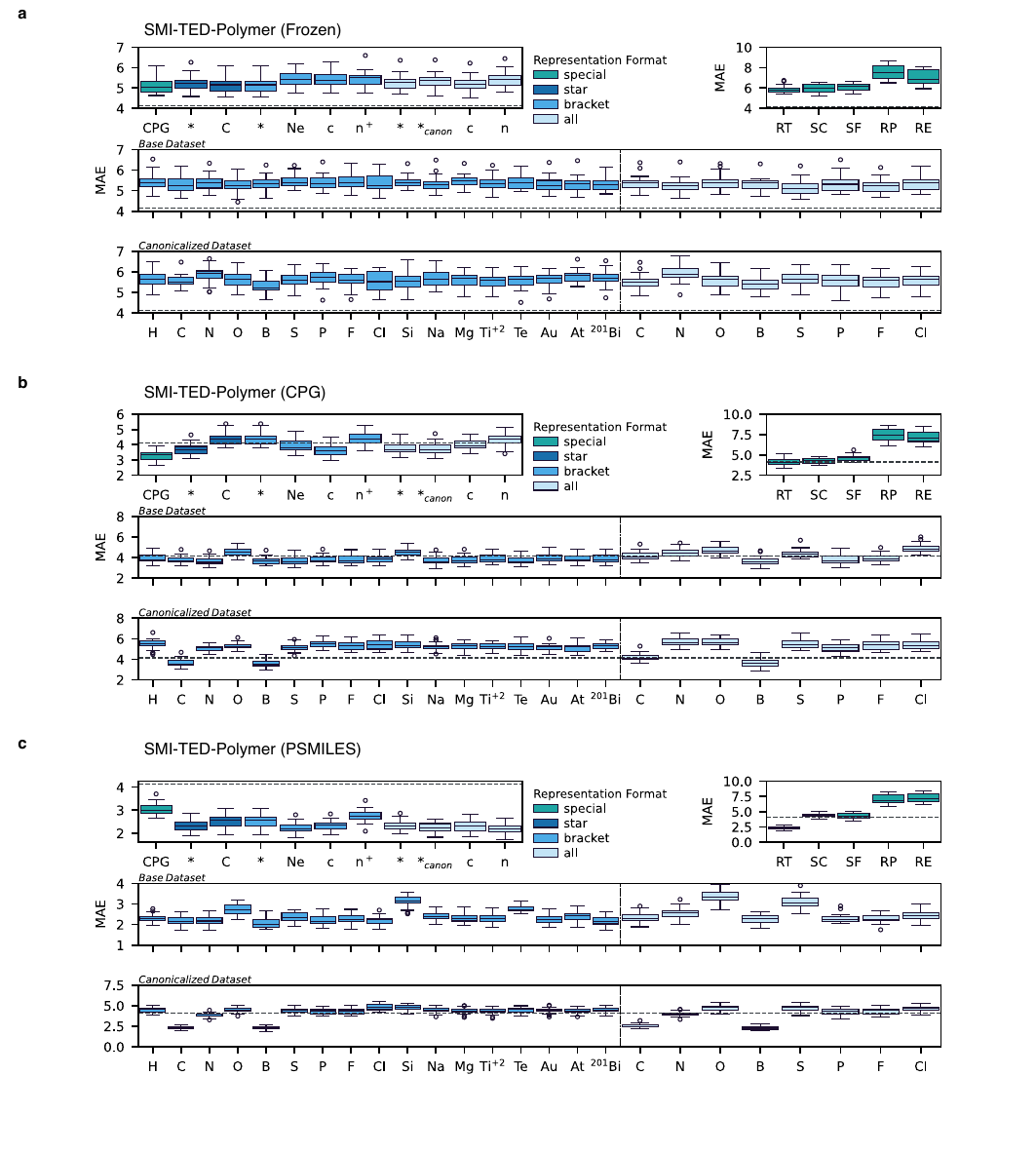}
\caption{MAE values from repeated 5-fold cross-validation experiments with the CO\textsubscript{2}:N\textsubscript{2} Gas Selectivity dataset using SMI-TED-Polymer models with frozen weights (\textbf{a}) or fine-tuned with CPG (\textbf{b}) and PSMILES (\textbf{c}).}\label{netl_co2_n2_stp_mae}
\end{figure}

\begin{figure}[H]
\centering
\includegraphics[width=1\textwidth]{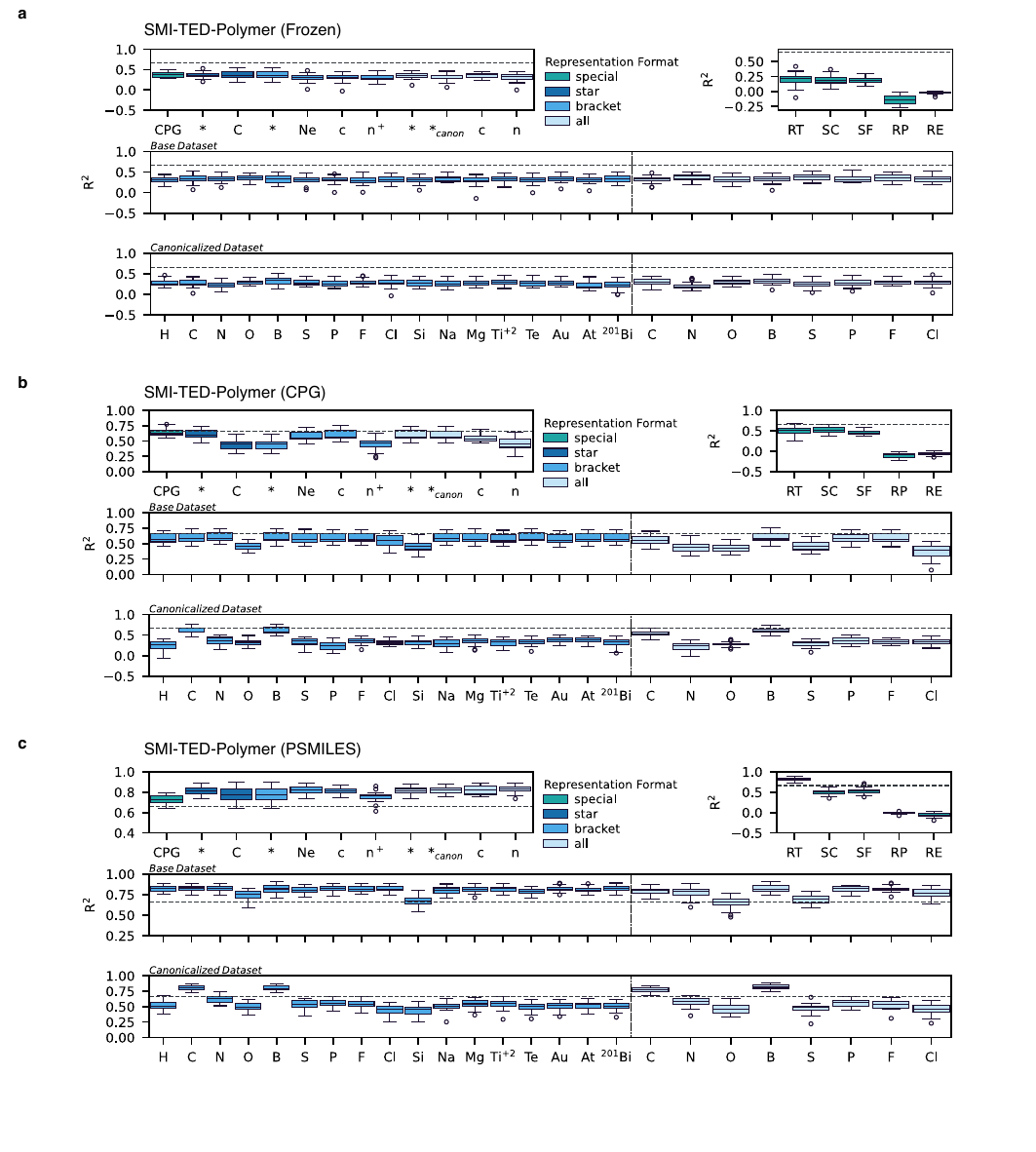}
\caption{R\textsuperscript{2} values from repeated 5-fold cross-validation experiments with the CO\textsubscript{2}:N\textsubscript{2} Gas Selectivity dataset using SMI-TED-Polymer models with frozen weights (\textbf{a}) or fine-tuned with CPG (\textbf{b}) and PSMILES (\textbf{c}).}\label{netl_co2_n2_stp_r2}
\end{figure}

\begin{figure}[H]
\centering
\includegraphics[width=1\textwidth]{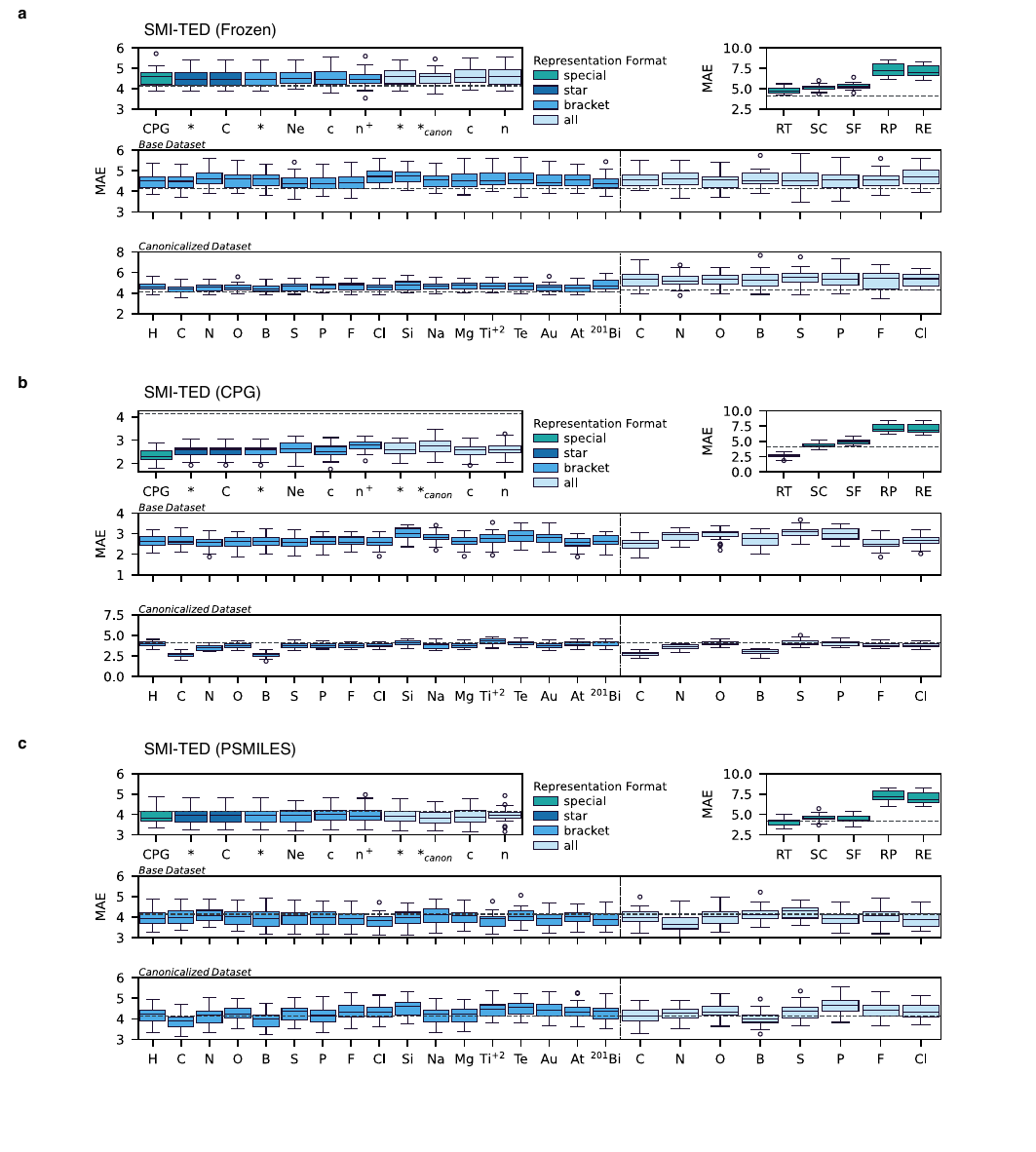}
\caption{MAE values from repeated 5-fold cross-validation experiments with the CO\textsubscript{2}:N\textsubscript{2} Gas Selectivity dataset using SMI-TED models with frozen weights (\textbf{a}) or fine-tuned with CPG (\textbf{b}) and PSMILES (\textbf{c}).}\label{netl_co2_n2_st_mae}
\end{figure}

\begin{figure}[H]
\centering
\includegraphics[width=1\textwidth]{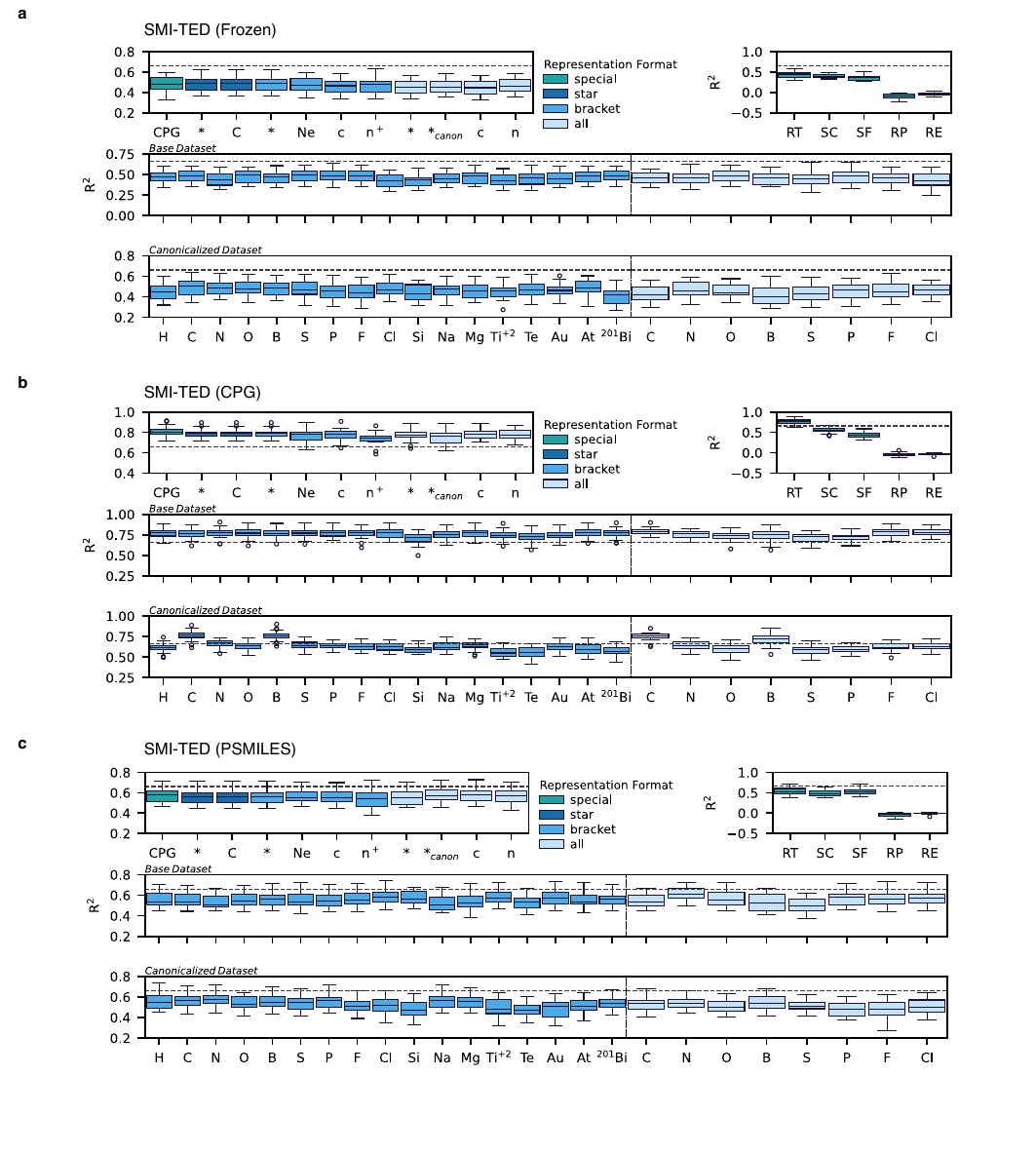}
\caption{R\textsuperscript{2} values from repeated 5-fold cross-validation experiments with the CO\textsubscript{2}:N\textsubscript{2} Gas Selectivity dataset using SMI-TED models with frozen weights (\textbf{a}) or fine-tuned with CPG (\textbf{b}) and PSMILES (\textbf{c}).}\label{netl_co2_n2_st_r2}
\end{figure}